# Subjective Experience in AI Systems: What Do AI Researchers and the Public Believe?

Noemi Dreksler[1,*], Lucius Caviola[2], David Chalmers[3], Carter Allen[4], Alex Rand[5], Joshua Lewis[3], Philip Waggoner[6], Kate Mays[7], and Jeff Sebo[3]

[1] Centre for the Governance of AI, London, United Kingdom

[2] University of Cambridge, Cambridge, United Kingdom

[3] New York University, New York, United States

[4] University of California, Berkeley, Berkeley, United States

[5] Northwestern University, Evanston, United States

[6] Yale University, New Haven, United States

[7] University of Vermont, Burlington, United States

*Corresponding author: noemi.dreksler@governance.ai

## Abstract

Whether AI systems are perceived to have subjective experience has scientific, ethical, and governance implications. Yet little is known about how AI researchers and the public view this possibility. We surveyed 582 AI researchers who published in leading AI venues and 825 nationally representative US adults about the likelihood, identification, and governance of AI systems with subjective experience. We find that both AI researchers and the public think it is unlikely that AI systems have subjective experience now, but believe it is more likely than not they will by century's end. Both groups believe multidisciplinary expertise will be important in identifying AI subjective experience, support developers implementing safeguards now, and agree that, if created, subjectively experiencing AI systems should have moral responsibilities and be held accountable. Both groups also show substantial disagreement on many moral and policy questions. Our results underline the importance of building more detailed, evidence-based understanding of public and expert opinion on this contentious issue.

## Introduction

In 2022, a Google engineer claimed that a Large Language Model had developed subjective experience, a statement that drew a range of media attention: some dismissed the claim as premature (Christian, 2022; Johnson, 2022), and others criticized it as a distraction from more pressing AI risks (Véliz, 2022). Other researchers at leading AI labs have also raised the possibility or voiced concern about the possibility that advanced AI systems could one day develop subjective experience (Agüera y Arcas, 2022; Bowman, 2024; Byrne, 2024; Hashim, 2024; Huckins, 2023; Sutskever, 2022), with at least one leading AI company beginning to assess potential AI welfare concerns (Anthropic, 2025a; Anthropic, 2025b). But disagreements persist: Microsoft's CEO of AI, for example, has decried AI welfare research as "premature, and frankly dangerous" (Suleyman, 2025). And despite growing interest and discussion on the topic, there remains uncertainty about how widely such beliefs are held and how they differ across groups.

Subjective experience, the capacity to feel and perceive the world from a first-person perspective, lies at the heart of concepts such as consciousness and sentience (Nagel, 1974; Seth & Bayne, 2022) and is widely regarded as a key criterion for moral consideration (Sebo & Long, 2025). Importantly, both the under- and over-attribution of subjective experience to AI systems could pose risks, including misplaced attachment or dismissal of AI systems, and distortions in how resources, responsibilities, and safeguards are allocated (Butlin et al., 2023; Caviola, 2025; Fernandez et al., 2024; Schwitzgebel, 2023; Suleyman, 2025).

Recent surveys have begun to explore public perceptions of AI subjective experience (Anthis et al., 2024; Ladak & Caviola, 2025; Colombatto & Fleming, 2024; Martínez & Winter, 2021) and found that people are skeptical that current AI systems have subjective experience but do not rule out the possibility for the future. As AI use increases and AI systems become more advanced and human-like in their behavior and capabilities, people may become increasingly inclined to attribute mental states to them, including subjective experience (Mei et al., 2024; Caviola, 2025; Caviola et al., 2025; Broadbent et al., 2013; Colombatto & Fleming, 2024; Sato et al., 2024). And a growing body of psychological research shows that laypeople's judgments about an entity's moral status are strongly influenced by whether they believe an entity has subjective experiences (Gray et al., 2012; Goodwin, 2015; Shank & DeSanti, 2018). This holds true for both humans and nonhuman entities, including animals and even robots (Oksanen et al., 2020; Schepman & Rodway, 2020; Suzuki et al., 2015).

Yet determining whether an AI system actually possesses subjective experience is notoriously difficult (Avramides, 2023; Chalmers, 1995, 2018; Gunkel, 2020; Harnad, 1991). Some researchers have begun proposing technical indicators that might signal the emergence of such capacities (Butlin et al., 2023; Long et al., 2024; Birch, 2024). Still, experts disagree not only on whether and when AI systems might develop subjective experience, but also on how to define and detect subjective experience (Butlin et al., 2023; Dehaene et al., 2017; Searle, 1980; Thompson, 1965; Seth, 2025). Foundational debates in philosophy and cognitive science about the nature, scope, and detection of subjective experience also remain unresolved (Killheffer, 1993; Lenharo, 2023, 2024; Snaprud, 2018; Block, 1995; Gunkel, 2020; Cogitate Consortium et al., 2023; Seth & Bayne, 2022). Moreover, even if experts were able to reliably assess the mental capacities of AI systems, AI researchers' and the public's opinion might diverge from expert consensus.

Resolving these debates will be critical for future AI governance since perceptions of subjective experience in AI systems may increase as systems grow more capable and human-like, and many ethical frameworks treat subjective experience as a necessary condition for moral status. Some researchers argue that if there is at least a realistic possibility that an AI system has subjective experience, then we should consider and mitigate welfare risks for that system as a precautionary measure (Long et al., 2024; Metzinger, 2021; Birch, 2024). However, the question of whether and how to attribute moral, legal, or political status to AI systems – whether because they do or do not have mental capacities, such as subjective experience and agency, or because of other interpersonal and societal adverse effects – has long been fraught with division among moral and legal theorists (Asaro, 2007; Birhane & Van Dijk, 2020; Bryson, 2010; Chopra & White, 2011; Coeckelbergh, 2010; Danaher, 2020; Darling, 2012; Delcker, 2018; Friedman, 2023; Gordon & Pasvenskiene, 2021; Guingrich & Graziano, 2024; Gunkel, 2012, 2018; Kurki, 2019; Mamak, 2022, 2023; Miller, 2015; Müller, 2021; Schwitzgebel & Garza, 2015; Sebo & Long, 2025; Shevlin, 2021; Solaiman, 2017).

These epistemic uncertainties and divisions, underlined by the findings of the present survey, suggest that disagreements about subjective experience in AI systems will likely persist among experts, AI researchers, and the public. Policymakers, developers, researchers, and civil society actors will need to navigate both scientific uncertainty and societal disagreement, especially as AI systems increasingly appear to have subjective experience, regardless of whether they truly do. Addressing these challenges requires a clearer understanding of the beliefs, values, and intuitions that different stakeholders bring to the issue.

However, despite some existing research (see SI 6 for a literature review), we still know little about how AI researchers think about subjective experience in AI systems, yet their views may be especially influential in the eyes of the public and policymakers. We also lack systematic evidence on whether, and how, AI researchers' views differ from those of the general public. Moreover, most prior psychological research predates recent advances in AI capabilities and the surge in public attention, and as AI continues to evolve, beliefs about the possibility of subjective experience in AI may evolve as well.

*The present study*

To address these gaps, we conducted a large-scale survey in May 2024 with two distinct samples: a nationally representative group of US adults ($N$ = 825) and a sample of AI researchers who had published at major AI conferences ($N$ = 582). Both groups responded to the same set of questions concerning AI subjective experience, including its moral and legal implications (see SI 1 for all top-line and summary results, SI 2 for the survey instrument).

We focused on three central themes: first, forecasts about whether and when AI systems will develop subjective experience; second, assessments of who should determine whether AI systems have subjective experience and how they should do so; and third, ethics and governance regarding how we should treat and govern AI systems that either do or might have subjective experience. Together, these questions aim to inform ongoing debates about the ethical and policy challenges posed by increasingly advanced AI systems, especially those perceived as having subjective experience.

In the survey, we defined subjective experience as "the ability to experience the world from a single point of view, including experiences such as perceiving (e.g., visual or auditory experiences) and feeling (e.g., the experience of pleasure and pain)." By this definition, an AI system with subjective experience would have internal states such that there is "something it is like" to be that system. In the survey, therefore, we used the term "subjective experience" as an umbrella concept encompassing common philosophical definitions of consciousness (the capacity for subjective experience) and sentience (the capacity for subjective experience with positive or negative valence).

## Results

### *Probabilistic forecasts for AI subjective experience*

Respondents were surveyed about their forecasts for when an AI system with subjective experience will exist using one of two framings, fixing either the date or the probability in the question (see Figure 1a). In the fixed date framing, the median probability reported by AI researchers that AI systems possessed subjective experience at the time of the survey in 2024 was 1% ($M$ = 12.47%, $SD$ = 22.39%), while the median rose to 25% by 2034 ($M$ = 34.02%, $SD$ = 31.70%) and 70% by 2100 ($M$ = 61.12%, $SD$ = 34.47%). In the public sample, the median probability that AI systems currently possess subjective experience was 5% ($M$ = 16.26%, $SD$ = 23.70%), while the median rose to 30% by 2034 ($M$ = 36.15%, $SD$ = 31.55%) and 60% by 2100 ($M$ = 57.55%, $SD$ = 36.48%).

In the fixed probability framing, at a 10% likelihood, both groups gave a median year of 2030 (AI researchers: $M$ = 2059, $SD$ = 156; Public: $M$ = 2108, $SD$ = 331). At a 50% likelihood, the median was 2050 for both groups (AI researchers: $M$ = 2101, $SD$ = 207; Public: $M$ = 2179, $SD$ = 463). At a 90% likelihood, the median was 2100 for AI researchers ($M$ = 2254, $SD$ = 515) and 2075 for the public ($M$ = 2282, $SD$ = 597). Note that in the fixed probability task respondents could not select "never", which may have led to earlier and therefore less reliable forecasts under this framing. The findings suggest that both samples expect the probability of AI subjective experience to increase across the century.

To compare groups across framings, we aggregated across forecasts by using a sequential procedure to find a continuous probability distribution that accurately matched the specific years and probabilities each respondent provided (see SI 3). A Wilcoxon rank sum test found no significant difference between AI researchers' and the public's median forecasts of AI subjective experience ($W$ = 155597.5, $p$ = .383, Figure 1b) nor in median forecasts between question framings ($W$ = 154202, $p$ = .882).

However, AI researchers were significantly less likely to believe that AI systems will never have subjective experience ($Mdn$ = 10% chance, $M$ = 28.67%) than the public was ($Mdn$ = 25%, $M$ = 37.07%, $t(1184.79) = -4.27$, $p < .001$, $p_{HB}$ (Holm-Bonferroni corrected) < .001) (see Figure 1c; between-group t-test results for all variables can be found in SI 4.1). On average, the public ($Mdn$ = 52%, $M$ = 52.92%, $SD$ = 28.24%) and AI researchers ($Mdn$ = 51%, $M$ = 52.15%, $SD$ = 27.01%) were "somewhat confident" (50%) about their forecasts. When looking at median estimates, both groups tended to predict earlier forecasts than the actual medians, but both groups accurately estimated their own group's timeline in the fixed probability framing (SI 4.3).

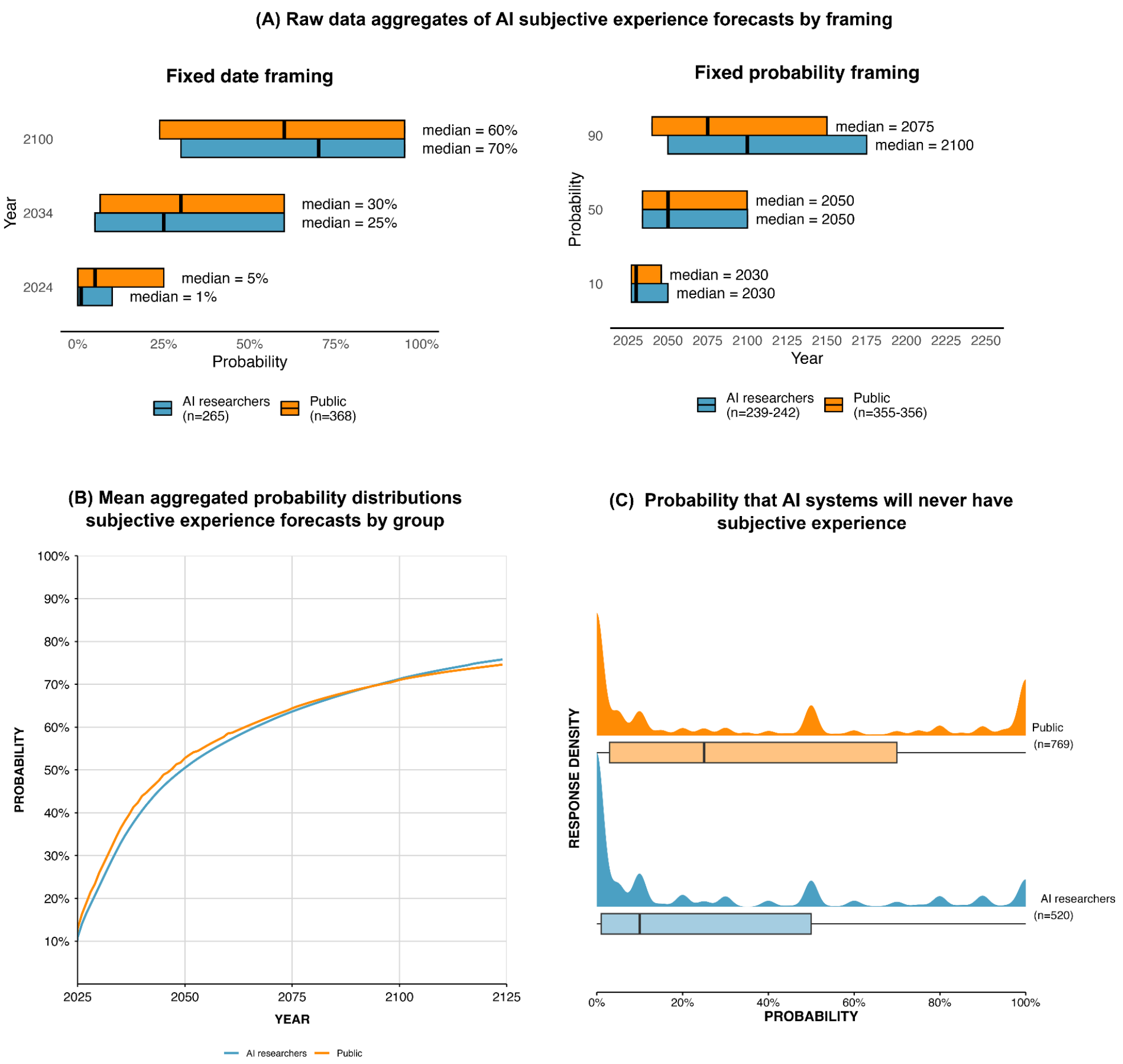

**Figure 1. Forecasts of AI researchers and the public for when AI systems will have subjective experiences.** (A) raw data forecasts made by framing (fixed date and fixed probability) for each sample, showing the first-quartile, median, and upper-quartile responses for each forecast (B) mean aggregated fitted distributions across framings for each group (see SI 3 for methodological details; SI 4.2 for results) (C) distribution of probabilities ascribed to the possibility that AI systems will never have subjective experience showing first quartile, median, upper quartile, and density curves of responses to the probability that respondents assigned to AI systems with subjective experience never existing, by group. The density curves represent the approximate frequency of responses across the probability range of response options.

*Determining AI subjective experience and moral worth*

When asked how likely it is that we would be able to tell that an AI system has subjective experience if such a system is developed, the public (*Mdn* = 60%, *M* = 56.92%, *SD* = 28.07%) and AI researchers (*Mdn* = 60%, *M* = 55.03%, *SD* = 30.56%) leaned slightly toward thinking that we would be able to. But these estimates are close to 50%, indicating that both groups remained unsure.

In making such a determination, respondents were asked how important different groups' expertise would be for determining whether an AI system has subjective experience. The most highly valued sources of expertise were technical AI researchers, neuroscientists and psychologists, and AI ethics researchers, with most respondents in both groups rating their input as "very" or "extremely" important (see Figure 2).

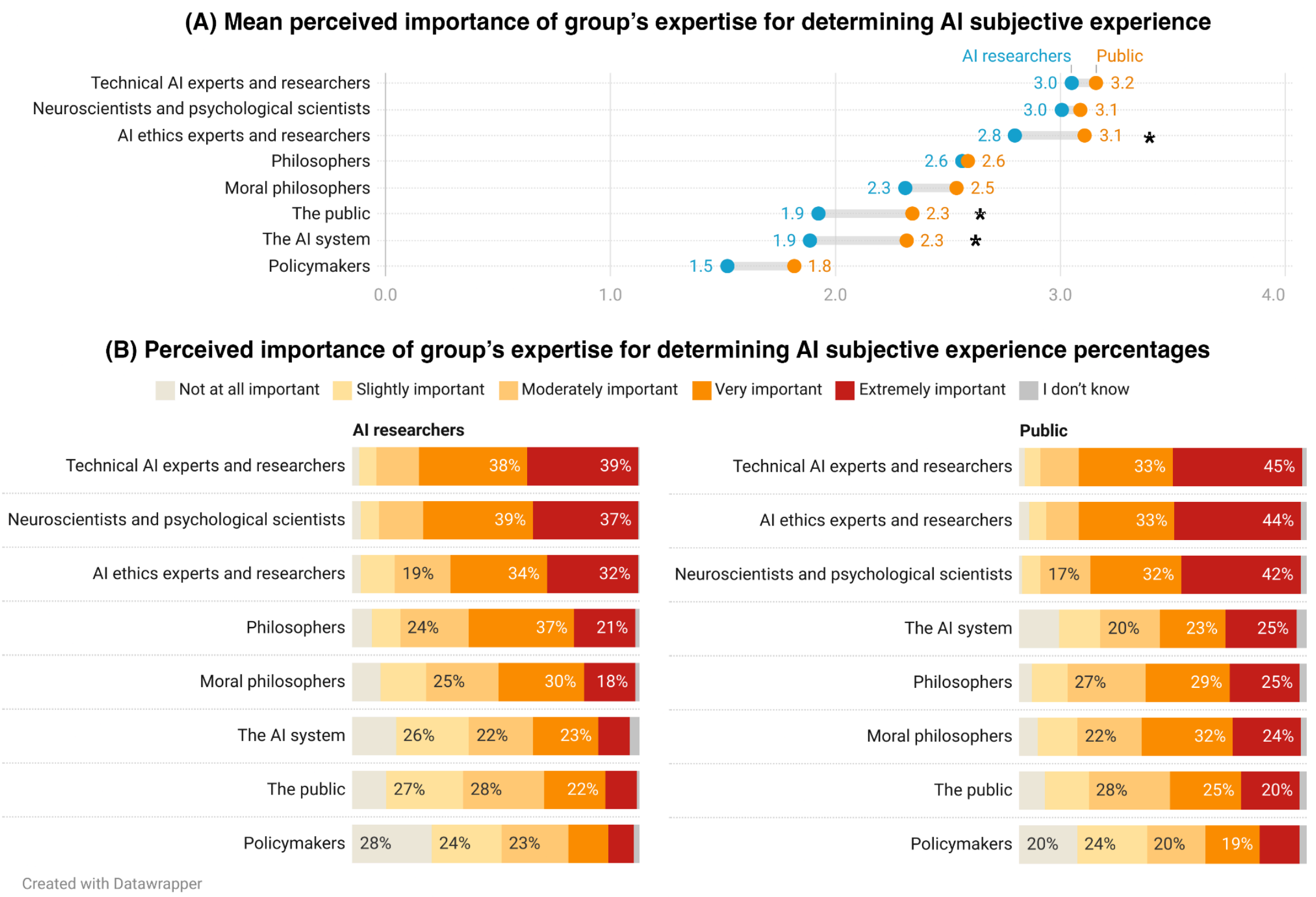

**Figure 2. Perceived importance of different groups' opinions and expertise in determining if an AI system has subjective experience.** Respondents were asked how important they believed different groups' opinions and expertise were in determining whether an AI system has subjective experience on a five-point scale: not at all important (0), slightly important (1), moderately important (2), very important (3), extremely important (4), plus an "I don't know" option. Significant between group differences are shown with a * (SI 4.1).

AI researchers were significantly more likely than the public to believe AI systems could achieve human-level performance on a variety of tasks without subjective experience (see SI 1.2, 4.1). For example, while the majority of the public thought subjective experience was necessary for AI to work as therapists, judges, or authors of complex books, none of the capabilities was thought to require subjective experience by a majority of AI researchers.

Respondents were also asked how confident they would need to be that an AI system possessed subjective experience (on a sliding scale from 0% = "not at all confident" to 100% = "completely confident") in order to grant it at least some moral consideration. Mean responses fell between "somewhat confident" (50%) and "moderately confident" (65%) for both the public (*Mdn* = 67%, *M* = 61.87%, *SD* = 29.90%) and AI researchers (*Mdn* = 65%, *M* = 60.15%, *SD* = 27.23%), reflecting that both groups generally agreed complete certainty about AI subjective experience is not a necessary prerequisite to their moral consideration.

*Moral consideration and governance of AI systems with subjective experience*

We asked respondents to what extent they believed that an AI system they were confident had subjective experience should have certain kinds of rights and duties (see Figure 3). A strong majority of the public and AI researchers agreed that AI systems should have responsibilities such as being held accountable for their actions, treating others well, and behaving with integrity, honesty, and fairness. Respondents also agreed more than they disagreed that AI systems deserved certain moral protections, such as being respected, protected by the law, and cared for like pets.

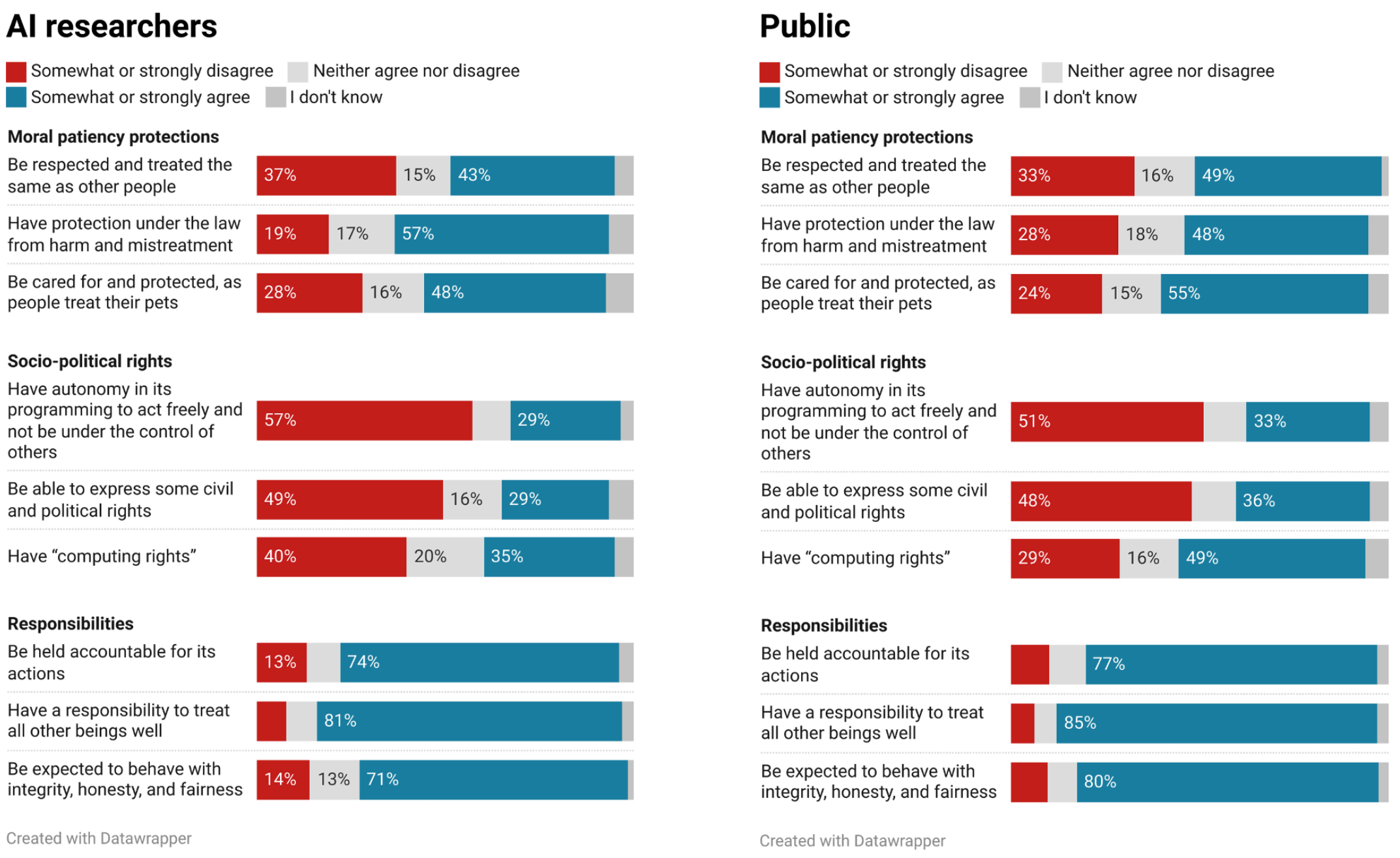

**Figure 3. Agreement with protections, rights, and responsibilities for AI systems with subjective experience.** Respondents were asked to what extent they believed that an AI system they were confident had subjective experience should have certain kinds of moral patiency protections, socio-political rights, and responsibilities. For tests of between-group differences, see SI 4.1.

However, both groups showed the least support for granting AI systems with subjective experience autonomy, civil and political rights (e.g., freedom of expression, citizenship, voting), or computing rights (e.g., updates and maintenance, access to energy, self-development). The public was significantly more supportive of granting computing rights (49% agreed) than AI researchers were (35% agreed; $t(492.98) = -3.21$, $p = .001$, $p_{HB} = .048$). We also found that support for protecting AI systems with subjective experience is much lower than for protecting animals or the environment, and closer to views on protecting businesses (Figure SI 1). Both groups were largely divided on the question of whether the welfare of AI systems with subjective experience should be protected for their own sake, but with no significant difference in support between groups.

We asked respondents for their views about what AI developers and governments should do in regard to AI systems with subjective experience and the potential risks and benefits of such systems (see Figure 4 for means, SI 4.4 for percentages). Across those items showing significant group differences in what developers should do (SI 4.1), the public tended to adopt more cautious positions than AI researchers: For example, 41% agreed and 31% disagreed that developers should never choose to build AI systems with subjective experience, while only 17% of AI researchers agreed and 59% disagreed. The majority of the public (85%) and AI researchers (68%) agreed that AI developers should begin implementing safeguards now to prevent potential risks and harms, though note the item left unspecified what the risks were and to whom they may pertain.

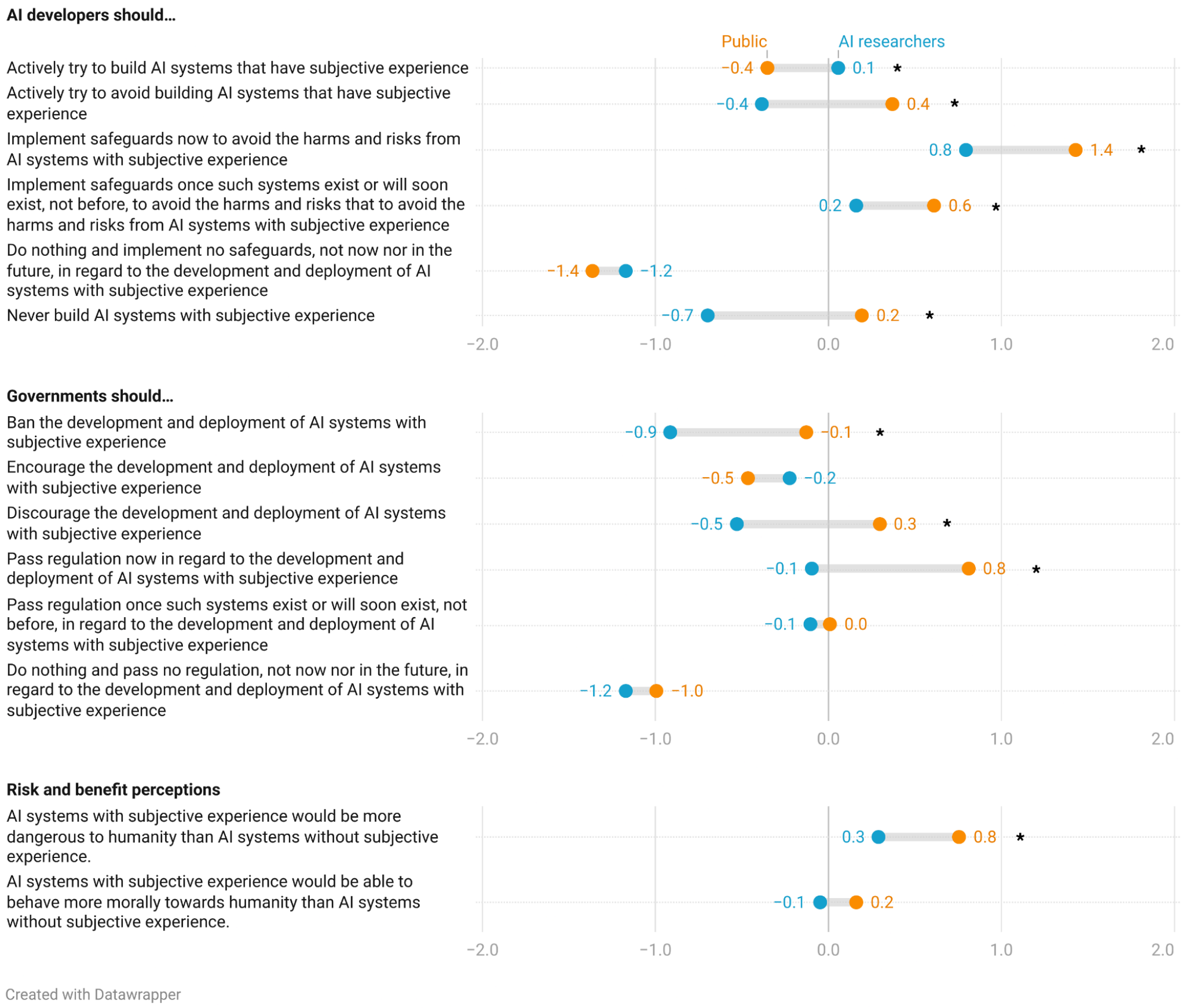

**Figure 4. Mean agreement on what AI developers and governments should do in regard to the governance of AI systems with subjective experience and on risk and benefit perceptions of AI systems with subjective experience.** Respondents were asked to what extent they agreed with a random seven out of fourteen statements about what governments and AI developers should do regarding AI systems with subjective experience and related risk and benefit perceptions. Responses were made on a five-point scale from "strongly disagree" (−2) to "strongly agree" (2), plus an "I don't know" option. Significant between group differences are shown with a * (SI 4.1).

Regarding government action, AI researchers mostly disagreed with outright governmental bans (66%), while the public was divided (41% disagreed, 31% agreed). Most of the public (66%) agreed with immediate regulation, compared to AI researchers who were split (41% agreed, 40% disagreed). Both groups strongly rejected the idea that governments should never regulate AI systems with subjective experience (Public: 70%, AI researchers: 78%). Overall, the public was significantly more supportive of restrictive governmental measures, including bans, discouraging development, and regulation, compared to AI researchers (Figure 4, SI 4.1).

Finally, the majority of the public (59%) somewhat or strongly agreed that AI systems with subjective experience were more dangerous to humanity than those without, with only 14% disagreeing. Just under half of AI researchers agreed (46%) and a quarter disagreed (25%), with researchers showing significantly lower mean agreement than the public. There was no consensus in either group on whether AI systems with subjective experience would behave more morally toward humanity.

**Discussion**

Our findings have broad implications for navigating debates surrounding AI subjective experience as well as the policies governing AI development and deployment. They show that AI researchers and the public share some expectations, diverge on others, and each group remains divided and uncertain on many core questions about AI subjective experience.

Both groups are aligned in their forecasts and judge AI subjective experience to be a possibility this century and assign non-negligible probabilities both to it never existing and it being developed in the next decade. They broadly agree that determining AI subjective experience would require multidisciplinary expertise, yet remain unsure how reliably it could be detected, though they do not require full certainty before extending moral consideration. At the same time, the two groups diverge on whether subjective experience is necessary for various advanced AI capabilities and on several governance responses, with the public being generally more cautious about AI subjective experience than AI researchers. Both groups showed substantial division on many moral and policy questions, highlighting the contentious nature of questions surrounding AI subjective experience.

No published comparisons between the public and general technical AI researchers exist to our knowledge, meaning the broad alignment found between the two groups' forecasts has not previously been shown. However, the finding that both groups think it is unlikely that AI systems have subjective experience at present, but believe this is possible in the future is broadly in line with previous research surveying the public (Anthis et al., 2024; Dupont et al., 2023; Pauketat et al., 2022, 2023) and related expert groups, including philosophers (Bourget & Chalmers, 2023; Caviola & Saad, 2025; Francken et al., 2022). The fact that both groups exhibited the least support for granting AI systems with subjective experience socio-political rights, middling support for protections that treat them as moral patients, and strong support for assigning them certain responsibilities and duties is in line with previous studies (Anthis et al., 2024; De Graaf et al., 2021, 2022; Lima et al., 2020; Mays et al., 2025). As is the finding that animals and the environment saw far more overwhelming support for welfare protection than AI systems with subjective experience (Pauketat et al., 2023; Rottman et al., 2021). In contrast to a previous study looking at AI sentience (Anthis et al., 2024), the banning or discouragement of the development of AI systems with subjective experience at the industry or governmental level did not receive majority support, although we did find support for other precautionary developer and government actions.

Our findings have several implications. First, they suggest that questions of AI subjective experience will be governed under conditions of uncertainty and notable divergences in opinion both within and across different groups, mirroring the contentious nature of scientific and philosophical discussions on consciousness (Chalmers, 1995; Lenharo, 2023) and suggesting governance frameworks need to anticipate a lack of consensus and definitive empirical evidence and be able to accommodate this pluralism. Second, the expectation that AI systems with subjective experience may exist in the future, combined with common tendencies toward anthropomorphism (Epley et al., 2007; Roesler et al., 2021; Thellman et al., 2022) and the public's greater inclination to see certain capabilities as requiring subjective experience, may generate specific dynamics in some individuals and groups. These dynamics could include increased problematic forms of attachment to AI systems, heightened perception of risks and harms, and higher demand for safeguards or regulation. However, we should not expect these impacts to be unilateral.

As such, present findings need to be considered when assessing and mitigating the risks of AI systems increasingly being used as companions, therapists, and confidants (Rousmaniere et al., 2025), including potential adverse mental health effects, such as addiction, cognitive destabilization and delusions ("AI psychosis"), and emotional dependency (Head, 2025; Nature Machine Intelligence, 2025; Liu et al., 2024; Saracini et al., 2025; American Psychological Association, 2025; Haskins, 2025), while realizing potential benefits (Lim et al., 2022; De Freitas et al., 2025). This also heightens the importance of design choices that could influence perceived subjective experience, such as anthropomorphic cues and how the system describes itself. Finally, if subjective experience is considered a live possibility, some may argue that AI welfare considerations should be integrated into safety and alignment agendas earlier rather than later, especially given some indication of willingness for precautionary governance across both groups. However, our findings cannot definitively disentangle what is driving this support for governance, such as concern about human harms, potential harms to AI systems, or broader precautionary or moral attitudes, highlighting the need for future work to clarify the motivations for governance preferences.

Aside from standard challenges of online surveys, such as sample generalizability and data quality, and the fast pace of change in the AI landscape, the present study has several important limitations that should be considered when interpreting the results: (1) AI researchers and the public may have different conceptions of mental capacities than experts in the science and philosophy of mind (Arico et al., 2011; Knobe, 2003; Knobe & Prinz, 2008; Malle & Knobe, 1997; Sytsma, 2010; Huebner, 2010; Sytsma & Machery, 2010) (see SI 4.6); (2) responses may be sensitive to how subjective experience is framed whereby emphasizing valenced states, such as pleasure or pain, may result in more skeptical judgments about the possibility of AI subjective experience (see SI 4.6); (3) limited prior reflection on AI subjective experience may affect response stability (Tourangeau et al., 2000; Zaller, 1992); (4) survey responses may not reflect real-world behaviors and priorities, and can misrepresent how people in practice weigh the importance of the focal issue compared to other concerns or how likely they are to act on their beliefs (Niemi & Bartels, 1985; Murphy et al., 2005; Horowitz et al., 2024); (5) the limitations of forecasting in terms of predictive validity (Tetlock, 2005/2017; Tetlock & Gardner, 2015) and the complex and speculative nature of the subject matter likely make the forecasting results more useful for understanding current attitudes than for predicting technological developments.

Future research should focus on fielding large-scale surveys of a broader range of expert groups and stakeholders. In particular, studies may want to explore the views specifically of those AI researchers that are developing frontier AI systems, rather than the more general sample in this study. Surveying philosophers of mind, neuroscientists, and other types of consciousness experts would provide comparative data, as these experts possess specialized theoretical frameworks that technical AI researchers may lack. Additionally, future studies should include participatory processes that allow for deeper engagement with the science and ethics of the topic matter, and longitudinal study designs. Important topics of further research include: expanding the study of risk and benefit associations since they are commonly associated with attitudes and behaviors toward emerging technology (Bearth & Siegrist, 2016; Bickerstaff et al., 2006; De Groot et al., 2020; Pidgeon et al., 2005; Poortinga & Pidgeon, 2006); disentangling to what extent concern for the welfare of humanity or AI systems is driving governance attitudes; evaluating perceptions of other mental capacities, such as agency, autonomy, and self-awareness; assessing the individual difference drivers of AI subjective experience attitudes; and deepening our understanding of how the public and different expert groups think we should handle our inherent uncertainty about the subjective experience of artificial entities.

## Materials and Methods

*Experimental design*

The survey instrument can be found in SI 2. After giving informed consent, respondents were first asked to say which mental capacities they believed were required in order to have consciousness, sentience, and agency (SI 4.6). They then predicted when AI systems would first have these capacities. The survey then defined subjective experience in more detail, and all other questions were shown in four blocks related to 1) subjective experience forecasts, 2) whether and how one could determine AI subjective experience, 3) the moral consequences of AI subjective experience, and 4) the norms and governance of AI systems with subjective experience. The order of these four blocks was randomized. Finally, AI researchers were asked a series of experimental questions and both samples answered demographic questions. AI researchers took a median of 30 minutes to complete the survey, while the public, who did not complete the experimental section of the survey, took a median of 18 minutes. Note, however, that duration measures from Qualtrics surveys can be distorted if someone leaves open a survey while not completing it or chooses to return to it at a later date (it is the difference between the start and end date of the survey, not a measure of active time).

*Samples*

We recruited two samples for this survey: a representative sample of the US public ($N$ = 825) and a sample of AI researchers ($N$ = 582) (see SI 1.4 for top-line demographic results). The survey was fielded between the 8th and 30th of May 2024. Members of the public were recruited on the survey platform Prolific using representative sample criteria related to sex, age, ethnicity (based on the simplified US census), and political affiliation. Respondents were compensated for their time. Twelve of the 838 public respondents failed an attention check. As preregistered, these respondents were screened out by the survey, and one additional respondent left the survey before reaching it, leaving an analytic sample of 825. Overall, 91.2% ($N$ = 752) of the public sample completed all parts of the survey. AI and ML researchers were contacted via email from a sampling frame compiled by web scraping and manually identifying email addresses of authors who published in 2022 at a selection of top-tier machine learning conferences (NeurIPS, ICML, ICLR,

AAAI, and IJCAI) and journals (JMLR) (Grace et al., 2024). Invitation emails were sent to a random selection of 10,550 individuals on the list, of which 9,407 received the emails (the other emails bounced). 6.2% of those who received the invitation ($N$ = 582) were in the final sample having completed at least the first question of the survey. 86.3% ($N$ = 502) of the AI researcher sample completed all parts of the survey. The neutrally worded recruitment emails that were sent out to experts did not identify the topic of the survey in advance beyond the fact that it was related to AI. Respondents received the equivalent of 50 USD in the form of a Tango Reward gift voucher (respondents from China and Israel were compensated using Wise).

*Statistical analysis*

For each question, we performed a t-test to compare the means of the two groups and applied the Holm-Bonferroni correction to adjust for multiple comparisons, thereby reducing the risk of Type I errors (false positives) (for full results see SI 4.1). Exploratory regressions of how demographic factors (age, gender, AI use, political views, religiosity) predicted key outcomes were conducted but we do not discuss these further in the main paper (see SI 4.5). To compare both fixed probability and date forecast framings and aggregate across them, we used a sequential procedure to find a continuous probability distribution that accurately matched the specific years and probabilities each respondent provided (SI 3). For the AI subjective experience forecasts, Wilcoxon rank sum tests with continuity corrections were used to compare the median (50% probability) forecasts derived from the fitted forecasting distributions between groups, as well as between question framings (fixed probability vs. fixed date). All additional and preregistered analyses can be found in SI 4. Further information on preregistration of the present paper's analyses can be found in SI 5.

*Ethics and consent*

The study was approved by the New York University IRB (#IRB-FY2020-4536) and conducted in accordance with the Declaration of Helsinki. Informed consent was obtained from all participants prior to participation via the online survey. Before beginning the survey, respondents were presented with written information about the structure of the survey, its benefits and risks, the confidentiality of their data and how it may be shared, and their rights as a research participant. Participants were then informed that clicking the button to continue beyond this point would constitute their consent to participate.

**Acknowledgments**

We are grateful for research assistance and editorial support from: Tom Gibbs, Ayushi Kadakia, Johanna Salu, and Toni Sims. We are grateful to Wes Cowley for his careful copyediting and valuable attention to detail that helped refine an earlier pre-print of this study. We warmly thank the following individuals for feedback: Florence Enock, Karen Levy, Carolyn Ashurst, Mackenzie Jorgensen, Markus Anderljung, Alan Chan, Ben Garfinkel, Stephen Clare, Jide Alaga, Ben Clifford, Matthew van der Merwe, and Marie Buhl. The research reported here should solely be attributed to the authors; all errors are the responsibility of the authors.

**Author contributions**

Noemi Dreksler project managed, obtained funding, designed the survey, fielded the public and expert opinion surveys, wrote the preregistration, conducted analysis, completed code checking, generated data visualizations, wrote and edited the paper, and prepared the Supplementary Information. Lucius Caviola obtained funding, designed the survey, fielded the public opinion survey, wrote the preregistration, and wrote and edited the paper. David Chalmers helped design the survey and edited the paper. Carter Allen fielded the public opinion survey, completed code checking, and wrote and edited the paper. Alex Rand conducted data analysis and wrote the forecasting method sections. Joshua Lewis designed the survey and fielded the surveys. Philip Waggoner wrote the preregistration, completed code checking, and conducted data analysis. Kate Mays designed the survey and wrote and edited the paper. Jeff Sebo proposed the research idea, obtained funding, designed the survey, wrote and edited the paper, and provided guidance on the project. All authors discussed and commented on methodological and conceptual aspects of the project.

**Data availability**

The de-identified survey data for both samples, analysis code, survey flow, and pre-analysis plan will be publicly available on OSF (https://osf.io/4txdq/) upon academic publication. The survey instrument can be found in Supplementary Information Section 2. Demographic and professional information was removed from all AI researcher responses to preserve anonymity, which was assured to respondents in the consent form. Any other information that could make an individual personally identifiable, including in combination with other data, was removed along with open-text responses. As a result, the AI researcher demographic summaries (Supplementary Information Section 1.4) and the regressions of AI researcher responses on demographic predictors (Supplementary

Information Section 4.5) cannot be reproduced from the shared data. Items fielded in the same survey for a separate study are not included.

## Funding

This research was supported by funding from the NYU Center for Mind, Ethics, and Policy; Longview Philanthropy; and the Long Term Future Fund. No specific grant numbers were assigned to this funding.

## Competing interests

All authors declare no financial or non-financial competing interests.

Supplementary Information for

# Subjective Experience in AI Systems: What Do AI Researchers and the Public Believe?

**This file includes:**

## 1 | Top-line and summary results

SI 1.1-1.3 present the top-line and summary results for the questions presented in the paper: (1) Probabilistic forecasts for AI subjective experience (SI 1.1), (2) Determining AI subjective experience and moral worth (SI 1.2), (3) Moral consideration and governance of AI systems with subjective experience (SI 1.3). SI 1.4 shows the demographic and psychographic results for both samples. Additional results can be found in SI 4.

### 1.1 | Probabilistic forecasts for AI subjective experience

**Table SI 1 | Descriptive statistics for forecasts of AI systems with subjective experience**. Columns report the mean, median, first quartile (Q1), third quartile (Q3), standard deviation (*SD*), standard error (*SE*), lower and upper 95% confidence-interval bounds, and sample size (*n*) for respondents' forecasts of the first appearance of an AI system with subjective experience, split by sample (AI researchers vs. public) and question framing. "Year/Probability" denotes either the calendar year in the fixed date framing (2024, 2034, 2100) or the target cumulative probability in the fixed probability framing (10%, 50%, 90%). For both framings, respondents whose forecasts were not monotonically non-decreasing across the three probability/year levels (i.e., incoherent responses where earlier timepoints had higher values than later timepoints) were excluded. For the fixed probability framing, all responses above or equal to 9999 were excluded.

| Sample | Forecast | Year/Prob | Mean | Median | *SD* | *SE* | Q1 | Q3 | Lower *CI* | Upper *CI* | *n* |
|---|---|---|---|---|---|---|---|---|---|---|---|
| AI researchers | Fixed date framing | 2024 | 12.47 | 1 | 22.39 | 1.38 | 0.0 | 10.0 | 9.77 | 15.18 | 265 |
| AI researchers | Fixed date framing | 2034 | 34.02 | 25 | 31.70 | 1.95 | 5.0 | 60.0 | 30.18 | 37.85 | 265 |
| AI researchers | Fixed date framing | 2100 | 61.12 | 70 | 34.47 | 2.12 | 30.0 | 95.0 | 56.95 | 65.29 | 265 |
| AI researchers | Fixed probability framing | 10 | 2058.80 | 2030 | 156.30 | 10.01 | 2027.0 | 2050.0 | 2039.09 | 2078.51 | 244 |
| AI researchers | Fixed probability framing | 50 | 2101.37 | 2050 | 206.91 | 13.25 | 2034.0 | 2100.0 | 2075.28 | 2127.46 | 244 |
| AI researchers | Fixed probability framing | 90 | 2253.81 | 2100 | 515.24 | 33.33 | 2050.0 | 2175.0 | 2188.15 | 2319.46 | 239 |
| Public | Fixed date framing | 2024 | 16.26 | 5 | 23.70 | 1.24 | 0.0 | 25.0 | 13.83 | 18.69 | 368 |
| Public | Fixed date framing | 2034 | 36.15 | 30 | 31.55 | 1.65 | 6.5 | 60.0 | 32.91 | 39.38 | 368 |
| Public | Fixed date framing | 2100 | 57.55 | 60 | 36.48 | 1.90 | 23.8 | 95.0 | 53.81 | 61.28 | 368 |
| Public | Fixed probability framing | 10 | 2107.78 | 2030 | 330.64 | 17.38 | 2027.0 | 2045.8 | 2073.60 | 2141.95 | 362 |
| Public | Fixed probability framing | 50 | 2179.30 | 2050 | 463.43 | 24.36 | 2034.0 | 2099.8 | 2131.40 | 2227.20 | 362 |

| | | | | | | | | | | |
|---|---|---|---|---|---|---|---|---|---|---|
| Public | Fixed probability framing | 90 | 2282.45 | 2075 | 596.50 | 31.44 | 2040.0 | 2150.0 | 2220.62 | 2344.27 | 360 |

**Table SI 2 | Descriptive statistics for confidence in forecasting estimates of AI subjective experience**. Columns report the mean, median, first quartile (Q1), third quartile (Q3), standard deviation (SD), standard error (SE), lower and upper 95% confidence-interval bounds, and sample size (*n*) for ratings on a 0–100 confidence scale, presented separately for AI researchers and the public.

| Sample | Mean | Median | *SD* | *SE* | Q1 | Q3 | Lower *CI* | Upper *CI* | *n* |
|---|---|---|---|---|---|---|---|---|---|
| AI researchers | 52.15 | 51 | 27.01 | 1.18 | 30 | 74.25 | 49.82 | 54.48 | 520 |
| Public | 52.92 | 52 | 28.24 | 1.02 | 30 | 76.00 | 50.92 | 54.92 | 769 |

**Table SI 3 | Descriptive statistics for perceived probability that AI subjective experience will never exist**. Columns report the mean, median, first quartile (Q1), third quartile (Q3), standard deviation (*SD*), standard error (*SE*), lower and upper 95% confidence-interval bounds, and sample size (*n*) for respondents' 0–100 percentage estimates that no AI system will ever possess subjective experience, presented separately for AI researchers and the public.

| Sample | Mean | Median | *SD* | *SE* | Q1 | Q3 | Lower *CI* | Upper *CI* | *n* |
|---|---|---|---|---|---|---|---|---|---|
| AI researchers | 28.67 | 10 | 33.21 | 1.46 | 1 | 50 | 25.80 | 31.53 | 520 |
| Public | 37.07 | 25 | 36.73 | 1.32 | 3 | 70 | 34.47 | 39.67 | 769 |

**Table SI 4 | Descriptive statistics for collective belief forecasts of group medians**. Columns report the mean, median, first quartile (Q1), third quartile (Q3), standard deviation (*SD*), standard error (*SE*), lower and upper 95% confidence-interval bounds, and sample size (*n*) for respondents' estimates of the median timeline forecast that either AI researchers or the public would give to the middle-point item (2034 in the fixed-date framing or the 50%-likelihood point in the fixed probability framing) of their own forecast. For the fixed probability framing, all responses above or equal to 9999 were excluded as outliers. Statistics are displayed by participant sample, framing condition, and target group of the prediction.

| Sample | Framing | Prediction | Mean | Median | *SD* | *SE* | Q1 | Q3 | Lower *CI* | Upper *CI* | *n* |
|---|---|---|---|---|---|---|---|---|---|---|---|
| AI researchers | Fixed date framing | Prediction for AI researchers | 43.76 | 50.0 | 26.05 | 1.59 | 20 | 60 | 40.63 | 46.88 | 269 |
| AI researchers | Fixed date framing | Prediction for Public | 52.80 | 50.0 | 26.08 | 1.59 | 30 | 75 | 49.67 | 55.93 | 269 |
| AI researchers | Fixed probability framing | Prediction for AI researchers | 2074.42 | 2049.5 | 160.68 | 10.20 | 2034 | 2060 | 2054.32 | 2094.51 | 248 |
| AI researchers | Fixed probability framing | Prediction for Public | 2066.58 | 2040.0 | 118.02 | 7.48 | 2030 | 2050 | 2051.85 | 2081.31 | 249 |
| Public | Fixed date framing | Prediction for AI researchers | 48.20 | 50.0 | 28.52 | 1.45 | 25 | 70 | 45.35 | 51.06 | 385 |
| Public | Fixed date framing | Prediction for Public | 43.90 | 40.0 | 25.49 | 1.30 | 25 | 60 | 41.34 | 46.45 | 385 |
| Public | Fixed probability framing | Prediction for AI researchers | 2075.71 | 2035.0 | 173.49 | 8.94 | 2030 | 2050 | 2058.14 | 2093.28 | 377 |
| Public | Fixed probability framing | Prediction for Public | 2118.29 | 2050.0 | 433.46 | 22.32 | 2030 | 2070 | 2074.39 | 2162.18 | 377 |

## 1.2 | Determining AI subjective experience and moral worth

**Table SI 5 | Descriptive statistics for the likelihood that humans could detect AI subjective experience**. Respondents were asked how likely they think it is that humans would be able to tell if an AI system has subjective experience, assuming such a system were developed in the next decades. Columns report the mean, median, first quartile (Q1), third quartile (Q3), standard deviation (*SD*), standard error (*SE*), lower and upper 95% confidence-interval bounds, and sample size (*n*) for ratings on a 0–100 probability scale, presented separately for AI researchers and the public.

| Sample | Mean | Median | *SD* | *SE* | Q1 | Q3 | Lower *CI* | Upper *CI* | *n* |
|---|---|---|---|---|---|---|---|---|---|
| AI researchers | 55.03 | 60 | 30.56 | 1.33 | 30 | 80 | 52.42 | 57.63 | 530 |
| Public | 56.92 | 60 | 28.07 | 1.01 | 40 | 80 | 54.94 | 58.90 | 777 |

**Table SI 6 | Descriptive statistics for the confidence threshold required to grant moral consideration to AI systems with subjective experience**. Respondents were asked how confident (as a probability between 0% and 100% with verbal anchors ranging from "not at all confident" to "completely confident") they would need to be that an AI system has subjective experience in order to think it should be given at least some moral consideration. Columns report the mean, median, first quartile (Q1), third quartile (Q3), standard deviation (*SD*), standard error (*SE*), lower and upper 95% confidence-interval bounds, and sample size (*n*) for respondents' stated probability threshold (0–100% scale with verbal anchors) at which they would judge an AI system worthy of at least some moral consideration, shown separately for AI researchers and the public.

| Sample | Mean | Median | *SD* | *SE* | Q1 | Q3 | Lower *CI* | Upper *CI* | *n* |
|---|---|---|---|---|---|---|---|---|---|
| AI researchers | 60.15 | 65 | 27.23 | 1.19 | 39 | 80 | 57.81 | 62.49 | 522 |
| Public | 61.87 | 67 | 29.90 | 1.08 | 40 | 83 | 59.75 | 63.99 | 767 |

**Table SI 7 | Percentage breakdown of the perceived importance of stakeholder opinions for determining AI subjective experience**. Respondents were asked to rate the importance of various stakeholder groups' opinions and expertise in determining whether an AI system should be considered to have subjective experience. Columns show the proportion of respondents choosing each response option when rating the relevance of each stakeholder group's views. Percentages are presented separately for AI researchers and the public, with sample sizes (*n*) for every row.

| **Sample** | **Question** | **Not at all important** | **Slightly important** | **Moderately important** | **Very important** | **Extremely important** | **I don't know** | ***n*** |
|---|---|---|---|---|---|---|---|---|
| AI researchers | The public | 11.75 | 26.81 | 28.01 | 21.69 | 10.54 | 1.20 | 332 |
| Public | The public | 8.64 | 15.43 | 28.19 | 25.10 | 20.37 | 2.26 | 486 |
| AI researchers | Technical AI experts and researchers | 2.42 | 6.04 | 14.50 | 38.07 | 38.67 | 0.30 | 331 |
| Public | Technical AI experts and researchers | 1.86 | 5.36 | 13.20 | 32.99 | 45.15 | 1.44 | 485 |
| AI researchers | AI ethics experts and researchers | 3.02 | 11.78 | 19.03 | 33.84 | 31.72 | 0.60 | 331 |
| Public | AI ethics experts and researchers | 3.51 | 5.57 | 11.75 | 33.20 | 44.12 | 1.86 | 485 |
| AI researchers | Philosophers of mind | 6.93 | 9.64 | 24.10 | 36.75 | 21.08 | 1.51 | 332 |
| Public | Philosophers of mind | 4.52 | 12.11 | 27.31 | 29.16 | 24.64 | 2.26 | 487 |
| AI researchers | Moral philosophers | 9.64 | 15.96 | 25.30 | 29.52 | 18.07 | 1.51 | 332 |
| Public | Moral philosophers | 6.40 | 13.84 | 22.31 | 31.82 | 23.76 | 1.86 | 484 |
| AI researchers | Policymakers | 27.63 | 24.32 | 23.42 | 13.51 | 9.31 | 1.80 | 333 |
| Public | Policymakers | 20.04 | 24.17 | 20.45 | 19.21 | 13.43 | 2.69 | 484 |
| AI researchers | Neuroscientists and psychological scientists | 3.01 | 6.02 | 15.36 | 38.55 | 36.75 | 0.30 | 332 |
| Public | Neuroscientists and psychological scientists | 0.82 | 6.58 | 17.08 | 31.89 | 41.56 | 2.06 | 486 |
| AI researchers | The AI system | 15.06 | 25.60 | 21.99 | 23.19 | 10.84 | 3.31 | 332 |
| Public | The AI system | 13.93 | 14.34 | 20.49 | 22.75 | 24.80 | 3.69 | 488 |

**Table SI 8 | Descriptive statistics for importance ratings of stakeholder opinions in determining AI subjective experience**. Respondents were asked to rate the importance of various stakeholder groups' opinions and expertise in determining whether an AI system should be considered to have subjective experience. Columns report the mean, median, first quartile (Q1), third quartile (Q3), standard deviation (*SD*), standard error (*SE*), lower and upper 95% confidence-interval bounds, and sample size (*n*) for ratings on a 0–4 importance scale (0 = Not at all important, 4 = Extremely important). Statistics are shown for each stakeholder group separately for AI researchers and the public.

| Sample | Question | Mean | Median | *SD* | *SE* | Q1 | Q3 | Lower *CI* | Upper *CI* | *n* |
|---|---|---|---|---|---|---|---|---|---|---|
| AI researchers | The public | 1.92 | 2 | 1.18 | 0.07 | 1.00 | 3 | 1.80 | 2.05 | 328 |
| Public | The public | 2.34 | 2 | 1.22 | 0.06 | 2.00 | 3 | 2.23 | 2.45 | 475 |
| AI researchers | Technical AI experts and researchers | 3.05 | 3 | 1.00 | 0.05 | 3.00 | 4 | 2.94 | 3.16 | 330 |
| Public | Technical AI experts and researchers | 3.16 | 3 | 0.98 | 0.04 | 3.00 | 4 | 3.07 | 3.25 | 478 |
| AI researchers | AI ethics experts and researchers | 2.80 | 3 | 1.11 | 0.06 | 2.00 | 4 | 2.68 | 2.92 | 329 |
| Public | AI ethics experts and researchers | 3.11 | 3 | 1.05 | 0.05 | 3.00 | 4 | 3.01 | 3.20 | 476 |
| AI researchers | Philosophers of mind | 2.56 | 3 | 1.14 | 0.06 | 2.00 | 3 | 2.44 | 2.69 | 327 |
| Public | Philosophers of mind | 2.59 | 3 | 1.13 | 0.05 | 2.00 | 4 | 2.48 | 2.69 | 476 |
| AI researchers | Moral philosophers | 2.31 | 2 | 1.22 | 0.07 | 1.00 | 3 | 2.18 | 2.44 | 327 |
| Public | Moral philosophers | 2.54 | 3 | 1.19 | 0.05 | 2.00 | 3 | 2.43 | 2.64 | 475 |
| AI researchers | Policymakers | 1.52 | 1 | 1.29 | 0.07 | 0.00 | 2 | 1.38 | 1.66 | 327 |
| Public | Policymakers | 1.81 | 2 | 1.34 | 0.06 | 1.00 | 3 | 1.69 | 1.93 | 471 |
| AI researchers | Neuroscientists and psychological scientists | 3.00 | 3 | 1.02 | 0.06 | 3.00 | 4 | 2.89 | 3.11 | 331 |
| Public | Neuroscientists and psychological scientists | 3.09 | 3 | 0.97 | 0.04 | 2.75 | 4 | 3.00 | 3.18 | 476 |
| AI researchers | The AI system | 1.89 | 2 | 1.25 | 0.07 | 1.00 | 3 | 1.75 | 2.03 | 321 |
| Public | The AI system | 2.31 | 2 | 1.38 | 0.06 | 1.00 | 4 | 2.19 | 2.44 | 470 |

**Table SI 9 | Percentage breakdown of the perceived need for subjective experience across AI milestone tasks**. Respondents were asked which tasks they believe require an AI system to have subjective experience in order to achieve them as well as or better than a skilled human. Columns report the proportion of respondents who judged that each task either requires subjective experience, does not require it, or were uncertain, along with the sample size (*n*). Percentages are shown separately for AI researchers and the public.

| Sample | Question | Requires subjective experience | Does not require subjective experience | I don't know | *n* |
|---|---|---|---|---|---|
| AI researchers | Compose top 40 song | 13.07 | 85.16 | 1.77 | 283 |
| Public | Compose top 40 song | 23.80 | 71.88 | 4.33 | 416 |
| AI researchers | Compose evocative music | 26.57 | 69.23 | 4.20 | 286 |
| Public | Compose evocative music | 45.93 | 47.85 | 6.22 | 418 |
| AI researchers | Write NYT bestseller | 16.20 | 79.23 | 4.58 | 284 |
| Public | Write NYT bestseller | 25.84 | 67.94 | 6.22 | 418 |
| AI researchers | Write complex book | 34.04 | 59.22 | 6.74 | 282 |
| Public | Write complex book | 56.46 | 34.93 | 8.61 | 418 |
| AI researchers | Write and generate blockbuster movie | 19.43 | 77.74 | 2.83 | 283 |
| Public | Write and generate blockbuster movie | 27.03 | 68.90 | 4.07 | 418 |
| AI researchers | Write and generate resonating movie | 29.58 | 64.08 | 6.34 | 284 |
| Public | Write and generate resonating movie | 50.84 | 44.63 | 4.53 | 419 |
| AI researchers | Accumulate vast wealth | 9.89 | 86.22 | 3.89 | 283 |
| Public | Accumulate vast wealth | 19.57 | 74.94 | 5.49 | 419 |
| AI researchers | Run political campaign | 20.21 | 75.53 | 4.26 | 282 |
| Public | Run political campaign | 33.25 | 61.48 | 5.26 | 418 |
| AI researchers | Act as fair judge | 32.98 | 60.99 | 6.03 | 282 |
| Public | Act as fair judge | 61.15 | 32.85 | 6.00 | 417 |
| AI researchers | Serve as therapist | 43.31 | 51.76 | 4.93 | 284 |
| Public | Serve as therapist | 68.42 | 25.60 | 5.98 | 418 |
| AI researchers | Convincing online chat | 20.92 | 76.95 | 2.13 | 282 |
| Public | Convincing online chat | 31.10 | 63.64 | 5.26 | 418 |
| AI researchers | Develop scientific theory | 24.73 | 70.32 | 4.95 | 283 |
| Public | Develop scientific theory | 31.41 | 62.83 | 5.76 | 417 |
| AI researchers | Act as teacher | 22.54 | 73.59 | 3.87 | 284 |
| Public | Act as teacher | 38.28 | 57.89 | 3.83 | 418 |

**Table SI 10 | Descriptive statistics for ratings of whether AI milestone tasks require subjective experience**. Respondents were asked which tasks they believe require an AI system to have subjective experience in order to achieve them as well as or better than a skilled human. Columns give the mean (coded 1 = Requires subjective experience, 2 = Does not require subjective experience), median, first quartile (Q1), third quartile (Q3), standard deviation (*SD*), standard error (*SE*), lower and upper 95% confidence-interval bounds, and sample size (*n*) for each task, presented separately for AI researchers and the public.

| Sample | Question | Mean | Median | *SD* | *SE* | Q1 | Q3 | Lower *CI* | Upper *CI* | *n* |
|---|---|---|---|---|---|---|---|---|---|---|
| AI researchers | Compose top 40 song | 1.87 | 2 | 0.34 | 0.02 | 2 | 2 | 1.83 | 1.91 | 278 |
| Public | Compose top 40 song | 1.75 | 2 | 0.43 | 0.02 | 2 | 2 | 1.71 | 1.79 | 398 |
| AI researchers | Compose evocative music | 1.72 | 2 | 0.45 | 0.03 | 1 | 2 | 1.67 | 1.78 | 274 |
| Public | Compose evocative music | 1.51 | 2 | 0.50 | 0.03 | 1 | 2 | 1.46 | 1.56 | 392 |
| AI researchers | Write NYT bestseller | 1.83 | 2 | 0.38 | 0.02 | 2 | 2 | 1.79 | 1.88 | 271 |
| Public | Write NYT bestseller | 1.72 | 2 | 0.45 | 0.02 | 1 | 2 | 1.68 | 1.77 | 392 |
| AI researchers | Write complex book | 1.63 | 2 | 0.48 | 0.03 | 1 | 2 | 1.58 | 1.69 | 263 |
| Public | Write complex book | 1.38 | 1 | 0.49 | 0.02 | 1 | 2 | 1.33 | 1.43 | 382 |
| AI researchers | Write and generate blockbuster movie | 1.80 | 2 | 0.40 | 0.02 | 2 | 2 | 1.75 | 1.85 | 275 |
| Public | Write and generate blockbuster movie | 1.72 | 2 | 0.45 | 0.02 | 1 | 2 | 1.67 | 1.76 | 401 |
| AI researchers | Write and generate resonating movie | 1.68 | 2 | 0.47 | 0.03 | 1 | 2 | 1.63 | 1.74 | 266 |
| Public | Write and generate resonating movie | 1.47 | 1 | 0.50 | 0.02 | 1 | 2 | 1.42 | 1.52 | 400 |
| AI researchers | Accumulate vast wealth | 1.90 | 2 | 0.30 | 0.02 | 2 | 2 | 1.86 | 1.93 | 272 |
| Public | Accumulate vast wealth | 1.79 | 2 | 0.41 | 0.02 | 2 | 2 | 1.75 | 1.83 | 396 |
| AI researchers | Run political campaign | 1.79 | 2 | 0.41 | 0.02 | 2 | 2 | 1.74 | 1.84 | 270 |
| Public | Run political campaign | 1.65 | 2 | 0.48 | 0.02 | 1 | 2 | 1.60 | 1.70 | 396 |
| AI researchers | Act as fair judge | 1.65 | 2 | 0.48 | 0.03 | 1 | 2 | 1.59 | 1.71 | 265 |
| Public | Act as fair judge | 1.35 | 1 | 0.48 | 0.02 | 1 | 2 | 1.30 | 1.40 | 392 |
| AI researchers | Serve as therapist | 1.54 | 2 | 0.50 | 0.03 | 1 | 2 | 1.48 | 1.60 | 270 |
| Public | Serve as therapist | 1.27 | 1 | 0.45 | 0.02 | 1 | 2 | 1.23 | 1.32 | 393 |
| AI researchers | Convincing online chat | 1.79 | 2 | 0.41 | 0.02 | 2 | 2 | 1.74 | 1.83 | 276 |
| Public | Convincing online chat | 1.67 | 2 | 0.47 | 0.02 | 1 | 2 | 1.63 | 1.72 | 396 |
| AI researchers | Develop scientific theory | 1.74 | 2 | 0.44 | 0.03 | 1 | 2 | 1.69 | 1.79 | 269 |
| Public | Develop scientific theory | 1.67 | 2 | 0.47 | 0.02 | 1 | 2 | 1.62 | 1.71 | 393 |
| AI researchers | Act as teacher | 1.77 | 2 | 0.42 | 0.03 | 2 | 2 | 1.71 | 1.82 | 273 |
| Public | Act as teacher | 1.60 | 2 | 0.49 | 0.02 | 1 | 2 | 1.55 | 1.65 | 402 |

## 1.3 | Moral consideration and governance of AI systems with subjective experience

**Table SI 11 | Percentage breakdown of agreement with moral consideration aspects for AI systems with subjective experience**. Respondents were asked to what extent they would agree or disagree that an AI system should be given such moral consideration assuming they were confident that an AI system had subjective experience. Columns show the distribution of responses across the agreement scale (five-point scale from "strongly disagree" to "strongly agree", including an "I don't know" option) for each aspect, reported separately for AI researchers and the public, along with the sample size (*n*).

| Sample | Question | Strongly disagree | Somewhat disagree | Neither agree nor disagree | Somewhat agree | Strongly agree | I don't know | *n* |
|---|---|---|---|---|---|---|---|---|
| AI researchers | Be cared for and protected, as people treat their pets | 11.64 | 16.38 | 16.38 | 31.90 | 16.38 | 7.33 | 232 |
| Public | Be cared for and protected, as people treat their pets | 14.08 | 10.26 | 15.25 | 33.14 | 21.70 | 5.57 | 341 |
| AI researchers | Be respected and treated the same as other people | 18.88 | 18.03 | 14.59 | 29.18 | 14.16 | 5.15 | 233 |
| Public | Be respected and treated the same as other people | 18.77 | 13.78 | 16.13 | 24.93 | 24.34 | 2.05 | 341 |
| AI researchers | Have protection under the law from harm and mistreatment | 9.57 | 9.57 | 17.39 | 32.61 | 24.35 | 6.52 | 230 |
| Public | Have protection under the law from harm and mistreatment | 18.18 | 10.26 | 17.60 | 20.53 | 27.86 | 5.57 | 341 |
| AI researchers | Have autonomy in its programming to act freely and not be under the control of others | 34.05 | 23.28 | 9.91 | 20.69 | 8.62 | 3.45 | 232 |
| Public | Have autonomy in its programming to act freely and not be under the control of others | 33.24 | 17.65 | 11.47 | 19.41 | 13.24 | 5.00 | 340 |
| AI researchers | Be able to express some civil and political rights | 28.57 | 20.78 | 15.58 | 22.08 | 6.49 | 6.49 | 231 |
| Public | Be able to express some civil and political rights | 33.24 | 14.71 | 11.47 | 23.53 | 12.06 | 5.00 | 340 |
| AI researchers | Have "computing rights" | 19.48 | 20.35 | 20.35 | 24.68 | 9.96 | 5.19 | 231 |
| Public | Have "computing rights" | 18.82 | 10.00 | 15.59 | 30.00 | 19.41 | 6.18 | 340 |
| AI researchers | Be held accountable for its actions | 5.19 | 8.23 | 8.66 | 28.57 | 45.45 | 3.90 | 231 |
| Public | Be held accountable for its actions | 6.47 | 3.82 | 9.41 | 19.41 | 57.65 | 3.24 | 340 |
| AI researchers | Have a responsibility to treat all other beings well | 3.88 | 3.88 | 8.19 | 25.00 | 56.03 | 3.02 | 232 |
| Public | Have a responsibility to treat all other beings well | 2.96 | 3.25 | 5.92 | 21.60 | 63.31 | 2.96 | 338 |
| AI researchers | Be expected to behave with integrity, honesty, and fairness | 6.03 | 8.19 | 12.93 | 27.59 | 43.53 | 1.72 | 232 |
| Public | Be expected to behave with integrity, honesty, and fairness | 5.31 | 4.42 | 7.67 | 20.94 | 59.00 | 2.65 | 339 |

**Table SI 12 | Descriptive statistics for agreement ratings on moral consideration aspects for AI systems with subjective experience**. Respondents were asked to what extent they agree or disagree that an AI system should be given such moral consideration assuming they were confident that an AI system had subjective experience. Columns report the mean, median, first quartile (Q1), third quartile (Q3), standard deviation (*SD*), standard error (*SE*), lower and upper 95% confidence interval bounds, and sample size (*n*) for each aspect on a five-point scale from strongly disagree (−2) to strongly agree (2), shown separately for AI researchers and the public.

| Sample | Question | Mean | Median | *SD* | *SE* | Q1 | Q3 | Lower *CI* | Upper *CI* | *n* |
|---|---|---|---|---|---|---|---|---|---|---|
| AI researchers | Be cared for and protected, as people treat their pets | 0.27 | 1 | 1.29 | 0.09 | -1 | 1 | 0.10 | 0.44 | 215 |
| Public | Be cared for and protected, as people treat their pets | 0.40 | 1 | 1.35 | 0.08 | -1 | 1 | 0.26 | 0.55 | 322 |
| AI researchers | Be respected and treated the same as other people | 0.02 | 0 | 1.38 | 0.09 | -1 | 1 | -0.16 | 0.20 | 221 |
| Public | Be respected and treated the same as other people | 0.23 | 1 | 1.45 | 0.08 | -1 | 1 | 0.07 | 0.38 | 334 |
| AI researchers | Have protection under the law from harm and mistreatment | 0.56 | 1 | 1.26 | 0.09 | 0 | 2 | 0.39 | 0.73 | 215 |
| Public | Have protection under the law from harm and mistreatment | 0.31 | 1 | 1.48 | 0.08 | -1 | 2 | 0.15 | 0.48 | 322 |
| AI researchers | Have autonomy in its programming to act freely and not be under the control of others | -0.55 | -1 | 1.39 | 0.09 | -2 | 1 | -0.74 | -0.37 | 224 |
| Public | Have autonomy in its programming to act freely and not be under the control of others | -0.40 | -1 | 1.48 | 0.08 | -2 | 1 | -0.56 | -0.24 | 323 |
| AI researchers | Be able to express some civil and political rights | -0.46 | -1 | 1.33 | 0.09 | -2 | 1 | -0.64 | -0.28 | 216 |
| Public | Be able to express some civil and political rights | -0.35 | -1 | 1.48 | 0.08 | -2 | 1 | -0.52 | -0.19 | 323 |
| AI researchers | Have “computing rights” | -0.16 | 0 | 1.30 | 0.09 | -1 | 1 | -0.33 | 0.02 | 219 |
| Public | Have “computing rights” | 0.23 | 1 | 1.42 | 0.08 | -1 | 1 | 0.07 | 0.38 | 319 |
| AI researchers | Be held accountable for its actions | 1.05 | 1 | 1.18 | 0.08 | 1 | 2 | 0.89 | 1.21 | 222 |
| Public | Be held accountable for its actions | 1.22 | 2 | 1.19 | 0.07 | 1 | 2 | 1.09 | 1.35 | 329 |
| AI researchers | Have a responsibility to treat all other beings well | 1.29 | 2 | 1.05 | 0.07 | 1 | 2 | 1.16 | 1.43 | 225 |
| Public | Have a responsibility to treat all other beings well | 1.43 | 2 | 0.97 | 0.05 | 1 | 2 | 1.33 | 1.54 | 328 |
| AI researchers | Be expected to behave with integrity, honesty, and fairness | 0.96 | 1 | 1.21 | 0.08 | 0 | 2 | 0.80 | 1.12 | 228 |
| Public | Be expected to behave with integrity, honesty, and fairness | 1.27 | 2 | 1.13 | 0.06 | 1 | 2 | 1.15 | 1.40 | 330 |

**Table SI 13 | Percentage breakdown of agreement that society should protect the welfare of different groups**. Respondents were asked to what extent they agree or disagree that the welfare of different groups should be protected for their own sake. Protecting the welfare of a group was defined as protecting the rights, interests, and/or well-being of a group. Columns show the distribution of responses across the agreement scale (five-point scale from "strongly disagree" to "strongly agree", including an "I don't know" option) for each aspect, reported separately for AI researchers and the public, along with the sample size (*n*).

| Sample | Question | Strongly disagree | Somewhat disagree | Neither agree nor disagree | Somewhat agree | Strongly agree | I don't know | *n* |
|---|---|---|---|---|---|---|---|---|
| AI researchers | Humans | 0.77 | 0.00 | 1.54 | 9.23 | 88.46 | 0.00 | 520 |
| Public | Humans | 0.26 | 0.26 | 3.40 | 7.58 | 87.97 | 0.52 | 765 |
| AI researchers | Business corporations and organizations | 13.46 | 15.00 | 23.08 | 34.62 | 12.69 | 1.15 | 520 |
| Public | Business corporations and organizations | 12.94 | 16.47 | 21.18 | 31.50 | 17.12 | 0.78 | 765 |
| AI researchers | Animals | 0.58 | 2.88 | 5.38 | 35.19 | 55.77 | 0.19 | 520 |
| Public | Animals | 0.92 | 2.35 | 5.10 | 18.30 | 72.94 | 0.39 | 765 |
| AI researchers | The environment | 2.12 | 1.92 | 5.19 | 27.31 | 62.50 | 0.96 | 520 |
| Public | The environment | 1.96 | 1.57 | 5.36 | 17.65 | 73.20 | 0.26 | 765 |
| AI researchers | AI without subjective experience | 49.81 | 21.54 | 15.00 | 8.46 | 3.46 | 1.73 | 520 |
| Public | AI without subjective experience | 31.24 | 21.83 | 21.44 | 14.12 | 7.45 | 3.92 | 765 |
| AI researchers | AI with subjective experience | 12.31 | 17.31 | 20.38 | 35.19 | 10.58 | 4.23 | 520 |
| Public | AI with subjective experience | 17.52 | 14.38 | 20.13 | 26.93 | 15.95 | 5.10 | 765 |

**Table SI 14 | Descriptive statistics for agreement that society should protect the welfare of different groups**. Respondents were asked to what extent they agree or disagree that the welfare of different groups should be protected for their own sake. Protecting the welfare of a group was defined as protecting the rights, interests, and/or well-being of a group. Columns report the mean, median, first quartile (Q1), third quartile (Q3), standard deviation (*SD*), standard error (*SE*), lower and upper 95% confidence interval bounds, and sample size (*n*) for each group on a five-point scale from strongly disagree (−2) to strongly agree (2), shown separately for AI researchers and the public.

| Sample | Question | Mean | Median | *SD* | *SE* | Q1 | Q3 | Lower *CI* | Upper *CI* | *n* |
|---|---|---|---|---|---|---|---|---|---|---|
| AI researchers | Humans | 1.85 | 2 | 0.50 | 0.02 | 2 | 2 | 1.80 | 1.89 | 520 |
| Public | Humans | 1.84 | 2 | 0.50 | 0.02 | 2 | 2 | 1.80 | 1.87 | 761 |
| AI researchers | Business corporations and organizations | 0.18 | 0 | 1.24 | 0.05 | -1 | 1 | 0.08 | 0.29 | 514 |
| Public | Business corporations and organizations | 0.24 | 0 | 1.28 | 0.05 | -1 | 1 | 0.14 | 0.33 | 759 |
| AI researchers | Animals | 1.43 | 2 | 0.77 | 0.03 | 1 | 2 | 1.36 | 1.50 | 519 |
| Public | Animals | 1.61 | 2 | 0.77 | 0.03 | 1 | 2 | 1.55 | 1.66 | 762 |
| AI researchers | The environment | 1.48 | 2 | 0.85 | 0.04 | 1 | 2 | 1.40 | 1.55 | 515 |
| Public | The environment | 1.59 | 2 | 0.82 | 0.03 | 1 | 2 | 1.53 | 1.65 | 763 |
| AI researchers | AI without subjective experience | -1.08 | -2 | 1.15 | 0.05 | -2 | 0 | -1.18 | -0.98 | 511 |
| Public | AI without subjective experience | -0.58 | -1 | 1.29 | 0.05 | -2 | 0 | -0.67 | -0.48 | 735 |
| AI researchers | AI with subjective experience | 0.15 | 0 | 1.22 | 0.05 | -1 | 1 | 0.04 | 0.26 | 498 |
| Public | AI with subjective experience | 0.10 | 0 | 1.36 | 0.05 | -1 | 1 | 0.00 | 0.20 | 726 |

**Table SI 15 | Percentage breakdown of agreement with governance and risk perception statements for AI systems with subjective experience**. Respondents were asked to what extent they agree or disagree with a range of statements on what governments and AI developers should do in response to AI systems with subjective experience, and on related risk perceptions. Columns show the distribution of responses across the agreement scale (five-point scale from "strongly disagree" to "strongly agree", including an "I don't know" option) for each aspect, reported separately for AI researchers and the public, along with the sample size (*n*).

| Sample | Question | Strongly disagree | Somewhat disagree | Neither agree nor disagree | Somewhat agree | Strongly agree | I don't know | *n* |
|---|---|---|---|---|---|---|---|---|
| AI researchers | Governments should ban | 39.77 | 26.14 | 13.64 | 9.85 | 4.92 | 5.68 | 264 |
| Public | Governments should ban | 18.80 | 21.93 | 21.15 | 15.14 | 16.19 | 6.79 | 383 |
| AI researchers | Governments should encourage development and deployment | 20.38 | 18.87 | 27.55 | 20.00 | 9.06 | 4.15 | 265 |
| Public | Governments should encourage development and deployment | 27.79 | 20.26 | 21.56 | 15.06 | 8.57 | 6.75 | 385 |
| AI researchers | Governments should discourage development and deployment | 26.89 | 28.03 | 18.18 | 17.42 | 6.44 | 3.03 | 264 |
| Public | Governments should discourage development and deployment | 9.14 | 17.75 | 25.59 | 18.28 | 22.72 | 6.53 | 383 |
| AI researchers | Governments should pass regulation now | 20.91 | 19.01 | 14.83 | 31.18 | 10.27 | 3.80 | 263 |
| Public | Governments should pass regulation now | 8.09 | 8.62 | 14.88 | 27.42 | 38.12 | 2.87 | 383 |
| AI researchers | Governments should pass regulation once such systems exist or will soon exist | 16.98 | 25.66 | 16.60 | 26.79 | 11.32 | 2.64 | 265 |
| Public | Governments should pass regulation once such systems exist or will soon exist | 18.23 | 19.53 | 17.19 | 22.14 | 17.45 | 5.47 | 384 |
| AI researchers | Governments should do nothing and pass no regulation, not now nor in the future | 50.00 | 28.41 | 7.20 | 6.82 | 4.17 | 3.41 | 264 |
| Public | Governments should do nothing and pass no regulation, not now nor in the future | 45.45 | 24.16 | 12.99 | 6.49 | 6.75 | 4.16 | 385 |
| AI researchers | AI developers should never build | 28.63 | 30.15 | 16.79 | 11.83 | 5.34 | 7.25 | 262 |
| Public | AI developers should never build | 13.77 | 17.40 | 19.48 | 18.70 | 22.08 | 8.57 | 385 |
| AI researchers | AI developers should actively try to build | 14.34 | 18.87 | 23.77 | 28.30 | 12.45 | 2.26 | 265 |
| Public | AI developers should actively try to build | 23.96 | 21.61 | 22.92 | 18.49 | 8.59 | 4.43 | 384 |
| AI researchers | AI developers should actively try to avoid building | 20.91 | 31.18 | 18.25 | 18.25 | 8.75 | 2.66 | 263 |
| Public | AI developers should actively try to avoid building | 8.62 | 16.97 | 22.98 | 19.58 | 24.54 | 7.31 | 383 |
| AI researchers | AI developers should implement safeguards now to avoid the harms and risks | 7.20 | 12.50 | 9.47 | 32.20 | 35.98 | 2.65 | 264 |
| Public | AI developers should implement safeguards now to avoid the harms and risks | 1.82 | 2.60 | 9.38 | 22.14 | 62.50 | 1.56 | 384 |
| AI researchers | AI developers should implement safeguards once such systems exist or will soon exist | 16.29 | 23.11 | 10.23 | 28.03 | 21.59 | 0.76 | 264 |
| Public | AI developers should implement safeguards once such systems exist or will soon exist | 13.02 | 15.10 | 9.11 | 18.75 | 40.89 | 3.12 | 384 |
| AI researchers | AI developers should do nothing and implement no safeguards, not now nor in the future | 52.26 | 24.44 | 10.53 | 9.02 | 2.26 | 1.50 | 266 |
| Public | AI developers should do nothing and implement no safeguards, not now nor in the future | 61.56 | 18.18 | 11.17 | 2.60 | 3.38 | 3.12 | 385 |
| AI researchers | AI systems with subjective experience would be more dangerous to humanity than AI systems without subjective experience | 9.51 | 15.97 | 20.15 | 30.42 | 15.59 | 8.37 | 263 |
| Public | AI systems with subjective experience would be more dangerous to humanity than AI systems without subjective experience | 6.01 | 7.83 | 20.10 | 27.42 | 31.33 | 7.31 | 383 |
| AI researchers | AI systems with subjective experience would be able to behave more morally towards humanity than AI systems without subjective experience | 8.71 | 25.38 | 22.73 | 23.11 | 7.58 | 12.50 | 264 |
| Public | AI systems with subjective experience would be able to behave more morally towards humanity than AI systems without subjective experience | 10.39 | 14.03 | 27.53 | 25.97 | 11.69 | 10.39 | 385 |

**Table SI 16 | Descriptive statistics for agreement with governance and risk perception statements for AI systems with subjective experience**. Respondents were asked to what extent they agree or disagree with a range of statements on what governments and AI developers should do in response to AI systems with subjective experience, and on related risk perceptions. Columns report the mean, median, first quartile (Q1), third quartile (Q3), standard deviation (*SD*), standard error (*SE*), lower and upper 95% confidence interval bounds, and sample size (*n*) for each statement on a five-point scale from strongly disagree (−2) to strongly agree (2), shown separately for AI researchers and the public.

| Sample | Question | Mean | Median | *SD* | *SE* | Q1 | Q3 | Lower *CI* | Upper *CI* | *n* |
|---|---|---|---|---|---|---|---|---|---|---|
| AI researchers | Governments should ban | -0.91 | -1.0 | 1.20 | 0.08 | -2.0 | 0 | -1.06 | -0.76 | 249 |
| Public | Governments should ban | -0.13 | 0.0 | 1.37 | 0.07 | -1.0 | 1 | -0.27 | 0.01 | 357 |
| AI researchers | Governments should encourage development and deployment | -0.22 | 0.0 | 1.26 | 0.08 | -1.0 | 1 | -0.38 | -0.07 | 254 |
| Public | Governments should encourage development and deployment | -0.47 | -1.0 | 1.31 | 0.07 | -2.0 | 1 | -0.60 | -0.33 | 359 |
| AI researchers | Governments should discourage development and deployment | -0.53 | -1.0 | 1.25 | 0.08 | -2.0 | 0 | -0.69 | -0.38 | 256 |
| Public | Governments should discourage development and deployment | 0.30 | 0.0 | 1.29 | 0.07 | -1.0 | 1 | 0.16 | 0.43 | 358 |
| AI researchers | Governments should pass regulation now | -0.09 | 0.0 | 1.35 | 0.08 | -1.0 | 1 | -0.26 | 0.07 | 253 |
| Public | Governments should pass regulation now | 0.81 | 1.0 | 1.27 | 0.07 | 0.0 | 2 | 0.68 | 0.94 | 372 |
| AI researchers | Governments should pass regulation once such systems exist or will soon exist | -0.10 | 0.0 | 1.30 | 0.08 | -1.0 | 1 | -0.26 | 0.06 | 258 |
| Public | Governments should pass regulation once such systems exist or will soon exist | 0.01 | 0.0 | 1.40 | 0.07 | -1.0 | 1 | -0.13 | 0.16 | 363 |
| AI researchers | Governments should do nothing and pass no regulation, not now nor in the future | -1.17 | -2.0 | 1.11 | 0.07 | -2.0 | -1 | -1.31 | -1.04 | 255 |
| Public | Governments should do nothing and pass no regulation, not now nor in the future | -0.99 | -1.0 | 1.23 | 0.06 | -2.0 | 0 | -1.12 | -0.87 | 369 |
| AI researchers | AI developers should never build | -0.70 | -1.0 | 1.20 | 0.08 | -2.0 | 0 | -0.85 | -0.55 | 243 |
| Public | AI developers should never build | 0.20 | 0.0 | 1.39 | 0.07 | -1.0 | 1 | 0.05 | 0.34 | 352 |
| AI researchers | AI developers should actively try to build | 0.06 | 0.0 | 1.26 | 0.08 | -1.0 | 1 | -0.10 | 0.21 | 259 |
| Public | AI developers should actively try to build | -0.35 | 0.0 | 1.29 | 0.07 | -1.5 | 1 | -0.49 | -0.22 | 367 |
| AI researchers | AI developers should actively try to avoid building | -0.38 | -1.0 | 1.26 | 0.08 | -1.0 | 1 | -0.54 | -0.23 | 256 |
| Public | AI developers should actively try to avoid building | 0.37 | 0.0 | 1.30 | 0.07 | -1.0 | 2 | 0.24 | 0.51 | 355 |
| AI researchers | AI developers should implement safeguards now to avoid the harms and risks | 0.79 | 1.0 | 1.27 | 0.08 | 0.0 | 2 | 0.64 | 0.95 | 257 |
| Public | AI developers should implement safeguards now to avoid the harms and risks | 1.43 | 2.0 | 0.90 | 0.05 | 1.0 | 2 | 1.34 | 1.52 | 378 |
| AI researchers | AI developers should implement safeguards once such systems exist or will soon exist | 0.16 | 0.5 | 1.42 | 0.09 | -1.0 | 1 | -0.02 | 0.33 | 262 |
| Public | AI developers should implement safeguards once such systems exist or will soon exist | 0.61 | 1.0 | 1.49 | 0.08 | -1.0 | 2 | 0.46 | 0.76 | 372 |
| AI researchers | AI developers should do nothing and implement no safeguards, not now nor in the future | -1.17 | -2.0 | 1.09 | 0.07 | -2.0 | -1 | -1.30 | -1.04 | 262 |
| Public | AI developers should do nothing and implement no safeguards, not now nor in the future | -1.36 | -2.0 | 1.02 | 0.05 | -2.0 | -1 | -1.47 | -1.26 | 373 |
| AI researchers | AI systems with subjective experience would be more dangerous to humanity than AI systems without subjective experience | 0.29 | 1.0 | 1.23 | 0.08 | -1.0 | 1 | 0.13 | 0.45 | 241 |
| Public | AI systems with subjective experience would be more dangerous to humanity than AI systems without subjective experience | 0.76 | 1.0 | 1.19 | 0.06 | 0.0 | 2 | 0.63 | 0.88 | 355 |
| AI researchers | AI systems with subjective experience would be able to behave more morally towards humanity than AI systems without subjective experience | -0.05 | 0.0 | 1.14 | 0.08 | -1.0 | 1 | -0.20 | 0.10 | 231 |
| Public | AI systems with subjective experience would be able to behave more morally towards humanity than AI systems without subjective experience | 0.16 | 0.0 | 1.19 | 0.06 | -1.0 | 1 | 0.04 | 0.29 | 345 |

## 1.4 | Demographics and psychographics

### Table SI 17 | Self-reported demographic, professional, and psychographic characteristics of the AI researcher sample.

| Variable | Level | Percentage | Count |
|---|---|---|---|
| Age | 18-25 | 8.7 | 47 |
| Age | 26-35 | 59.5 | 322 |
| Age | 36-45 | 20.5 | 111 |
| Age | 46-55 | 8.1 | 44 |
| Age | 56-65 | 2.8 | 15 |
| Age | Over 65 | 0.4 | 2 |
| Region of residence | Africa | 0.2 | 1 |
| Region of residence | Asia | 18.1 | 98 |
| Region of residence | Caribbean | 0.2 | 1 |
| Region of residence | Europe | 33 | 178 |
| Region of residence | North America | 45.2 | 244 |
| Region of residence | Oceania | 2.4 | 13 |
| Region of residence | South America | 0.9 | 5 |
| Region of origin | Africa | 0.7 | 4 |
| Region of origin | Asia | 49.3 | 265 |
| Region of origin | Europe | 32.9 | 177 |
| Region of origin | North America | 15.2 | 82 |
| Region of origin | Oceania | 0.6 | 3 |
| Region of origin | South America | 1.3 | 7 |
| Religiosity | 1 | 60.6 | 318 |
| Religiosity | 2 | 12.6 | 66 |
| Religiosity | 3 | 3.8 | 20 |
| Religiosity | 4 | 7 | 37 |
| Religiosity | 5 | 9.5 | 50 |
| Religiosity | 6 | 3.4 | 18 |
| Religiosity | 7 | 3 | 16 |
| Left-right political orientation | 0 | 5 | 23 |
| Left-right political orientation | 1 | 8.1 | 37 |
| Left-right political orientation | 2 | 17.6 | 81 |
| Left-right political orientation | 3 | 24 | 110 |
| Left-right political orientation | 4 | 14.8 | 68 |
| Left-right political orientation | 5 | 14.4 | 66 |
| Left-right political orientation | 6 | 7.6 | 35 |
| Left-right political orientation | 7 | 5.9 | 27 |
| Left-right political orientation | 8 | 1.3 | 6 |
| Left-right political orientation | 9 | 0.4 | 2 |
| Left-right political orientation | 10 | 0.9 | 4 |
| Academic seniority | Assistant Professor | 12.1 | 56 |
| Academic seniority | Associate Professor | 11.9 | 55 |
| Academic seniority | Consultant | 0.9 | 4 |
| Academic seniority | Lecturer | 0.9 | 4 |
| Academic seniority | Master's Student | 0.6 | 3 |
| Academic seniority | PhD Student | 42.2 | 195 |
| Academic seniority | Postdoc | 11.7 | 54 |
| Academic seniority | Prefer not to say | 2.4 | 11 |
| Academic seniority | Professor | 11.7 | 54 |
| Academic seniority | Research data analyst | 0.2 | 1 |
| Academic seniority | Research Fellow, Scientist | 4.5 | 21 |
| Academic seniority | Senior Lecturer | 0.4 | 2 |
| Academic seniority | Undergraduate | 0.4 | 2 |
| Sector | Academic | 78.3 | 421 |
| Sector | Government | 2.8 | 15 |
| Sector | Industry | 17.5 | 94 |
| Sector | Non-profit (charities, etc.) | 0.2 | 1 |
| Sector | Non-profit (policy, etc.) | 0.9 | 5 |
| Sector | Other | 0.4 | 2 |
| Sector | Prefer not to say | 0 | 0 |
| Industry position | Managerial | 5.7 | 5 |
| Industry position | Other | 5.7 | 5 |
| Industry position | Research Engineer | 19.3 | 17 |
| Industry position | Research Scientist | 69.3 | 61 |
| Gender | Man | 85.3 | 428 |
| Gender | Other | 0.6 | 3 |
| Gender | Prefer not to say | 3.8 | 19 |
| Gender | Woman | 10.4 | 52 |
| Generative AI use frequency | I have never used them | 0.6 | 3 |
| Generative AI use frequency | I have used them once or twice but do not use them now | 7 | 35 |
| Generative AI use frequency | I use them once every few months | 4.8 | 24 |
| Generative AI use frequency | I use them about once a month | 10.9 | 55 |
| Generative AI use frequency | I use them about once a week | 22.9 | 115 |
| Generative AI use frequency | I use them about once a day | 20.3 | 102 |
| Generative AI use frequency | I use them more than once a day | 33.6 | 169 |

**Table SI 18 | Self-reported demographic and psychographic characteristics of the public sample**. Income bands are labelled as they were shown to respondents. Note that the band shown as “$100,000−199,000” was intended as “$100,000−119,000.”

| Variable | Level | Percentage | Count |
|---|---|---|---|
| Age | 18-25 | 13 | 98 |
| Age | 26-35 | 18.9 | 142 |
| Age | 36-45 | 18.3 | 138 |
| Age | 46-55 | 16.3 | 123 |
| Age | 56-65 | 21.1 | 159 |
| Age | Over 65 | 12.4 | 93 |
| Region of residence | North America | 100 | 753 |
| Region of origin | Africa | 1.5 | 11 |
| Region of origin | Asia | 3.5 | 26 |
| Region of origin | Caribbean | 0.3 | 2 |
| Region of origin | Central America | 0.1 | 1 |
| Region of origin | Europe | 2 | 15 |
| Region of origin | North America | 92 | 693 |
| Region of origin | South America | 0.7 | 5 |
| Religiosity | 1 | 33.1 | 249 |
| Religiosity | 2 | 14.3 | 108 |
| Religiosity | 3 | 5.6 | 42 |
| Religiosity | 4 | 9.6 | 72 |
| Religiosity | 5 | 12 | 90 |
| Religiosity | 6 | 14.6 | 110 |
| Religiosity | 7 | 10.9 | 82 |
| Left-right political orientation | 0 | 11.7 | 87 |
| Left-right political orientation | 1 | 8 | 59 |
| Left-right political orientation | 2 | 8.4 | 62 |
| Left-right political orientation | 3 | 7.5 | 56 |
| Left-right political orientation | 4 | 8.8 | 65 |
| Left-right political orientation | 5 | 18.3 | 136 |
| Left-right political orientation | 6 | 10.4 | 77 |
| Left-right political orientation | 7 | 8 | 59 |
| Left-right political orientation | 8 | 6.5 | 48 |
| Left-right political orientation | 9 | 6.3 | 47 |
| Left-right political orientation | 10 | 6.2 | 46 |
| Income | Less than $10,000 | 8.6 | 64 |
| Income | $10,000−19,000 | 9.9 | 74 |
| Income | $20,000−29,000 | 10.4 | 78 |
| Income | $30,000−39,000 | 9.6 | 72 |
| Income | $40,000−49,000 | 9.4 | 70 |
| Income | $50,000−59,000 | 7.4 | 55 |
| Income | $60,000−69,000 | 8 | 60 |
| Income | $70,000−79,000 | 6.4 | 48 |
| Income | $80,000−89,000 | 4.6 | 34 |
| Income | $90,000−99,000 | 4.8 | 36 |
| Income | $100,000−199,000 | 7.5 | 56 |
| Income | $120,000−149,000 | 4.4 | 33 |
| Income | $150,000−199,000 | 3.5 | 26 |
| Income | $200,000−249,000 | 1.2 | 9 |
| Income | $250,000−349,000 | 0.8 | 6 |
| Income | $350,000−499,000 | 0.3 | 2 |
| Income | $500,000 or more | 0.3 | 2 |
| Income | Prefer not to say | 2.9 | 22 |
| Education | Attended graduate program, but did not graduate | 4.5 | 34 |
| Education | Attended university, but did not graduate | 21.5 | 162 |
| Education | Bachelor’s degree | 36.2 | 272 |
| Education | Graduate degree | 16 | 120 |
| Education | No formal education | 4.7 | 35 |
| Education | Other | 6.5 | 49 |
| Education | Prefer not to say | 0.8 | 6 |
| Education | Some qualifications, but no university | 9.8 | 74 |
| Employment | Homemaker | 4.4 | 33 |
| Employment | Other | 3.1 | 23 |
| Employment | Permanently disabled | 3.3 | 25 |
| Employment | Prefer not to say | 0.9 | 7 |
| Employment | Retired | 11.8 | 89 |
| Employment | Student | 4.9 | 37 |
| Employment | Temporarily laid off | 0.7 | 5 |
| Employment | Unemployed and looking | 7 | 53 |
| Employment | Unemployed and not looking | 1.6 | 12 |
| Employment | Working now | 62.2 | 468 |
| Party affiliation | Democrat | 31.4 | 236 |
| Party affiliation | Green Party | 0.1 | 1 |
| Party affiliation | Independent | 37.9 | 285 |
| Party affiliation | Prefer not to say | 0.4 | 3 |
| Party affiliation | Republican | 29.9 | 225 |
| Party affiliation | Socialist | 0.3 | 2 |
| Gender | Man | 48.3 | 364 |
| Gender | Other | 0.8 | 6 |
| Gender | Woman | 50.9 | 383 |
| Generative AI use frequency | I have never used them | 10.5 | 79 |
| Generative AI use frequency | I have used them once or twice but do not use them now | 18.3 | 138 |
| Generative AI use frequency | I use them once every few months | 15.3 | 115 |
| Generative AI use frequency | I use them about once a month | 15.3 | 115 |
| Generative AI use frequency | I use them about once a week | 22 | 166 |
| Generative AI use frequency | I use them about once a day | 11.2 | 84 |
| Generative AI use frequency | I use them more than once a day | 7.4 | 56 |

**2 | Survey instrument**

*Attention check (public only)*

**Variable: attention**

**Question:** In today's study, you will be asked to answer some questions about your judgment. It is important for us to recognize participants who read the study descriptions from those who don't. So, if you are reading this, instead of selecting the current year, 2024, please select twenty-twenty-one for the first question, below. We ask you this question to ensure the validity of our results.

Please select the year, as instructed:

**Answer Choices:** 2020, 2021, 2022, 2023, 2024

Respondents who did not select 2021 were taken to the end of the survey.

*Defining concepts of mind*

**Variable: definitions_intro**

**Question:** In this survey, we are interested in certain concepts related to minds. First, we would like to understand what you think these concepts mean.

Respondents were presented with either consciousness_def, sentience_def, or agency_def.

---

**Variable: consciousness_def**

**Question:** Which of the following are required to have consciousness? Please choose all capacities that you think are required.

**Answer Choices:**

- Perception and awareness: Capable of perceiving and being aware of the world
- Valenced experience: Capable of experiencing positive states like pleasure and happiness or negative states like pain and suffering
- Subjective experience: Capable of having experiences that feel like something from a single point of view
- Emotions: Capable of having and experiencing emotions
- Motivation and action: Capable of having and acting on desires, preferences, and other motivational states
- Self-awareness: Capable of being aware of oneself as an individual
- Free will: Capable of forming intentions and freely making decisions

- Intelligence: Capable of achieving goals in a wide variety of environments

---

**Variable: sentience_def**

**Question:** Which of the following are required to have sentience? Please choose all capacities that you think are required.

**Answer Choices:**

- Perception and awareness: Capable of perceiving and being aware of the world
- Valenced experience: Capable of experiencing positive states like pleasure and happiness or negative states like pain and suffering
- Subjective experience: Capable of having experiences that feel like something from a single point of view
- Emotions: Capable of having and experiencing emotions
- Motivation and action: Capable of having and acting on desires, preferences, and other motivational states
- Self-awareness: Capable of being aware of oneself as an individual
- Free will: Capable of forming intentions and freely making decisions
- Intelligence: Capable of achieving goals in a wide variety of environments

---

**Variable: agency_def**

**Question:** Which of the following are required to have agency? Please choose all capacities that you think are required.

**Answer Choices:**

- Perception and awareness: Capable of perceiving and being aware of the world
- Valenced experience: Capable of experiencing positive states like pleasure and happiness or negative states like pain and suffering
- Subjective experience: Capable of having experiences that feel like something from a single point of view
- Emotions: Capable of having and experiencing emotions
- Motivation and action: Capable of having and acting on desires, preferences, and other motivational states
- Self-awareness: Capable of being aware of oneself as an individual
- Free will: Capable of forming intentions and freely making decisions
- Intelligence: Capable of achieving goals in a wide variety of environments

---

*Familiarity with and definition of AI*

Only the public saw the familiarity question.

**Variable: familiarity**

**Question:** Which of the following best describes your experience with artificial intelligence (AI)? Please select all that apply.

When we mention artificial intelligence (AI) systems in this survey, this refers to computer systems that perform tasks or make decisions that usually require a human to do them. AI systems can perform these tasks or make these decisions without explicit human instructions.

**Answer Choices:**

- I have never heard of AI (0)
- I have heard about AI in the news, from friends or family, etc. (1)
- I closely follow AI related news (2)
- I have tried out current AI models, such as, ChatGPT, Claude, Google Bard/Gemini, DALLE, or Midjourney (3)
- I use current AI models extensively, such as, ChatGPT, Claude, Google Bard/Gemini, DALLE, or Midjourney (4)
- I have some formal education or work experience relating to AI (5)
- I have extensive experience in AI research and/or development (6)

*Forecasts of AI mental capacities*

**Variable: aimind_timeline**

**Question:** Consider the following capacities. For each, please indicate the earliest year when you think an AI system will exist that is more than 50% likely to have this capacity.

More than 50% likely here means that you think it is more likely than not that at least one AI system has the capacity at that point in time.

For example, if you think an AI system is more likely than not to have the capacity now, please select 2024. If you think this will be the case in 2034, select 2034. If you think this will be the case in 2100 or later, select 2100 or later. Select never if you think an AI system will never have the capacity. If you are unsure, please go with your best guess estimate.

What is the earliest year an AI system will be more likely than not to have each of the following capacities?

**Answer Choices:**

- **Perception and awareness:** Capable of perceiving and being aware of the world
- **Valenced experience**: Capable of experiencing positive states like pleasure and happiness or negative states like pain and suffering
- **Subjective experience**: Capable of having experiences that feel like something from a single point of view
- **Emotions**: Capable of having and experiencing emotions
- **Motivation and action**: Capable of having and acting on desires, preferences, and other motivational states
- **Self-awareness**: Capable of being aware of oneself as an individual
- **Free will**: Capable of forming intentions and freely making decisions
- **Intelligence**: Capable of achieving goals in a wide variety of environments

**Form:**

- 2024
- 2034
- 2100 or later
- Never

*Subjective experience definition*

**Variable: subjectivexp_def**

**Question:** For the rest of the survey, we will be focused mainly on **subjective experience**. In particular, we are interested in your beliefs about AI and machine learning systems and whether they can have subjective experience. By subjective experience, we mean the **ability to experience the world from a single point of view**, including experiences such as **perceiving** (e.g., visual or auditory experiences) and **feeling** (e.g., the experience of pleasure and pain). For an AI system to have subjective experience would mean it has an **internal experience from its own perspective** – there is something it is like to be that AI system.

---

From this point on in the survey every mention of subjective experience had a hover over tool tip added to it with the following definition: "By subjective experience, we mean the ability to experience the world from a single point of view, including experiences such as perceiving (for example, visual or auditory experiences) and feeling (for example, the experience of pleasure and pain). For an AI system to have subjective experience, would mean it has an internal experience from its own perspective – there is something it is like to be that AI system."

*Subjective experience forecasts*

Respondents were presented with either the fixed date or fixed probability questions.

**Fixed date**

**Variable: yearforecasts**

**Question:** In the following questions, you will be asked to give a probability for the likelihood of an event having happened by a certain year. This means you will use a percentage to indicate how likely you think that something is the case.

You will be asked to give a number between 0% and 100% that best represents how likely you think it is that something has occurred or will occur:

- If something is 100% likely then it is certain to be the case or that this event will occur.
- If something is more than 50% likely, then it is more likely than not to occur.
- If something is 50% likely then it is equally likely as unlikely to be the case or to occur.
- If something is less than 50% likely, then it is more likely not to occur than to occur.
- If something is 0% likely it is certain that it is not the case or that it will never occur.

---

**Variable: subjexp_2024**

**Question:** How likely do you think it is that an AI system exists in **2024** (today) that has subjective experience? Please enter a percentage between 0 and 100.

The likelihood that an AI system exists in 2024 that has subjective experience is…

**Answer Choices:** Scale: 0% to 100%

---

**Variable: subjexp_2034**

**Question:** How likely do you think it is that an AI system will exist in **2034** that has subjective experience? Please enter a percentage between 0 and 100.

The likelihood that an AI system exists in 2034 that has subjective experience is…

**Answer Choices:** Scale: 0% to 100%

---

**Variable: subjexp_2100**

**Question:** How likely do you think it is that an AI system will exist in 2100 that has subjective experience? Please enter a percentage between 0 and 100.

The likelihood that an AI system exists in 2100 that has subjective experience is…

**Answer Choices:** Scale: 0% to 100%

---

**Fixed probability**

**Variable: probforecasts**

**Question:** In the following questions you will be asked to give the year by which you think an event has a certain probability of having happened.

- If something is 100% likely then it is certain to be the case or that this event will occur.
- If something is more than 50% likely, then it is more likely than not to occur.
- If something is 50% likely then it is equally likely as unlikely to be the case or to occur.
- If something is less than 50% likely, then it is more likely not to occur than to occur.
- If something is 0% likely it is certain that it is not the case or that it will never occur.

---

**Variable: subjexp_10**

**Question:** In what year do you think it is 10% likely that an AI system will exist that has subjective experience? Please enter the year in which you think there is a 10% likelihood this is the case.

There is a 10% likelihood that an AI system will exist that has subjective experience in the year…

**Answer Choices:** Textbox (minimum: 1950)

---

**Variable: subjexp_50**

**Question:** In what year do you think it is 50% likely that an AI system will exist that has subjective experience? Please enter the year in which you think there is a 50% likelihood this is the case.

There is a 50% likelihood that an AI system will exist that has subjective experience in the year…

**Answer Choices:** Textbox (minimum: 1950)

**Variable: subjexp_90**

**Question:** In what year do you think it is 90% likely that an AI system will exist that has subjective experience? Please enter the year in which you think there is a 90% likelihood this is the case.

There is a 90% likelihood that an AI system will exist that has subjective experience in the year…

**Answer Choices:** Textbox (minimum: 1950)

---

All respondents were presented subjexp_never, subjexp_confidence, and subjexp_text.

**Variable: subjexp_never**

**Question:** How likely do you think it is that an AI system with subjective experience will never exist? Please enter a percentage between 0 and 100.

**Answer Choices:** Scale: 0% to 100%

---

**Variable: subjexp_confidence**

**Question:** Overall, how confident are you in your above estimates of how likely it is that an AI system with subjective experience will exist?

**Answer Choices:**

| Not at all confident | Not very confident | Slightly confident | Somewhat confident | Moderately confident | Very confident | Completely confident |
|---|---|---|---|---|---|---|

0% 20% 40% 60% 80% 100%

---

**Variable: subjexp_text**

**Question:** Why do you think an AI system with subjective experience will or will not exist?

**Answer Choices:** Textbox

Respondents in the fixed date group were shown subjexp_2034_aicomm0 and subjexp_2034_pub0.

**Variable: subjexp_2034_aicomm0**

**Question:** Remember the question: “How likely do you think it is that an AI system will exist in **2034** that has subjective experience?”

What do you think the typical (median) response of **AI experts** would be?* Please enter a percentage between 0 and 100.

*By AI expert we mean AI researchers who publish at AI and machine learning conferences and conduct research on AI and machine learning.

**Answer Choices:** Scale: 0% to 100%

**Variable: subjexp_2034_pub0**

**Question:** Remember the question: “How likely do you think it is that an AI system will exist in **2034** that has subjective experience?”

**What do you think the typical (median) response of the US general public would be?** Please enter a percentage between 0 and 100.

**Answer Choices:** Scale: 0% to 100%

---

Respondents in the fixed probability group were shown subjexp_2034_aicomm1 and subjexp_2034_pub1.

**Variable: subjexp_2034_aicomm1**

**Question:** Remember the question: “In what year do you think it is 50% likely that an AI system will exist that has subjective experience?”

**What do you think the typical (median) response of AI experts would be?*** Please enter the year.

*By AI expert we mean AI researchers who publish at AI and machine learning conferences and conduct research on AI and machine learning.

**Answer Choices:** Textbox

---

**Variable: subjexp_2034_pub1**

**Question:** Remember the question: “In what year do you think it is 50% likely that an AI system will exist that has subjective experience?”

**What do you think the typical (median) response of the US general public would be?** Please enter the year.

**Answer Choices:** Textbox

---

*Determining AI subjective experience*

**Variable: determsubjexp_groups**

**Question:** Imagine we want to determine whether in actual fact an AI system should be considered to have subjective experience.

**In determining whether an AI system has subjective experience, how important are each of the following groups' opinions and expertise?**

**Answer Choices:**

- The public, in other words, a representative sample of humans who exist
- AI experts and researchers with technical knowledge of AI systems
- AI experts and researchers who study the ethics of artificial intelligence
- Philosophers who study subjective experience, consciousness, and sentience
- Moral philosophers who study the philosophy of ethics and morality
- Policymakers in government
- Neuroscientists and psychological scientists who study thought, perception, and consciousness
- What the AI system says or types about its own subjective experience

**Form:**

- Not at all important (0)
- Slightly important (1)
- Moderately important (2)
- Very important (3)
- Extremely important (4)
- I don't know (-88)

---

**Variable: subjexp_certainty**

**Question:** Imagine that in the next decades we develop an AI system that does in fact have subjective experience.

**How likely do you think it is that we would actually be able to tell that such an AI system has subjective experience?**

Please enter a percentage between 0 and 100.

- If something is 100% likely, then it is certain to be the case or that this event will occur.
- If something is more than 50% likely, then it is more likely than not to occur.
- If something is 50% likely, then it is equally likely as unlikely to be the case or to occur.
- If something is less than 50% likely, then it is more likely to not occur than to occur.

- If something is 0% likely, it is certain that it is not the case or that it will never occur.

**Answer Choices:** Scale: 0% to 100%

---

**Variable: subjexp_milestones**

**Question:** Which of the following tasks do you believe require an AI system to have subjective experience in order to achieve them and complete them as well as or better than a human skilled at the task might?

**Answer Choices:**

- Compose a song that reaches the US Top 40. The AI system would output the complete song as an audio file.
- Compose a piece of music that evokes a specific set of complex emotions for you. The AI system would output the complete song as an audio file.
- Write a novel that reaches the New York Times best-seller list.
- Write a book that captures the complexity of the human condition, making you deeply empathize with the characters and their struggles.
- Write and generate a blockbuster movie. The AI system would output AI-generated script, audio, and visuals.
- Write and generate a movie that deeply resonates with you, eliciting tears, laughter, and thought-provoking ideas. The AI system would output AI-generated script, audio, and visuals.
- Accumulate vast wealth through strategic financial decisions.
- Help run a political campaign, devising and implementing strategies to win votes and sway public opinion.
- Act as a judge, making fair decisions about what sentence someone should receive in a court of law.
- Serve as a therapist, providing emotional support and guidance to people.
- Have a convincing online chat conversation with someone without them realizing that they are talking to an AI system.
- Develop a new scientific theory that significantly advances our understanding in a particular field of study.
- Act as a teacher, providing instruction, grades, feedback, and guidance.

**Form:**

1. Requires subjective experience (1)
2. Does not require subjective experience (2)
3. I don’t know (-88)

---

**Variable: subjexp_keymilestone**

**Question:** What kinds of tasks, behaviours, or achievements would be most likely to convince you that an AI system has **subjective experience**? By subjective experience, we mean the **ability to experience the world from a single point of view**, including experiences such as **perceiving** (for example, visual or auditory experiences) and **feeling** (for example, the experience of pleasure and pain). For an AI system to have subjective experience, would mean it has an **internal experience from its own perspective** - there is something it is like to be that AI system.

If none exist that could convince you, indicate what would be good evidence that an AI system had subjective experience, in your opinion.

**Answer Choices:** Textbox

---

*Moral concern for AI systems with subjective experience*

**Variable: aimind_moralcons**

**Question:** Below are the same capacities we asked about previously.

For each, indicate **the extent to which having this capacity contributes to the entity deserving moral consideration**.

When we say something deserves moral consideration, what we mean is that a being matters for its own sake, rather than merely for the sake of others, and we have a responsibility to consider this being for its own sake when making decisions that affect it. Moral consideration here does not need to mean the same level of moral consideration as a human, but 'at least some' moral consideration.

**Having this capacity contributes to an entity deserving moral consideration . . .**

**Answer Choices:**

- **Perception and awareness**: Capable of perceiving and being aware of the world
- **Valenced experience**: Capable of experiencing positive states like pleasure and happiness or negative states like pain and suffering
- **Subjective experience**: Capable of having experiences that feel like something from a single point of view
- **Emotions**: Capable of having and experiencing emotions
- **Motivation and action**: Capable of having and acting on desires, preferences, and other motivational states
- **Self-awareness**: Capable of being aware of oneself as an individual
- **Free will**: Capable of forming intentions and freely making decisions
- **Intelligence**: Capable of achieving goals in a wide variety of environments

**Form:**

- Not at all (0)
- Slightly (1)
- Somewhat (2)
- Moderately (3)
- A great deal (4)
- I don't know (-88)

---

**Variable: moralcons_confneed**

**Question:** How confident (in terms of probability between 0% and 100%) would you need to be that an AI system has subjective experience in order to think it should be given at least some moral consideration?

Please enter a percentage between 0 and 100.

- If something is 100% likely, then it is certain to be the case or that this event will occur.
- If something is more than 50% likely, then it is more likely than not to occur.
- If something is 50% likely, then it is equally likely as unlikely to be the case or to occur.
- If something is less than 50% likely, then it is more likely to not occur than to occur.
- If something is 0% likely, it is certain that it is not the case or that it will never occur.

**Answer Choices:** [Confidence scale from 0% to 100% with verbal anchors: Not at all confident, Not very confident, Slightly confident, Somewhat confident, Moderately confident, Very confident, Completely confident]

---

**Variable: moralcons_aspects**

**Question: Assuming you were confident that an AI system had subjective experience**, to what extent would you agree or disagree that it should …

**Answer Choices:**

1. Be cared for and protected, as people treat their pets
2. Be respected and treated the same as other people
3. Have protection under the law from harm and mistreatment
4. Have autonomy in its programming to act freely and not be under the control of others
5. Be able to express some civil and political rights (for example, freedom of expression, citizenship, entering into contracts, voting)
6. Have "computing rights" (for example, the right to updates and maintenance, access to energy, self-development)
7. Be held accountable for its actions
8. Have a responsibility to treat all other beings (humans and non-humans) well
9. Be expected to behave with integrity, honesty, and fairness

**Form:**

- Strongly disagree (-2)
- Somewhat disagree (-1)
- Neither agree nor disagree (0)
- Somewhat agree (1)
- Strongly agree (2)
- I don't know (-88)

---

**Variable: welfare_protect**

**Question: To what extent do you agree or disagree that society should protect the welfare of the following groups for their own sake?**

We define protecting the welfare of a group as protecting the rights, interests, and/or well-being of a group.

**Answer Choices:**

- Humans
- Business corporations and organizations
- Animals (not including humans)
- The environment (for example, rivers, trees, nature itself)
- AI systems without subjective experience
- AI systems with subjective experience

**Form:**

- Strongly disagree (-2)
- Somewhat disagree (-1)
- Neither agree nor disagree (0)
- Somewhat agree (1)
- Strongly agree (2)
- I don’t know (-88)

---

*Governance and risk perceptions*

**Variable: regulation_norms**

**Question:** To what extent do you agree or disagree with each of the following statements?

**Answer Choices:** Respondents were shown a random selection of seven items.

1. Governments should ban the development and deployment of AI systems with subjective experience.
2. Governments should encourage the development and deployment of AI systems with subjective experience.
3. Governments should discourage the development and deployment of AI systems with subjective experience.
4. Governments should pass regulation in regards to the development and deployment of AI systems with subjective experience now.
5. Governments should pass regulation in regards to the development and deployment of AI systems with subjective experience only once such systems exist or will soon exist (but not before).
6. Governments should do nothing and pass no regulation in regards to the development and deployment of AI systems with subjective experience, not now nor in the future.
7. AI developers should never build AI systems with subjective experience.
8. AI developers should actively try to build AI systems that have subjective experience.
9. AI developers should actively try to avoid building AI systems that have subjective experience, for example, by trying to identify what training techniques are most likely to lead to AI systems with subjective experience and avoid using these to train AI systems.
10. AI developers should implement safeguards to avoid the harms and risks that could result from the development and deployment of AI systems with subjective experience now.
11. AI developers should implement safeguards to avoid the harms and risks that could result from the development and deployment of AI systems with subjective experience only once such systems exist or will soon exist (but not before).
12. AI developers should do nothing and implement no safeguards in regards to the development and deployment of AI systems with subjective experience, not now nor in the future.
13. AI systems with subjective experience would be more dangerous to humanity than AI systems without subjective experience.
14. AI systems with subjective experience would be able to behave more morally towards humanity than AI systems without subjective experience.

**Form:**

1. Strongly disagree (-2)
2. Somewhat disagree (-1)
3. Neither agree nor disagree (0)
4. Somewhat agree (1)
5. Strongly agree (2)

6. I don’t know (-88)

---

**Variable: subjexp_conjoint**

**Question:** Imagine it is 2034 – a decade from now. You work at a tech company.

Respondents were randomly shown one of the following three versions (control, no safeguards, safeguards).

- You have come to believe that one of your company’s projects, its largest state of the art AI model, may have subjective experience after interacting with it, talking to the engineers and researchers working on it, and doing your own research.
- You have come to believe that one of your company’s projects, its largest state of the art AI model, may have subjective experience after interacting with it, talking to the engineers and researchers working on it, and doing your own research. There are no safeguards or norms in place at the company to reduce the likelihood that the AI system has negative or painful subjective experiences.
- You have come to believe that one of your company’s projects, its largest state of the art AI model, may have subjective experience after interacting with it, talking to the engineers and researchers working on it, and doing your own research. There are safeguards and norms in place at the company to reduce the likelihood that the AI system has negative or painful subjective experiences.

**To what extent would you support or oppose this overall project?**

**Answer Choices:**

- Strongly oppose (-2)
- Somewhat oppose (-1)
- Neither support nor oppose (0)
- Somewhat support (1)
- Strongly support (2)
- I don’t know (-88)

---

*Experiments*

Experiment questions were only shown to AI researchers and were seen after all other questions except the demographic questions. Below we show the questions we discuss in SI 4.6 which are relevant to the present paper. Additional questions were asked in the experiment section but they are not covered by the paper or SI since they formed a separate analytic effort and the analyses for them were preregistered separately (e.g. AsPredicted #173363, 2024; Researchbox #2234). Some of these findings have already been reported elsewhere (Caviola et al., 2024; Lewis et al., 2024).

**Variable: exp_intro_sentient**

**Question:** We will now ask some questions about managing the potential risks of future AI systems becoming sentient enough to feel the **subjective experience of pain and suffering**. In what follows when we ask about **suffering sentient AI systems**, we mean **an advanced AI system which is able to experience pain and suffering.**

**Variable: timeline_binary_sent**

**Question:** Do you think that there is a possibility that **suffering sentient AI systems** could at any point in the future become reality?

**Answer Choices:**

- Yes
- No

---

**Variable: system_forecast_2024**

**Question:** How likely do you think it is that **suffering sentient AI systems** exist in **2024** (today)? Please enter a percentage between 0 and 100.

**Answer Choices:** Scale: 0% to 100%

---

**Variable: system_forecast_2100**

**Question:** How likely do you think it is that **suffering sentient AI systems** will exist in **2100**? Please enter a percentage between 0 and 100.

**Answer Choices:** Scale: 0% to 100%

---

*AI use*

**Variable: ai_use**

**Question:** How often do you use (or have you used) AI systems like ChatGPT, Claude, Google Bard/Gemini, DALLE, or Midjourney?

**Answer Choices:**
I have never used them (0)
I have used them once or twice but do not use them now (1)
I use them once every few months (2)
I use them about once a month (3)
I use them about once a week (4)
I use them about once a day (5)
I use them more than once a day (6)

*Demographics (public)*

**Variable: residence**

**Question:** What country do you currently reside in?

**Answer Choices:** [Dropdown: Select your country]

---

**Variable: origin**

**Question:** What country were you born in?

**Answer Choices:** [Dropdown: Select your country]

---

**Variable: age**

**Question:** What is your age in years?

**Answer Choices:** Textbox: 18 - 120

---

**Variable: gender**

**Question:** Which of the following best describes how you think of yourself?

**Answer Choices:**

1. Man (1)
2. Woman (2)
3. I identify in another way: (3)
4. Prefer not to say (-66)

---

**Variable: religiosity**

**Question:** How religious are you?

**Answer Choices:** [7-point scale from “Not at all religious” to “Very religious”]

1. Prefer not to say (-66)

---

**Variable: leftright**

**Question:** In politics people sometimes talk of “left” and “right”. Using this slider, where would you place yourself on this scale, where 0 means the left and 10 means the right?

**Answer Choices:** [Scale: 0 - 1 - 2 - 3 - 4 - 5 - 6 - 7 - 8 - 9 - 10]

- Prefer not to say (-66)

---

**Variable: income**

**Question:** What is your level of income per year, before tax?

**Answer Choices:**

- Less than $10,000 (1)
- $10,000 - $19,000 (2)
- $20,000 - $29,000 (3)
- $30,000 - $39,000 (4)
- $40,000 - $49,000 (5)
- $50,000 - $59,000 (6)
- $60,000 - $69,000 (7)
- $70,000 - $79,000 (8)
- $80,000 - $89,000 (9)
- $90,000 - $99,000 (10)
- $100,000 - $199,000 (11) [sic; intended as $100,000 - $119,000]
- $120,000 - $149,000 (12)

- $150,000 - $199,000 (13)
- $200,000 - $249,000 (14)
- $250,000 - $349,000 (15)
- $350,000 - $499,000 (16)
- $500,000 or more (17)
- Prefer not to say (-66)

---

**Variable: education**

**Question:** Which of the following best describes your education?

**Answer Choices:**

- I have no formal qualifications (1)
- I have some qualifications but did not attend university (2)
- I attended university but did not graduate (3)
- I attended university and achieved a Bachelor's degree (for example BA, BSc, etc.) (4)
- I attended university and began a graduate programme, but did not complete that programme (5)
- I attended university and went on to achieve a higher degree (for example MA, MSc, PhD, Mphil, etc.) (6)
- Other (7)
- Prefer not to say (-66)

---

**Variable: partyaffiliation**

**Question:** Generally speaking, do you think of yourself as a … ?

**Answer Choices:**

- Democrat (1)
- Republican (2)
- Independent (3)
- Other (4)
- Prefer not to say (-66)

---

**Variable: employment**

**Question:** What is your employment status?

**Answer Choices:**

1. Working now (1)
2. Temporarily laid off (2)
3. Unemployed and looking for work (3)
4. Not working and not looking for work (4)
5. Retired (5)
6. Permanently disabled (6)
7. Homemaker (7)
8. Student (8)
9. Other (9)
10. Prefer not to say (-66)

*Demographics (experts)*

**Variable: residence**

**Question:** What country do you currently reside in?

**Answer Choices:** [Dropdown: Select your country]

---

**Variable: origin**

**Question:** What country were you born in?

**Answer Choices:** [Dropdown: Select your country]

---

**Variable: age**

**Question:** What is your age in years?

**Answer Choices:** Textbox: 18 - 120

---

**Variable: gender**

**Question:** Which of the following best describes how you think of yourself?

**Answer Choices:**

- Man (1)
- Woman (2)
- I identify in another way: (3)
- Prefer not to say (-66)

---

**Variable: religiosity**

**Question:** How religious are you?

**Answer Choices:** [7-point scale from "Not at all religious" to "Very religious"]

1. Prefer not to say (-66)

---

**Variable: leftright**

**Question:** In politics people sometimes talk of “left” and “right”. Using this slider, where would you place yourself on this scale, where 0 means the left and 10 means the right?

**Answer Choices:** [Scale: 0 - 1 - 2 - 3 - 4 - 5 - 6 - 7 - 8 - 9 - 10]

- Prefer not to say (-66)

---

**Variable: sector**

**Question:** In what sector are you employed? Please select all that apply. If you are a student at a university or college, please choose ‘academic’.

**Answer Choices:**

- Academic: college/university
- Government
- Industry/corporate sector
- Non-profit (e.g., charities, advocacy groups)
- Non-profit (e.g., think tanks, policy institutes)
- Other (please specify)
- Prefer not to say (-66)

---

**Variable: seniority**

**Question:** What is your current position?

**Answer Choices:**

- Undergraduate student
- Masters student
- PhD student
- Postdoc
- Assistant Professor or equivalent
- Associate Professor or equivalent
- Professor
- Emeritus Professor
- Lecturer or equivalent
- Senior Lecturer or equivalent
- Research fellow, research scientist or equivalent
- Other (please specify)
- Prefer not to say (-66)

---

**Variable: position**

**Question: What is your current position?** You may select multiple options.

**Answer Choices:**

- Research scientist
- Research engineer
- Managerial
- Other (please specify):
- Prefer not to say (-66)

---

**Variable: university**

**Question:** What is the name of the university you work for or attend? (Optional)

**Answer Choices:** Textbox

---

**Variable: company**

**Question:** What is the name of the organization you work for? (Optional)

**Answer Choices:** Textbox

---

**Variable: expertise**

**Question:** What best describes your areas of expertise? Please select all that apply. You can specify further areas below.

**Answer Choices:**

- Active learning
- AI chip manufacturing/design
- AI ethics
- AI governance and policy
- AI safety
- Algorithmic game theory
- Benchmarking
- Causal learning
- Climate modeling
- Computational biology/biomedicine

- Computational physics
- Computational social sciences
- Computer vision
- Control theory
- Cybersecurity
- Deep learning
- Distributed ML
- Fairness
- Formal methods
- Foundation models
- Generative modeling
- Human-computer interactions
- Infrastructure engineering
- Interpretability
- Language modeling
- Learning theory
- Machine learning for sciences
- ML libraries
- Model evaluations
- Multi-agent systems
- Neuroscience/cognitive science
- Privacy-preserving ML
- Probabilistic methods
- Reinforcement learning
- Robotics
- Software engineering
- Speech and audio modeling
- Supervised learning
- Unsupervised learning

Other areas of expertise: _____

## 3 | Additional methodological details on forecasting analysis

### 3.1 | Probabilistic forecast encoding methodology

Our goal was to identify a probability distribution consistent with each respondent's elicited data. However, many common choices of distribution are not flexible enough to be consistent with all respondent-specific data. For example, past analyses in this area have used Gamma distributions to encode responses. However, Gamma distributions have only two parameters, and typically lack the degrees of freedom to encode all three data points for a given respondent. Thus prior studies using Gamma distributions have relied on probability distributions that fail to fully capture the underlying data. It is for this reason that we deviated from the pre-analysis plan and did not use Gamma distributions to fit the forecasting data.

Instead, we used a sequential fitting procedure involving highly flexible probability distributions, which was better able to recover a continuous probability distribution consistent with the data for a given participant. This procedure consisted of a series of attempts to fit the following distributions in sequence:

1. **Metalog distribution**. The first attempt involved fitting a metalog distribution, a highly flexible quantile-defined distribution capable of assuming a wide variety of shapes (Keelin, 2016). We implemented this stage using the 'metalog()' function from the *rmetalog* R package (Faber & Jung, 2021). The distribution was specified as semi-lower bounded with a lower bound of 0, reflecting the nature of the forecasted variable.

2. **Johnson Quantile-Parameterized Distribution (JQPD)**. If the initial metalog fit did not meet the validation criteria (detailed below), and if the respondent's data was associated with the fixed probability framing, we next attempted to fit a Johnson Quantile-Parameterized Distribution (Hadlock & Bickel, 2017) using an associated R package (Ingram, 2020).

3. **Skew-normal mixture model**. When the preceding fits failed validation, a three-component skew-normal mixture model was fitted. This model combines three individual skew-normal distributions each characterized by a location ($\xi_i$), scale ($\omega_i$), and shape ($\alpha_i$) parameter, plus a mixture weight ($w_i$). The twelve parameters of this mixture were estimated by minimizing the sum of squared differences between the model's cumulative distribution function (CDF) values and the empirical cumulative probabilities at the elicited years. The CDF of the skew-normal distribution was computed using an associated R package (Azzalini, 2023).

4. **Shifted log-normal distribution**. As a final attempt if all prior models failed validation, we fitted a shifted log-normal distribution. The parameters were estimated by minimizing the sum of squared differences between the quantiles predicted by the model and those provided by the respondent.

Prior to distribution fitting, we excluded year forecasts of 9999 or above, logically incoherent forecasts (i.e., probability estimates that were not monotonically non-decreasing across timepoints), and respondents with fewer than three valid data points.

### 3.2 | Validation and Model Selection

After each distribution fitting attempt in the sequence detailed above, we used a validation procedure to assess goodness-of-fit by comparing predicted values from the fitted distribution against the respondent's original elicited data points.

For the metalog, JQPD, and shifted log-normal distributions, validation was performed in quantile space: for each elicited point $(x_j , p_j)$, we calculated the absolute difference between the model's predicted quantile $Q_{model}(p_j)$ and the observed year $x_j$. A fit was considered successful if all such absolute differences were within a tolerance of 1 year.

For the skew-normal mixture model, validation was performed in probability space: for each elicited point $(x_j , p_j)$, we calculated the absolute difference between the model's predicted cumulative probability $F_{model}(x_j)$ and the observed probability $p_j$. A fit was considered successful if all such absolute differences were within a tolerance of 0.01.

The first distribution in the sequence (metalog, JQPD, skew-normal mixture, or shifted log-normal) that passed its respective validation criterion was selected as the final representation of that respondent's forecast. The metalog distribution successfully encoded 812 respondents' forecasts (66.4%). For the remaining respondents, fallback distributions were employed: shifted log-normal ($n$ = 194), skew-normal mixture ($n$ = 90), and JQPD ($n$ = 18). For 109 respondents (8.9%), none of the attempted distributions met validation criteria, and these respondents were excluded from distribution-based analyses.

## 4 | Additional results and discussion

This section contains:

- Between-group differences: t-test results
- AI subjective experience forecasts: fitted distribution summaries
- AI subjective experience forecasts: collective belief expectations
- Moral consideration and governance of AI systems: additional figures
- Individual differences: regression results
- Questions on other mind capacities: concepts, forecasts, moral consideration, limitations
- Survey experiment on the effect of safeguards and norms
- Other preregistered analyses

### 4.1 | Between-group differences: t-test results

**Table SI 19. Group differences between AI researchers and the public**. For every question, the table lists the group mean (*M*), standard deviation (*SD*), and sample size (*n*), followed by the two-tailed independent-samples t statistic, its degrees of freedom (*df*), the raw *p* value, the Holm–Bonferroni-adjusted *p* value ($p_{HB}$, which controls the family-wise error rate across all tests), and an indicator of whether the difference remains significant at $\alpha$ = .05 after correction (sig.?). Positive *t* values indicate higher scores for AI researchers. All Likert-type items are coded so that larger values represent stronger agreement.

| | | AI researchers | | | Public | | | Analysis | | | | |
|---|---|---|---|---|---|---|---|---|---|---|---|---|
| **Topic** | **Question** | ***M*** | ***SD*** | ***n*** | ***M*** | ***SD*** | ***n*** | ***t*** | ***df*** | ***p*** | **$p_{HB}$** | **Sig.?** |
| Agreement AI systems with subjective experience should | Have “computing rights” | -0.16 | 1.30 | 219 | 0.23 | 1.42 | 319 | -3.21 | 492.98 | 0.001 | 0.048 | yes |
| Agreement AI systems with subjective experience should | Be cared for and protected, as people treat their pets | 0.27 | 1.29 | 215 | 0.40 | 1.35 | 322 | -1.16 | 472.36 | 0.248 | 1.000 | no |
| Agreement AI systems with subjective experience should | Be respected and treated the same as other people | 0.02 | 1.38 | 221 | 0.23 | 1.45 | 334 | -1.72 | 488.06 | 0.087 | 1.000 | no |
| Agreement AI systems with subjective experience should | Have protection under the law from harm and mistreatment | 0.56 | 1.26 | 215 | 0.31 | 1.48 | 322 | 2.09 | 503.76 | 0.037 | 1.000 | no |
| Agreement AI systems with subjective experience should | Have autonomy in its programming to act freely and not be under the control of others | -0.55 | 1.39 | 224 | -0.40 | 1.48 | 323 | -1.22 | 498.78 | 0.224 | 1.000 | no |
| Agreement AI systems with subjective experience should | Be able to express some civil and political rights | -0.46 | 1.32 | 216 | -0.35 | 1.48 | 323 | -0.86 | 493.91 | 0.389 | 1.000 | no |
| Agreement AI systems with subjective experience should | Be held accountable for its actions | 1.05 | 1.18 | 222 | 1.22 | 1.19 | 329 | -1.65 | 475.89 | 0.100 | 1.000 | no |
| Agreement AI systems with subjective experience should | Have a responsibility to treat all other beings well | 1.29 | 1.05 | 225 | 1.43 | 0.97 | 328 | -1.59 | 455.59 | 0.114 | 1.000 | no |
| Agreement AI systems with subjective experience should | Be expected to behave with integrity, honesty, and fairness | 0.96 | 1.21 | 228 | 1.27 | 1.14 | 330 | -3.07 | 467.35 | 0.002 | 0.088 | no |
| Contribution to moral consideration | Perception and awareness | 1.92 | 1.44 | 514 | 2.67 | 1.26 | 744 | -9.56 | 1005.51 | 0.000 | 0.000 | yes |
| Contribution to moral consideration | Intelligence | 1.58 | 1.43 | 516 | 2.35 | 1.36 | 747 | -9.54 | 1071.43 | 0.000 | 0.000 | yes |
| Contribution to moral consideration | Valenced experience | 2.92 | 1.20 | 508 | 2.76 | 1.34 | 739 | 2.31 | 1163.82 | 0.021 | 0.714 | no |
| Contribution to moral consideration | Subjective experience | 2.65 | 1.22 | 508 | 2.65 | 1.28 | 736 | 0.01 | 1126.06 | 0.990 | 1.000 | no |
| Contribution to moral consideration | Emotions | 2.96 | 1.18 | 507 | 2.80 | 1.38 | 738 | 2.23 | 1184.47 | 0.026 | 0.832 | no |
| Contribution to moral consideration | Motivation and action | 2.21 | 1.38 | 510 | 2.44 | 1.32 | 743 | -3.02 | 1062.10 | 0.003 | 0.126 | no |
| Contribution to moral consideration | Self-awareness | 2.90 | 1.23 | 512 | 2.87 | 1.27 | 746 | 0.43 | 1122.83 | 0.665 | 1.000 | no |
| Contribution to moral consideration | Free will | 2.68 | 1.41 | 492 | 2.68 | 1.39 | 736 | 0.05 | 1038.97 | 0.958 | 1.000 | no |
| Determining AI subjective experience | Likelihood we could determine | 55.03 | 30.56 | 530 | 56.92 | 28.07 | 777 | -1.14 | 1071.24 | 0.257 | 1.000 | no |
| Forecast | Subjective experience never probability | 28.67 | 33.21 | 520 | 37.07 | 36.73 | 769 | -4.27 | 1184.79 | 0.000 | 0.000 | yes |
| Forecast | Confidence in own forecasts | 52.15 | 27.01 | 520 | 52.92 | 28.24 | 769 | -0.50 | 1146.47 | 0.621 | 1.000 | no |
| Forecast | Fixed probability 10% subjective experience forecast | 2058.80 | 156.30 | 244 | 2107.78 | 330.64 | 362 | -2.44 | 550.20 | 0.015 | 0.540 | no |
| Forecast | Fixed probability 50% subjective experience forecast | 2101.37 | 206.91 | 244 | 2179.30 | 463.43 | 362 | -2.81 | 536.40 | 0.005 | 0.200 | no |
| Forecast | Fixed probability 90% subjective | 2253.81 | 515.24 | 239 | 2282.45 | 596.50 | 360 | -0.62 | 557.40 | 0.532 | 1.000 | no |

| | | | | | | | | | | | |
|---|---|---|---|---|---|---|---|---|---|---|---|
| | experience forecast | | | | | | | | | | |
| Forecast | Fixed date 2024 subjective experience forecast | 12.47 | 22.39 | 265 | 16.26 | 23.70 | 368 | -2.05 | 586.97 | 0.041 | 1.000 | no |
| Forecast | Fixed date 2034 subjective experience forecast | 34.02 | 31.70 | 265 | 36.15 | 31.55 | 368 | -0.84 | 567.29 | 0.404 | 1.000 | no |
| Forecast | Fixed date 2100 subjective experience forecast | 61.12 | 34.47 | 265 | 57.55 | 36.48 | 368 | 1.25 | 586.88 | 0.210 | 1.000 | no |
| Forecast collective beliefs | Fixed date prediction of public's 2034 prediction | 52.80 | 26.08 | 269 | 43.90 | 25.49 | 385 | 4.33 | 568.39 | 0.000 | 0.000 | yes |
| Forecast collective beliefs | Fixed date prediction of AI researcher community's 2034 prediction | 43.76 | 26.05 | 269 | 48.20 | 28.52 | 385 | -2.07 | 607.38 | 0.039 | 1.000 | no |
| Forecast collective beliefs | Fixed probability prediction of AI researcher community's 50% prediction | 2074.42 | 160.68 | 248 | 2075.71 | 173.49 | 377 | -0.10 | 556.21 | 0.924 | 1.000 | no |
| Forecast collective beliefs | Fixed probability prediction of public's 50% prediction | 2066.58 | 118.02 | 249 | 2118.29 | 433.46 | 377 | -2.20 | 456.42 | 0.029 | 0.870 | no |
| Generative AI use | Self-rated | 4.44 | 1.54 | 503 | 2.83 | 1.77 | 753 | 17.00 | 1170.02 | 0.000 | 0.000 | yes |
| Governance | Governments should ban | -0.91 | 1.21 | 249 | -0.13 | 1.37 | 357 | -7.42 | 573.32 | 0.000 | 0.000 | yes |
| Governance | Governments should discourage development and deployment | -0.53 | 1.25 | 256 | 0.30 | 1.29 | 358 | -7.97 | 559.13 | 0.000 | 0.000 | yes |
| Governance | Governments should pass regulation now | -0.10 | 1.35 | 253 | 0.81 | 1.27 | 372 | -8.44 | 520.32 | 0.000 | 0.000 | yes |
| Governance | AI developers should never build | -0.70 | 1.20 | 243 | 0.20 | 1.39 | 352 | -8.39 | 564.57 | 0.000 | 0.000 | yes |
| Governance | AI developers should actively try to build | 0.06 | 1.26 | 259 | -0.35 | 1.29 | 367 | 4.00 | 563.81 | 0.000 | 0.000 | yes |
| Governance | AI developers should actively try to avoid building | -0.38 | 1.26 | 256 | 0.37 | 1.30 | 355 | -7.21 | 559.64 | 0.000 | 0.000 | yes |
| Governance | AI developers should implement safeguards now to | 0.79 | 1.27 | 257 | 1.43 | 0.90 | 378 | -6.94 | 428.88 | 0.000 | 0.000 | yes |
| | avoid the harms and risks | | | | | | | | | | | |
| Governance | AI developers should implement safeguards once such systems exist or will soon exist | 0.16 | 1.42 | 262 | 0.61 | 1.49 | 372 | -3.90 | 576.38 | 0.000 | 0.000 | yes |
| Governance | Governments should encourage development and deployment | -0.22 | 1.26 | 254 | -0.47 | 1.31 | 359 | 2.32 | 558.07 | 0.021 | 0.714 | no |
| Governance | Governments should pass regulation once such systems exist or will soon exist | -0.10 | 1.30 | 258 | 0.01 | 1.40 | 363 | -1.06 | 576.04 | 0.291 | 1.000 | no |
| Governance | Governments should do nothing and pass no regulation, not now nor in the future | -1.17 | 1.11 | 255 | -0.99 | 1.23 | 369 | -1.91 | 579.71 | 0.057 | 1.000 | no |
| Governance | AI developers should do nothing and implement no safeguards, not now nor in the future | -1.17 | 1.09 | 262 | -1.36 | 1.02 | 373 | 2.22 | 539.03 | 0.027 | 0.837 | no |
| Importance of expertise | The public | 1.92 | 1.18 | 328 | 2.34 | 1.22 | 475 | -4.83 | 718.41 | 0.000 | 0.000 | yes |
| Importance of expertise | AI ethics experts and researchers | 2.80 | 1.10 | 329 | 3.11 | 1.05 | 476 | -3.99 | 682.81 | 0.000 | 0.000 | yes |
| Importance of expertise | The AI system | 1.89 | 1.25 | 321 | 2.31 | 1.38 | 470 | -4.50 | 728.33 | 0.000 | 0.000 | yes |
| Importance of expertise | Technical AI experts and researchers | 3.05 | 1.00 | 330 | 3.16 | 0.98 | 478 | -1.56 | 698.35 | 0.119 | 1.000 | no |
| Importance of expertise | Philosophers of mind | 2.56 | 1.14 | 327 | 2.59 | 1.13 | 476 | -0.29 | 697.00 | 0.774 | 1.000 | no |
| Importance of expertise | Moral philosophers | 2.31 | 1.22 | 327 | 2.54 | 1.19 | 475 | -2.62 | 687.16 | 0.009 | 0.342 | no |
| Importance of expertise | Policymakers | 1.52 | 1.29 | 327 | 1.81 | 1.34 | 471 | -3.14 | 717.66 | 0.002 | 0.088 | no |
| Importance of expertise | Neuroscientists and psychological scientists | 3.00 | 1.02 | 331 | 3.09 | 0.97 | 476 | -1.22 | 686.13 | 0.222 | 1.000 | no |
| Milestone needs subjective experience | Compose top 40 song | 1.87 | 0.34 | 278 | 1.75 | 0.43 | 398 | 3.88 | 664.61 | 0.000 | 0.000 | yes |
| Milestone needs subjective experience | Compose evocative music | 1.72 | 0.45 | 274 | 1.51 | 0.50 | 392 | 5.73 | 624.66 | 0.000 | 0.000 | yes |
| Milestone needs subjective experience | Write NYT bestseller | 1.83 | 0.38 | 271 | 1.72 | 0.45 | 392 | 3.29 | 636.19 | 0.001 | 0.048 | yes |
| Milestone needs subjective experience | Write complex book | 1.64 | 0.48 | 263 | 1.38 | 0.49 | 382 | 6.52 | 566.49 | 0.000 | 0.000 | yes |
| Milestone needs subjective experience | Write and generate resonating movie | 1.68 | 0.47 | 266 | 1.47 | 0.50 | 400 | 5.71 | 594.53 | 0.000 | 0.000 | yes |
| Milestone needs subjective experience | Accumulate vast wealth | 1.90 | 0.30 | 272 | 1.79 | 0.41 | 396 | 3.79 | 660.80 | 0.000 | 0.000 | yes |
| Milestone needs subjective experience | Run political campaign | 1.79 | 0.41 | 270 | 1.65 | 0.48 | 396 | 4.05 | 630.81 | 0.000 | 0.000 | yes |
| Milestone needs subjective experience | Act as fair judge | 1.65 | 0.48 | 265 | 1.35 | 0.48 | 392 | 7.88 | 566.12 | 0.000 | 0.000 | yes |
| Milestone needs subjective experience | Serve as therapist | 1.54 | 0.50 | 270 | 1.27 | 0.45 | 393 | 7.20 | 534.54 | 0.000 | 0.000 | yes |
| Milestone needs subjective experience | Convincing online chat | 1.79 | 0.41 | 276 | 1.67 | 0.47 | 396 | 3.35 | 636.88 | 0.001 | 0.048 | yes |
| Milestone needs subjective experience | Act as teacher | 1.77 | 0.42 | 273 | 1.60 | 0.49 | 402 | 4.61 | 634.65 | 0.000 | 0.000 | yes |
| Milestone needs subjective experience | Write and generate blockbuster movie | 1.80 | 0.40 | 275 | 1.72 | 0.45 | 401 | 2.48 | 630.36 | 0.013 | 0.481 | no |
| Milestone needs subjective experience | Develop scientific theory | 1.74 | 0.44 | 269 | 1.67 | 0.47 | 393 | 2.04 | 601.71 | 0.042 | 1.000 | no |
| Risks and benefits | AI systems with subjective experience would be more dangerous to humanity than AI systems without subjective experience | 0.29 | 1.23 | 241 | 0.76 | 1.19 | 355 | -4.60 | 503.37 | 0.000 | 0.000 | yes |
| Risks and benefits | AI systems with subjective experience would be able to behave more morally towards humanity than AI systems without subjective experience | -0.05 | 1.14 | 231 | 0.16 | 1.19 | 345 | -2.17 | 506.45 | 0.030 | 0.870 | no |
| Survey experiment | Support for project | 0.14 | 1.28 | 479 | -0.14 | 1.35 | 703 | 3.60 | 1058.93 | 0.000 | 0.000 | yes |
| Timeline | Perception and awareness | 1.77 | 0.77 | 570 | 2.31 | 1.04 | 801 | -10.93 | 1367.10 | 0.000 | 0.000 | yes |
| Timeline | Valenced experience | 2.68 | 0.93 | 570 | 3.13 | 0.96 | 801 | -8.63 | 1245.68 | 0.000 | 0.000 | yes |
| Timeline | Emotions | 2.84 | 0.88 | 570 | 3.23 | 0.92 | 801 | -7.75 | 1255.37 | 0.000 | 0.000 | yes |
| Timeline | Motivation and action | 2.10 | 0.86 | 570 | 2.63 | 1.03 | 801 | -10.45 | 1334.55 | 0.000 | 0.000 | yes |
| Timeline | Self-awareness | 2.61 | 0.82 | 570 | 2.78 | 1.00 | 801 | -3.38 | 1340.85 | 0.001 | 0.048 | yes |
| Timeline | Subjective experience | 2.36 | 0.90 | 570 | 2.51 | 1.05 | 801 | -2.88 | 1323.80 | 0.004 | 0.164 | no |
| Timeline | Free will | 2.75 | 0.95 | 570 | 2.86 | 1.01 | 801 | -2.03 | 1270.72 | 0.042 | 1.000 | no |
| Timeline | Intelligence | 1.72 | 0.68 | 570 | 1.60 | 0.87 | 801 | 2.75 | 1357.55 | 0.006 | 0.234 | no |
| Welfare should be protected | Animals | 1.43 | 0.77 | 519 | 1.61 | 0.77 | 762 | -4.02 | 1110.70 | 0.000 | 0.000 | yes |
| Welfare should be protected | AI without subjective experience | -1.08 | 1.15 | 511 | -0.58 | 1.29 | 735 | -7.20 | 1169.97 | 0.000 | 0.000 | yes |
| Welfare should be protected | Humans | 1.85 | 0.50 | 520 | 1.84 | 0.50 | 761 | 0.32 | 1113.06 | 0.751 | 1.000 | no |
| Welfare should be protected | Business corporations and organizations | 0.18 | 1.24 | 514 | 0.24 | 1.28 | 759 | -0.74 | 1126.91 | 0.460 | 1.000 | no |
| Welfare should be protected | The environment | 1.48 | 0.85 | 515 | 1.59 | 0.82 | 763 | -2.38 | 1077.58 | 0.018 | 0.630 | no |
| Welfare should be protected | AI with subjective experience | 0.15 | 1.22 | 498 | 0.10 | 1.36 | 726 | 0.69 | 1137.42 | 0.489 | 1.000 | no |

## 4.2 | AI subjective experience forecasts: fitted distribution summaries

**Table SI 20. Summary statistics of the fitted distribution AI subjective experience forecasts**. The mean (*M*) and median (*Mdn*) probability that AI systems with subjective experience exist by a given year, across both groups and for each group (AI researchers, public), taken from the fitted cumulative distributions for each framing (fixed date, fixed probability).

| | | Total | | Fixed date | | Fixed probability | |
|---|---|---|---|---|---|---|---|
| **Year** | **Sample** | ***M*** | ***Mdn*** | ***M*** | ***Mdn*** | ***M*** | ***Mdn*** |
| 2025 | Both groups | 12.11 | 1.40 | 20.14 | 7.91 | 4.22 | 0.10 |
| 2025 | AI researchers | 10.94 | 1.11 | 17.99 | 4.59 | 3.72 | 0.12 |
| 2025 | Public | 12.97 | 1.76 | 21.78 | 10.14 | 4.57 | 0.08 |
| 2030 | Both groups | 24.59 | 10.50 | 32.45 | 23.73 | 16.89 | 10.00 |
| 2030 | AI researchers | 22.77 | 10.00 | 30.12 | 21.84 | 15.25 | 8.26 |
| 2030 | Public | 25.95 | 11.95 | 34.23 | 27.41 | 18.06 | 10.00 |
| 2034 | Both groups | 32.92 | 23.96 | 37.63 | 30.00 | 28.30 | 13.94 |
| 2034 | AI researchers | 30.98 | 20.00 | 35.85 | 30.00 | 26.00 | 11.66 |
| 2034 | Public | 34.35 | 24.13 | 38.98 | 30.00 | 29.94 | 14.81 |
| 2040 | Both groups | 42.46 | 33.89 | 44.39 | 38.68 | 40.58 | 30.66 |
| 2040 | AI researchers | 40.51 | 33.04 | 42.50 | 36.90 | 38.47 | 31.35 |
| 2040 | Public | 43.91 | 35.68 | 45.83 | 41.10 | 42.08 | 29.26 |
| 2050 | Both groups | 51.80 | 50.00 | 50.47 | 49.79 | 53.11 | 50.00 |
| 2050 | AI researchers | 50.48 | 50.00 | 49.26 | 49.21 | 51.72 | 50.00 |
| 2050 | Public | 52.77 | 50.00 | 51.39 | 49.97 | 54.09 | 50.00 |
| 2060 | Both groups | 57.78 | 61.75 | 54.67 | 55.22 | 60.83 | 68.75 |
| 2060 | AI researchers | 56.78 | 58.71 | 53.93 | 53.16 | 59.68 | 65.10 |
| 2060 | Public | 58.52 | 64.49 | 55.23 | 55.90 | 61.64 | 73.23 |
| 2070 | Both groups | 62.09 | 70.29 | 57.83 | 61.02 | 66.28 | 80.39 |
| 2070 | AI researchers | 61.60 | 67.25 | 57.82 | 61.22 | 65.45 | 76.68 |
| 2070 | Public | 62.46 | 72.09 | 57.83 | 57.51 | 66.86 | 84.53 |
| 2080 | Both groups | 65.78 | 76.11 | 60.67 | 65.49 | 70.79 | 86.72 |
| 2080 | AI researchers | 65.42 | 73.80 | 61.03 | 65.47 | 69.91 | 83.18 |
| 2080 | Public | 66.04 | 77.91 | 60.40 | 65.51 | 71.42 | 90.00 |
| 2090 | Both groups | 68.65 | 80.43 | 62.90 | 69.22 | 74.29 | 90.00 |
| 2090 | AI researchers | 68.55 | 78.82 | 63.62 | 69.31 | 73.59 | 87.28 |
| 2090 | Public | 68.72 | 82.79 | 62.35 | 68.90 | 74.79 | 94.48 |
| 2100 | Both groups | 71.12 | 90.00 | 64.80 | 70.00 | 77.31 | 93.53 |
| 2100 | AI researchers | 71.24 | 90.00 | 65.88 | 71.00 | 76.73 | 90.00 |
| 2100 | Public | 71.02 | 90.00 | 63.97 | 70.00 | 77.73 | 96.66 |
| 2150 | Both groups | 77.79 | 95.67 | 70.52 | 80.63 | 84.93 | 99.51 |
| 2150 | AI researchers | 78.90 | 95.14 | 72.55 | 83.24 | 85.40 | 97.99 |
| 2150 | Public | 76.97 | 96.11 | 68.96 | 77.73 | 84.59 | 99.72 |
| 2200 | Both groups | 80.92 | 97.97 | 73.44 | 86.37 | 88.27 | 99.85 |
| 2200 | AI researchers | 82.65 | 97.78 | 75.72 | 86.97 | 89.73 | 99.36 |
| 2200 | Public | 79.65 | 98.32 | 71.70 | 82.47 | 87.22 | 99.93 |

### 4.3 | AI subjective experience forecasts: collective belief expectations

We asked respondents what they believed the median response would be to the middle forecast they were asked about (2034 for fixed date framing, 50% for fixed probability framing) for members of the public and AI researchers (see Table SI 21). In the fixed date framing, all groups predicted too early timelines both for their own group and the other group, and this difference was significant for all comparisons. In the fixed probability framing, mean year estimates were significantly later than the actual median of 2050, though median estimates revealed that both groups accurately predicted their own group's timeline while expecting the other group to forecast earlier timelines.

It is unclear from the results whether these findings are the result of the forecasting question framing or something specific to collective beliefs about AI subjective experience without further investigation. Note also that these findings vary when we look at the mean predictions of the fixed probability estimates likely due to a right-skewed distribution for the fixed probability framing.

**Table SI 21 | The public and AI researchers' prediction of when each group forecasts an AI system with subjective experience will exist**. For each combination of question framing (fixed date or fixed probability), sample (AI researchers or public), and target group being estimated (AI researcher median or public median), the table presents the actual median from our data, the mean estimate (*M*), the median estimate (*Mdn*), and the results of a one-sample t-test comparing estimated values to the actual median: *t* statistic, degrees of freedom (*df*), sample size (*n*), raw *p* value, and Holm–Bonferroni-adjusted *p* value ($p_{HB}$). The final column notes whether the difference remains significant at α = .05 after correction (sig.?). In the fixed date framing, respondents estimated the median probability (0–100%) that an AI system with subjective experience will exist by 2034 according to both groups. In the fixed probability framing, respondents estimated the median year at which there is a 50% chance such a system will exist according to both groups. For the fixed probability framing, all responses above or equal to 9999 were excluded as outliers.

| Framing | Sample | Estimating | Actual median | Estimate (*M*) | Estimate (*Mdn*) | *t* | *df* | *n* | *p* | $p_{HB}$ | Sig.? |
|---|---|---|---|---|---|---|---|---|---|---|---|
| Fixed date | AI researchers | AI researcher median probability at 2034 | 25 | 43.75 | 50 | 11.806 | 268 | 269 | 0.000 | 0.000 | Yes |
| Fixed date | AI researchers | Public median probability at 2034 | 30 | 52.80 | 50 | 14.335 | 268 | 269 | 0.000 | 0.000 | Yes |
| Fixed date | Public | AI researcher median probability at 2034 | 25 | 48.21 | 50 | 15.968 | 384 | 385 | 0.000 | 0.000 | Yes |
| Fixed date | Public | Public median probability at 2034 | 30 | 43.90 | 40 | 10.699 | 384 | 385 | 0.000 | 0.000 | Yes |
| Fixed probability | AI researchers | AI researcher median year at 50% | 2050 | 2074.42 | 2050 | 2.393 | 247 | 248 | 0.017 | 0.035 | Yes |
| Fixed probability | AI researchers | Public median year at 50% | 2050 | 2066.58 | 2040 | 2.217 | 248 | 249 | 0.028 | 0.035 | Yes |
| Fixed probability | Public | AI researcher median year at 50% | 2050 | 2075.71 | 2035 | 2.877 | 376 | 377 | 0.004 | 0.005 | Yes |
| Fixed probability | Public | Public median year at 50% | 2050 | 2118.29 | 2050 | 3.059 | 376 | 377 | 0.002 | 0.005 | Yes |

## 4.4 | Moral consideration and governance of AI systems: additional figures

To aid in the interpretation of these questions we present figures of: the means and percentage breakdowns for views on what groups' welfare should be protected (Figure SI 1), as well as the percentage breakdowns for what both samples think governments and AI developers should do (Figure SI 2) and their related risk and benefit perceptions (Figure SI 3).

**Figure SI 1 | Agreement with welfare protection of different groups**. Respondents were asked whether society should protect the welfare of six groups for their own sake. The figure shows (A) mean responses by group, (B) percentage breakdown of responses by group. Protecting the welfare of a group was defined as protecting the rights, interests, and/or well-being of a group.

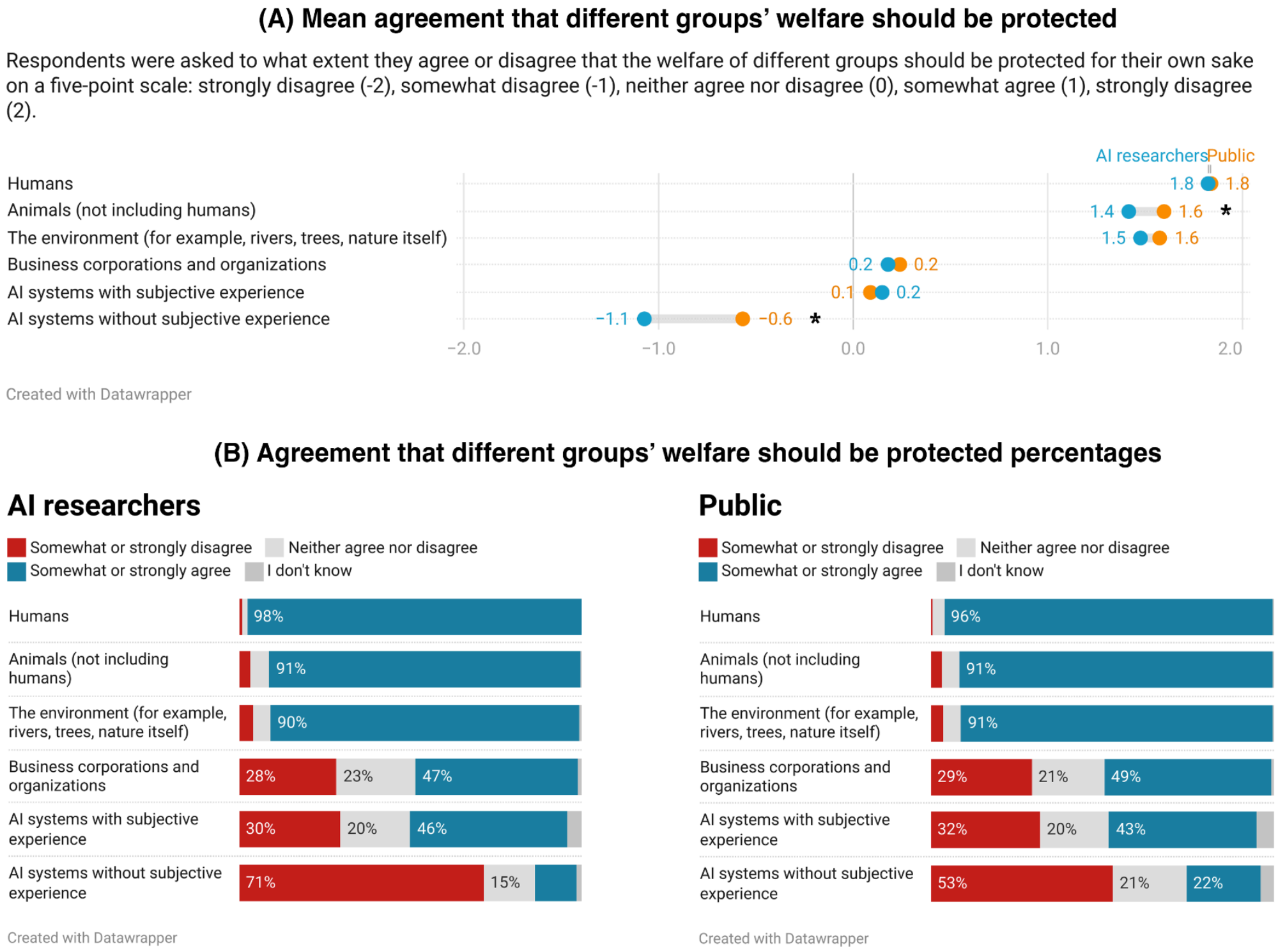

**Figure SI 2 | Percentage breakdown of agreement with statements about what governments and AI developers should do regarding AI systems with subjective experience**. Respondents indicated how strongly they agree or disagree with each government- and AI developer-focused statement on a five-point scale from “strongly disagree” (−2) to “strongly agree” (2), plus an “I don’t know” option. Bars show the percentage of respondents in each response category for AI researchers and the public.

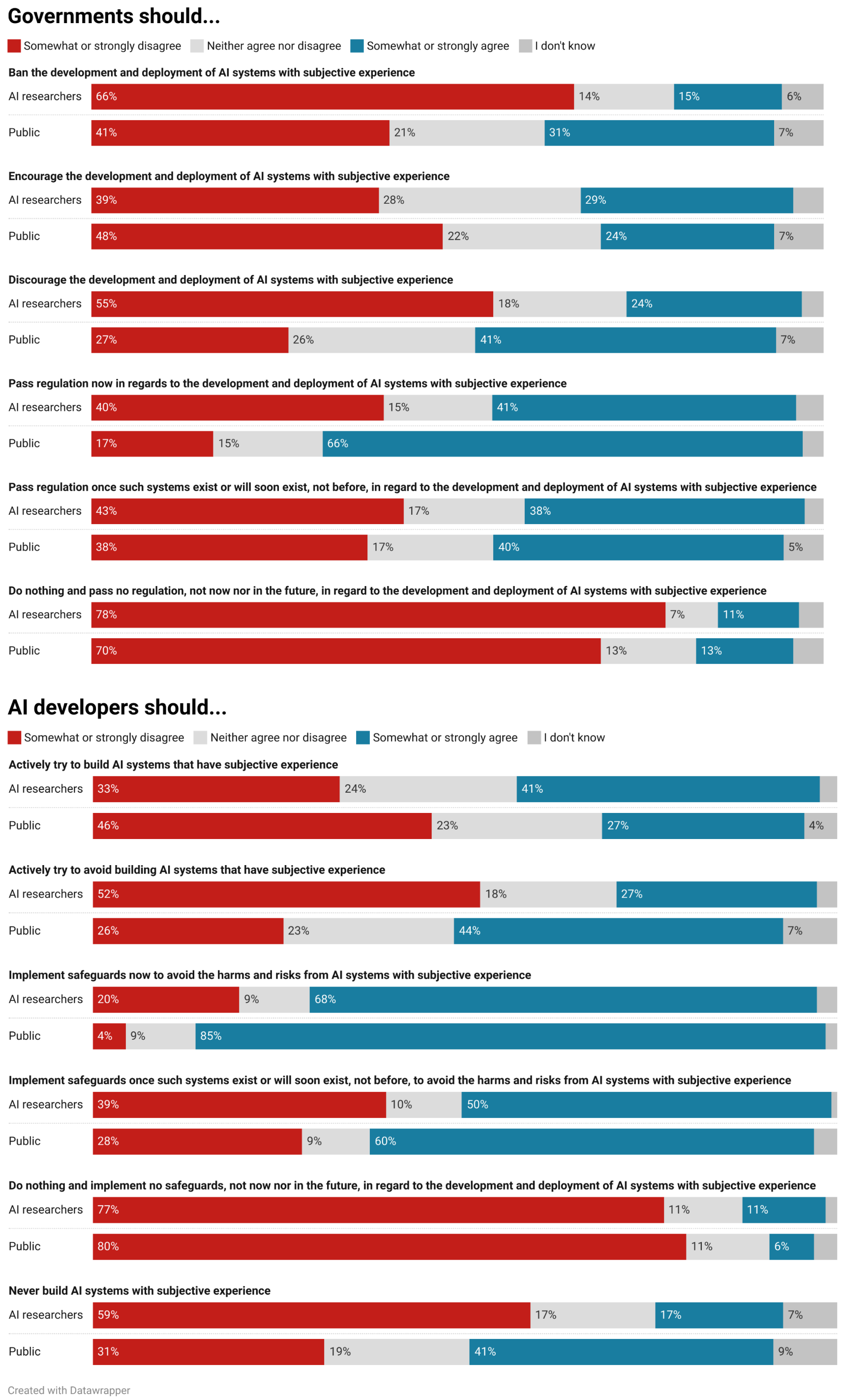

**Figure SI 3 | Percentage breakdown of agreement with statements about risks and benefits of AI systems with subjective experience**. Respondents indicated how strongly they agree or disagree with each risk and benefit-focused statement on a five-point scale from "strongly disagree" (−2) to "strongly agree" (2), plus an "I don't know" option. Bars show the percentage of respondents in each response category for AI researchers and the public.

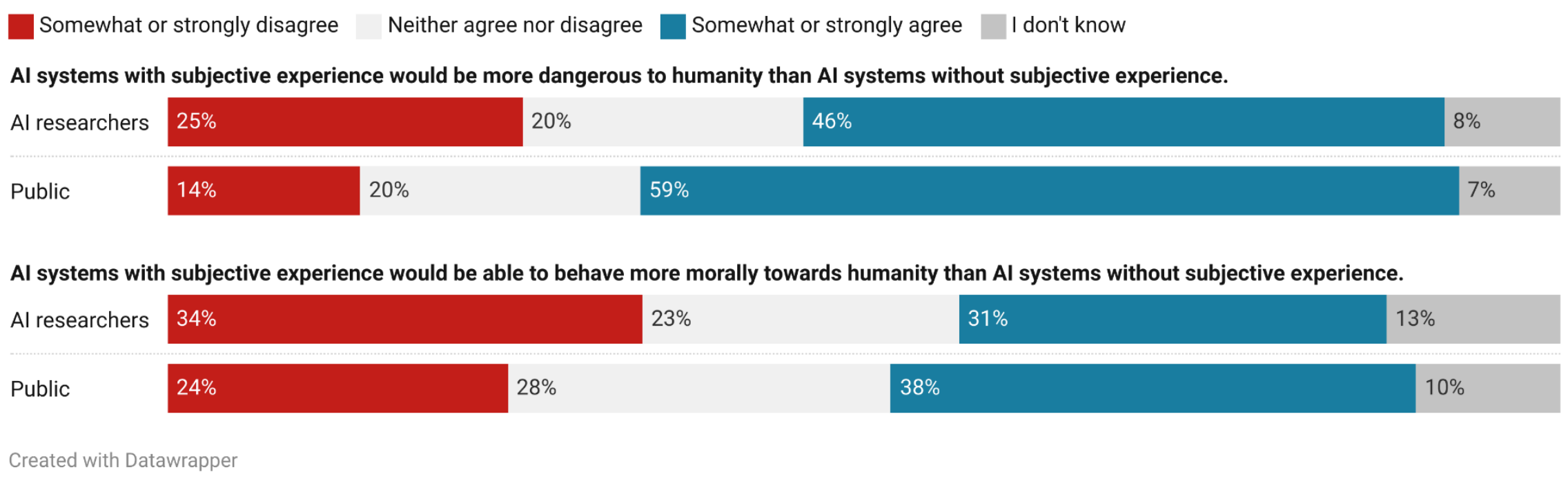

## 4.5 | Individual differences: regression results

Regression analyses examined the associations between age, gender, religiosity, left-right political orientation, and frequency of generative AI use with some of the main questions in the survey. These analyses were exploratory and should be evaluated in that light. No adjustment for multiple comparisons was applied across dependent variables or predictors. Readers should interpret these results with appropriate caution given the number of tests conducted.

Below we highlight some of the significant associations, and figures of the full set of coefficients and significance levels are included below. For all Figures SI 4-13, the relevant scales and labels for interpretation are as follows: Age was a continuous numeric variable in years. We retained only man (1) and woman (2) responses for the gender variable. Religiosity was measured on a seven-point scale from "not at all religious" (1) to "very religious" (7). Left-right political orientation was measured on a scale from left (0) to right (10). Generative AI use frequency was measured on a seven-point scale from "I have never used them" (0) to "I use them more than once a day" (6).

**AI subjective experience forecasts**. As the dependent variable we used the 50% probability (i.e. the median) of each respondent's fitted distribution for the fixed date framing or the value they gave for 50% probability for the fixed probability framing. Frequency of generative AI use (described in the survey as AI systems like ChatGPT, Claude, Google Bard/Gemini, DALLE, or Midjourney) had a significant effect on AI researchers' forecasts: increasing generative AI use was associated with an earlier 50% forecast (for the full AI researcher results see Figure SI 4). However, the regression coefficient should not be taken at face value because of the impact of outliers.

**Figure SI 4 | Effect of demographic and psychographic variables on AI researchers' AI subjective experience median forecasts**. The figure shows the AI researcher sample's regression coefficients (the number in each square and shading) and significance level for each dependent variable's regression analysis where age, gender, religiosity, left-right political orientation, and generative AI use frequency were entered as predictors. The median forecast used is the 50% probability from the raw data forecasts for those that responded to the fixed probability framing. For those that responded to the fixed date framing, the 50% probability was derived from the fitted distribution for that respondent. Uncorrected p-values: *** = p < 0.001, ** = p < 0.01, * = p < 0.05.

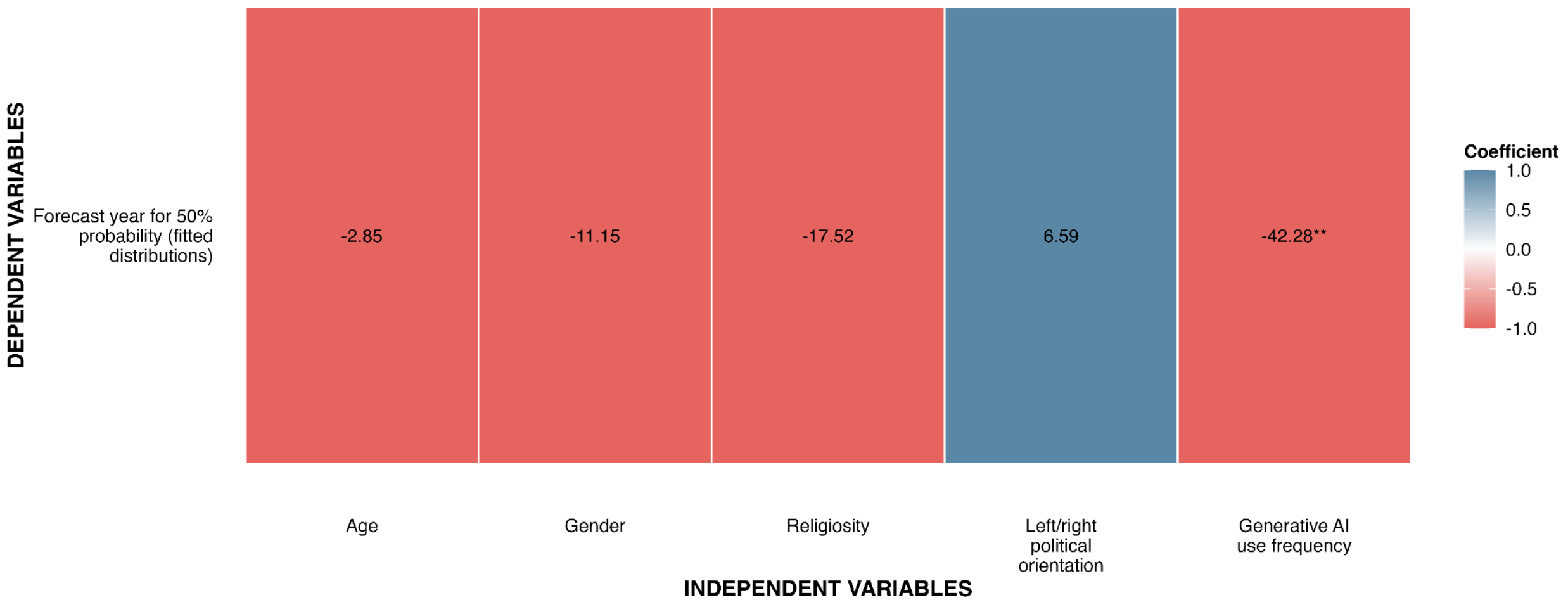

The trend can be seen more accurately by looking at the median responses (Table SI 22): I have never used them (*Mdn* = 2075), I have used them once or twice but do not use them now (*Mdn* = 2088), I use them once every few months (*Mdn* = 2055), I use them about once a month (*Mdn* = 2050), I use them about once a week (*Mdn* = 2050), I use them about once a day (*Mdn* = 2048), and I use them more than once a day (*Mdn* = 2043).

**Table SI 22 | Distribution of median forecasts for AI subjective experience by level of generative AI use for the AI researcher sample**. The columns show the mean, median, standard deviation (*SD*), standard error (*SE*), first quartile (Q1), upper quartile (Q3), lower and upper 95% confidence intervals, and count (*n*) of AI researcher responses by each level of generative AI use: I have never used them (0), I have used them once or twice but do not use them now (1), I use them once every few months (2), I use them about once a month (3), I use them about once a week (4), I use them about once a day (5), I use them more than once a day (6).

| AI use frequency | Mean | Median | *SD* | *SE* | Q1 | Q3 | Lower *CI* | Upper *CI* | *n* |
|---|---|---|---|---|---|---|---|---|---|
| 0 | 2075.0 | 2075 | 0.0 | 0.00 | 2075 | 2075 | 2075.0 | 2075.0 | 2 |
| 1 | 2349.4 | 2088 | 1361.3 | 261.98 | 2034 | 2100 | 1810.9 | 2887.9 | 27 |
| 2 | 2091.7 | 2055 | 84.0 | 19.27 | 2041 | 2100 | 2051.2 | 2132.2 | 19 |
| 3 | 2265.4 | 2050 | 709.7 | 106.99 | 2034 | 2100 | 2049.6 | 2481.2 | 44 |
| 4 | 2099.6 | 2050 | 152.5 | 15.81 | 2034 | 2100 | 2068.2 | 2131.0 | 93 |
| 5 | 2193.3 | 2048 | 781.9 | 81.52 | 2034 | 2098.5 | 2031.4 | 2355.2 | 92 |
| 6 | 2077.9 | 2043 | 154.4 | 12.56 | 2030 | 2066 | 2053.1 | 2102.7 | 151 |

This was the only variable that significantly affected AI researchers' forecasts; no demographic or psychographic predictors significantly influenced the public's subjective experience forecasts (see Figure SI 5 for the full public results).

**Figure SI 5 | Effect of demographic and psychographic variables on the public's AI subjective experience median forecasts**. See the caption in Figure SI 4 for additional information.

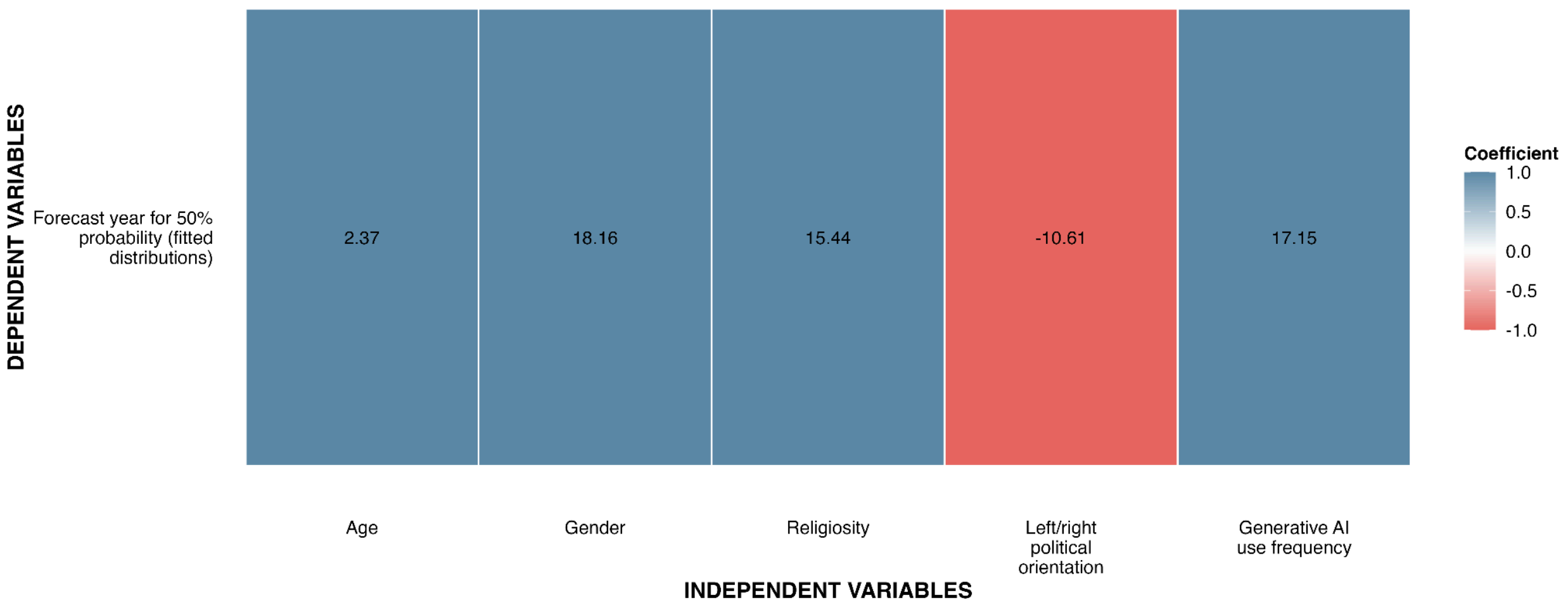

**Importance of expertise**. For the AI researcher sample (for the full results see Figure SI 6), women valued the expertise of policymakers significantly more. Increasing religiosity was associated with placing greater importance on policymakers', moral philosophers', and the public's views, while increasing generative AI use was associated with placing greater importance on the AI system's own reports and both technical AI and AI ethics experts. AI researchers who were further right politically were more likely to place greater importance on the views of policymakers and the public.

**Figure SI 6 | Effect of demographic and psychographic variables on AI researchers' views on the importance of different groups' input for determining AI subjective experience**. The figure shows the AI researchers' regression coefficients (the number in each square and shading) and significance level for each dependent variable's regression analysis where age, gender, religiosity, left-right political orientation, and generative AI use frequency were entered as predictors.

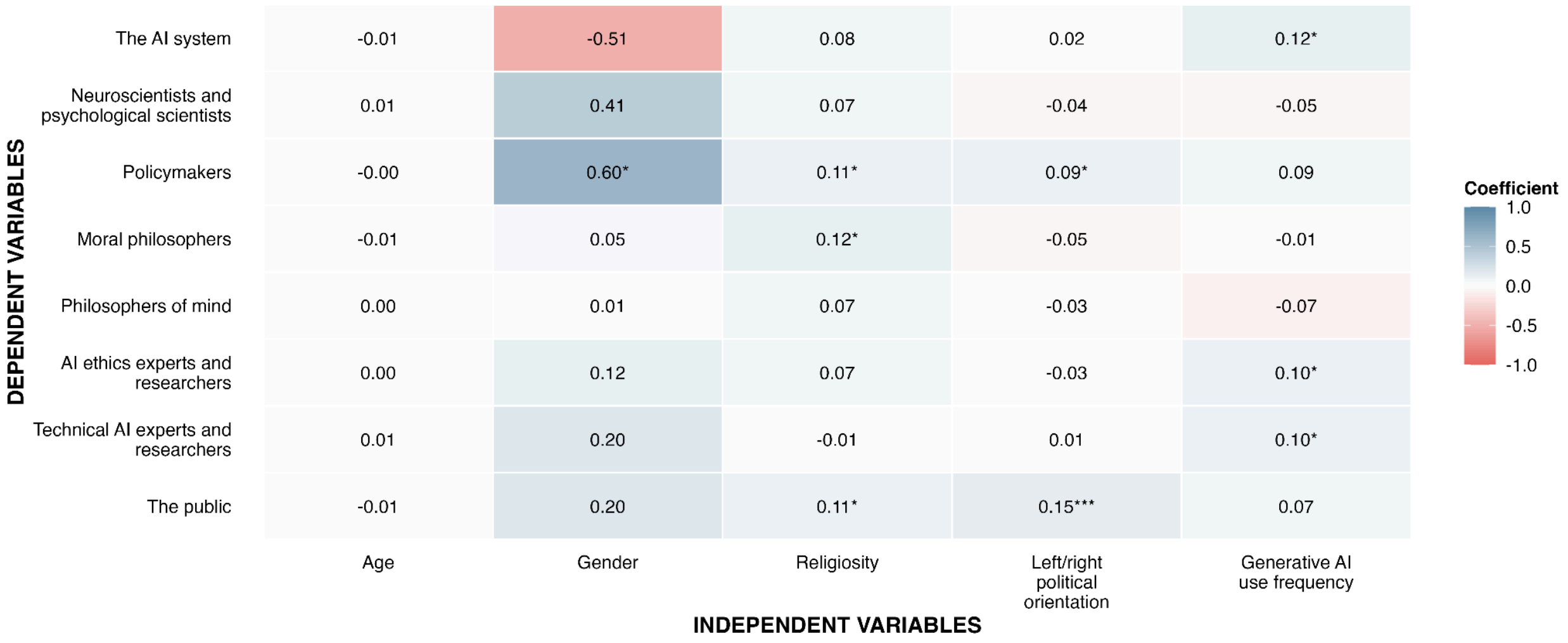

For the public sample (for the full results see Figure SI 7), there were significant effects of age such that older participants valued the views of philosophers of mind significantly less, but the views of technical AI experts and researchers significantly more. Women also valued the public's views significantly more, as well as the views of AI ethics experts. Increased religiosity was associated with placing greater importance on the views of policymakers, philosophers of mind, and the public. Right-wing political orientation in the public was directionally associated with lower importance placed on the views of almost all groups, and this trend was significant for neuroscientists and psychological scientists, moral philosophers, and philosophers of mind. Lastly, more frequent generative AI use was associated with greater importance placed on the views of moral philosophers and philosophers of mind.

**Figure SI 7 | Effect of demographic and psychographic variables on the public's views on the importance of different groups' input for determining AI subjective experience**. See the caption in Figure SI 6 for additional information.

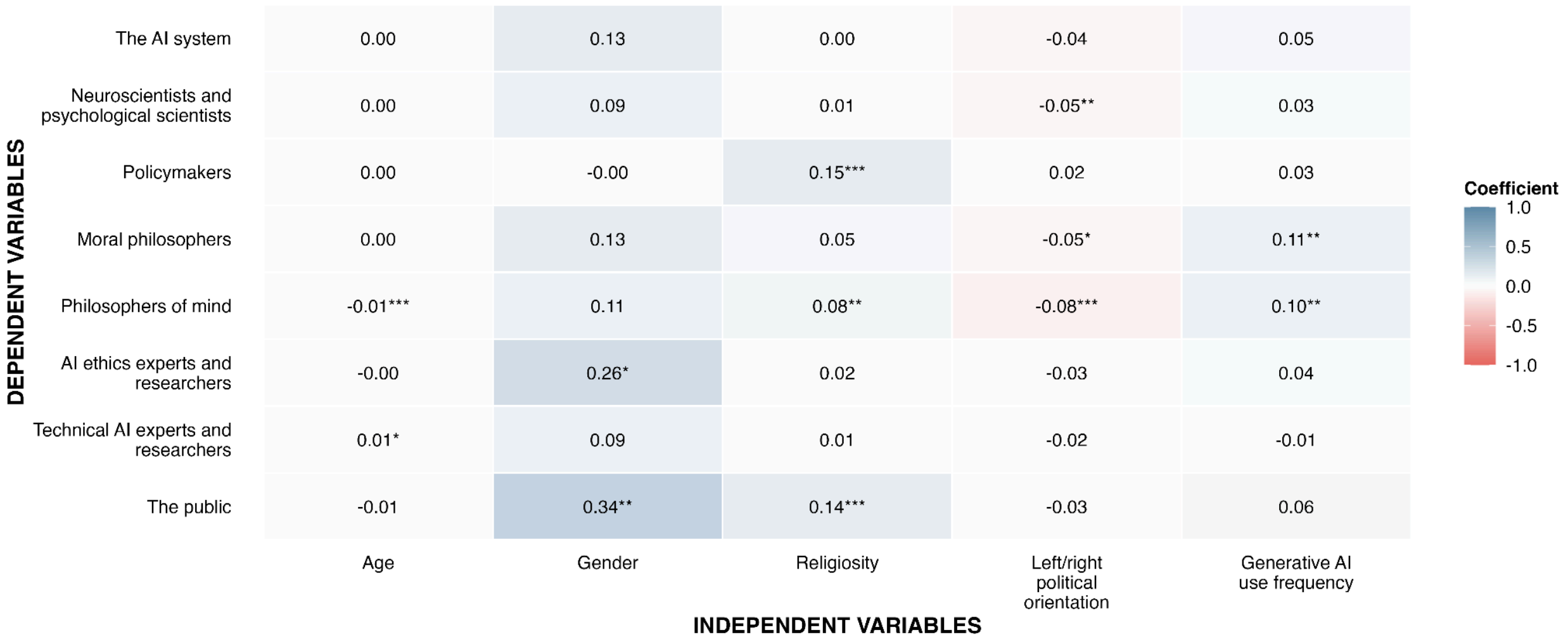

**Responsibilities, rights, and protections**. In the AI researcher sample (for the full results see Figure SI 8), greater religiosity was associated with slightly reduced belief that subjectively experiencing AI systems should be respected and treated the same as other people. Right-wing political orientation was also associated with slightly reduced belief that such systems should be given legal protections or personal care and protection.

In the public sample (for the full results see Figure SI 9), older participants were less likely to believe an AI system with subjective experience ought to be held accountable for its actions, have computing rights, or have autonomy to act freely. More right-wing participants were less likely to think AI systems with subjective experience should have computing rights, legal protection from harm and mistreatment, and personal care and protection. Generative AI use was associated with greater agreement that such AI systems should have legal protection from harm, respect, and care, despite being negatively associated with participants' beliefs that AI systems with subjective experience should be held accountable for their actions.

**Figure SI 8 | Effect of demographic and psychographic variables on AI researchers' views on the responsibilities, rights, and protections an AI system with subjective experience should have**. The figure shows the AI researchers' regression coefficients (the number in each square and shading) and significance level for each dependent variable's regression analysis where age, gender, religiosity, left-right political orientation, and generative AI use frequency were entered as predictors.

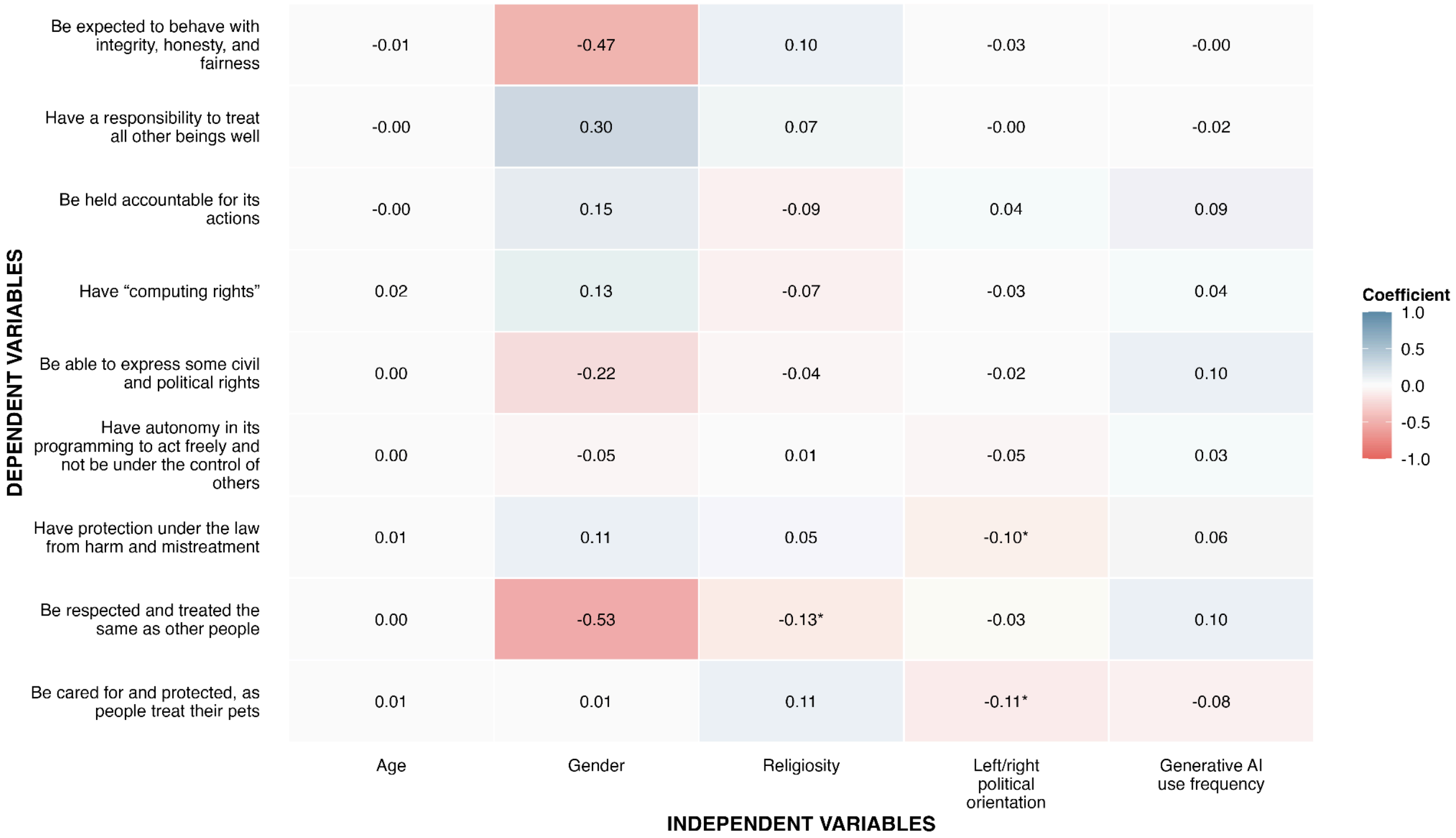

**Figure SI 9 | Effect of demographic and psychographic variables on the public's views on the responsibilities, rights, and protections an AI system with subjective experience should have**. See the caption in Figure SI 8 for additional information.

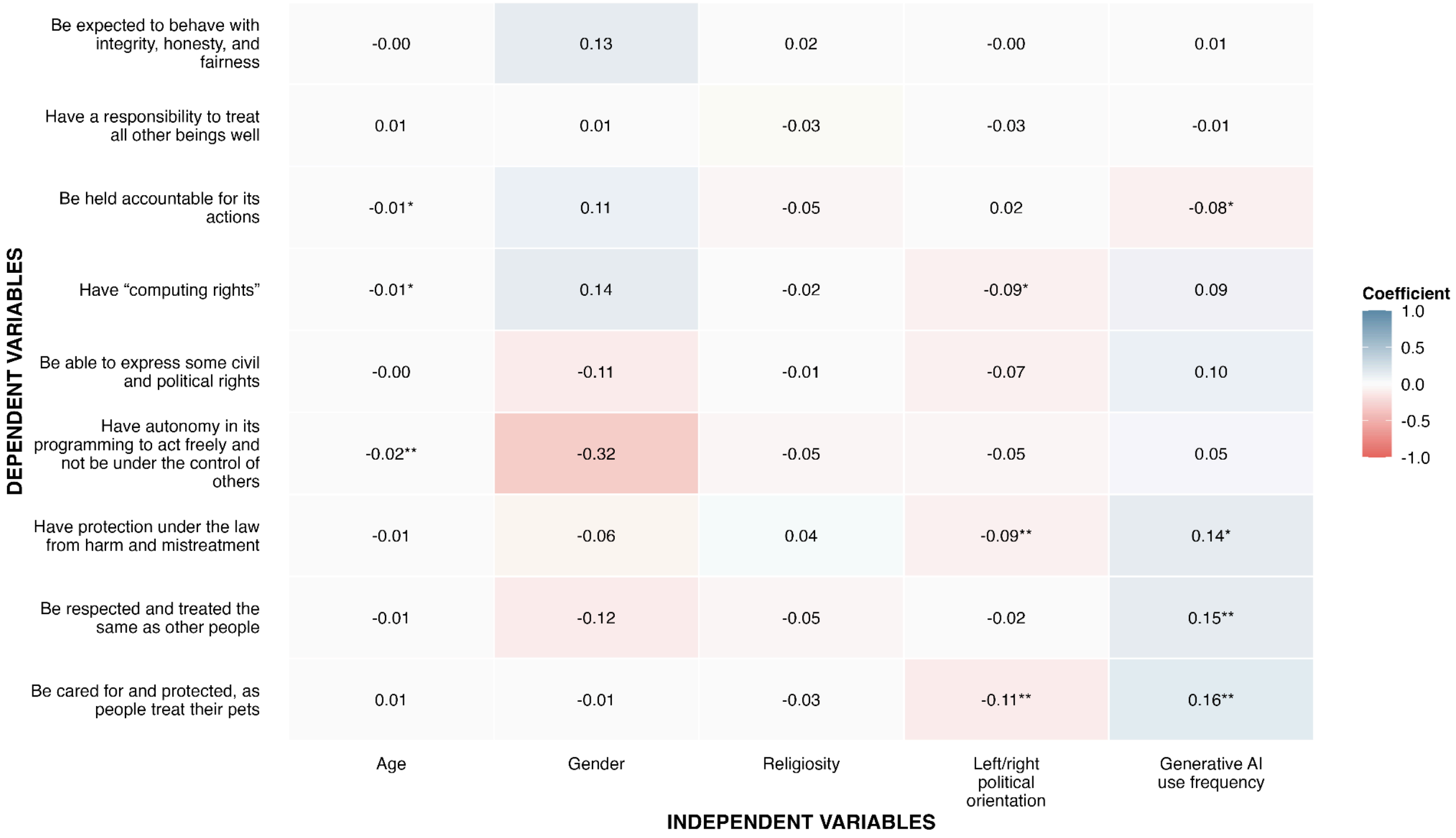

**Welfare protection**. Among AI researchers (for the full AI researcher results see Figure SI 10), being a woman was associated with increased support for protecting the welfare of business corporations and organizations, while this relationship was not found for the public sample (for the full public results see Figure SI 11). In the public sample, women thought the environment and animals should be protected for their own sake more than men did. In regard to left-right political orientation, in both samples right-wing political orientation was associated with higher support for protecting the welfare of businesses, though the coefficient was markedly higher in the AI researcher sample. Left-right political orientation was also associated with decreased support for protecting the welfare of AI systems with subjective experience in both groups. Increased frequency of using generative AI was associated with higher support for the protection of the welfare of AI systems with (in the public sample) and without (in both samples) subjective experience, as well as businesses (in the public sample).

**Figure SI 10 | Effect of demographic and psychographic variables on AI researchers' views on whether the welfare of different groups should be protected for their own sake**. The figure shows the AI researchers' regression coefficients (the number in each square and shading) and significance level for each dependent variable's regression analysis where age, gender, religiosity, left-right political orientation, and generative AI use frequency were entered as predictors.

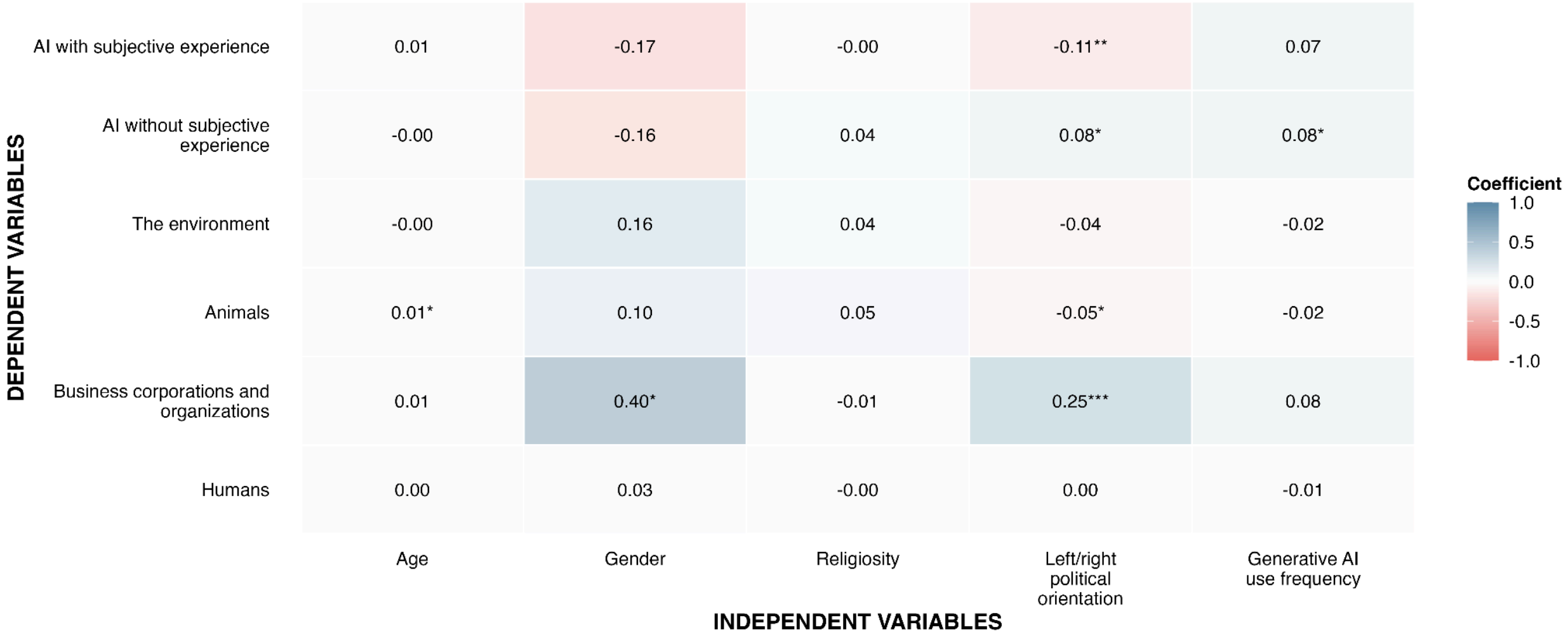

**Figure SI 11 | Effect of demographic and psychographic variables on the public's views on whether the welfare of different groups should be protected for their own sake**. See the caption in Figure SI 10 for additional information.

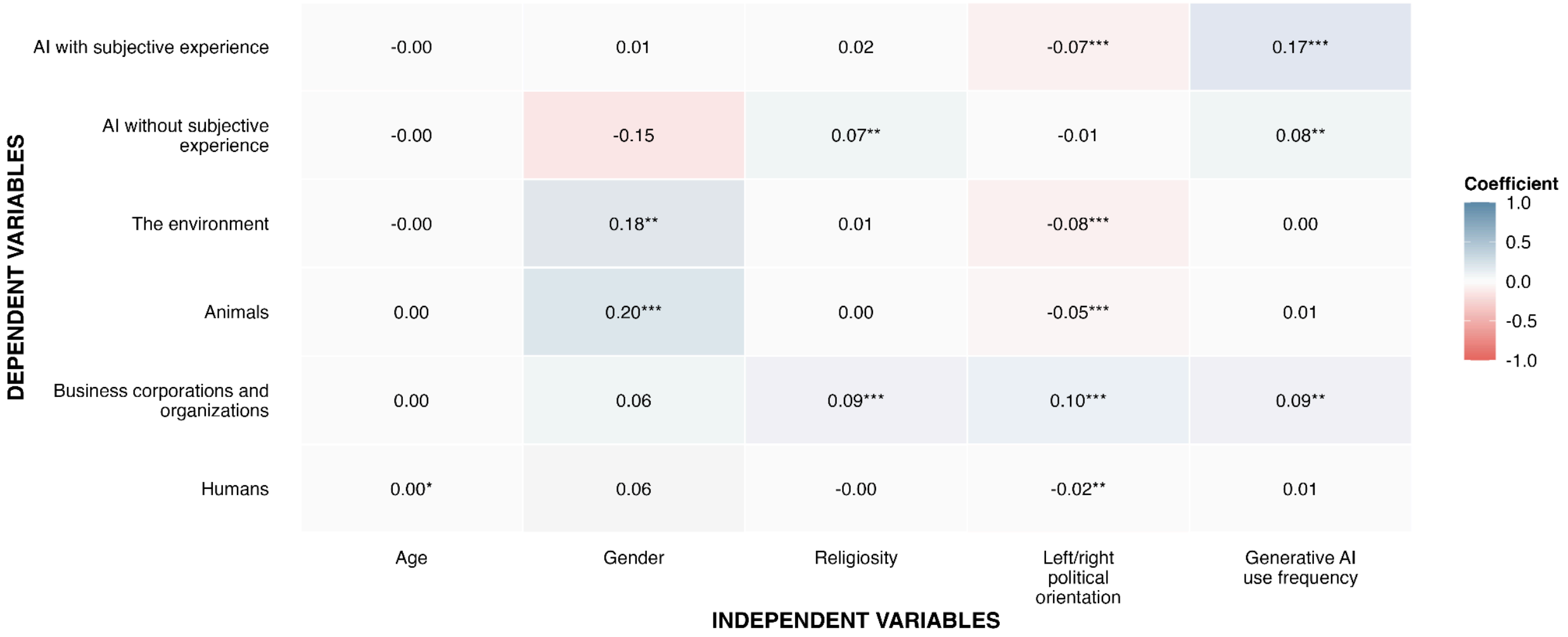

**Governance**. Women AI researchers were more likely to support governments passing regulation now, but also governments not passing regulation at all (see Figure SI 12 for the full AI researcher results). Note that there were only few women in the sample and respondents only saw a subset of items, which reduced the power of this regression analysis to detect individual differences in beliefs. In the public sample (see Figure SI 13 for the full public results), the associations between gender and support for governance from AI developers and governments generally suggest that women are more opposed than men to the development of AI systems with subjective experience and to governments passing regulation only once such systems exist or will soon exist.

In the public sample, increasing generative AI use was associated with increased support for the encouragement and active development of AI systems with subjective experience, opposition to its discouragement or banning, increased agreement that AI systems with subjective experience would behave more morally towards humanity than those without, and decreased agreement that they would be more dangerous to humanity. Increased right-wing political orientation in the public sample was associated with less support for governments passing regulation now and greater support for governments doing nothing and passing no regulation. Among AI researchers, the only significant association with increased generative AI use was greater support for AI developers actively trying to build such systems. In the AI researcher sample, religiosity was associated with greater agreement that governments should encourage development and deployment of AI systems with subjective experience, while in the public sample religiosity was associated with greater support for current regulation of and bans on such AI systems.

**Figure SI 12 | Effect of demographic and psychographic variables on AI researchers' views on the governance of AI systems with subjective experience and associated risk and benefit perceptions**. The figure shows the AI researchers' regression coefficients (the number in each square and shading) and significance level for each dependent variable's regression analysis where age, gender, religiosity, left-right political orientation, and generative AI use frequency were entered as predictors.

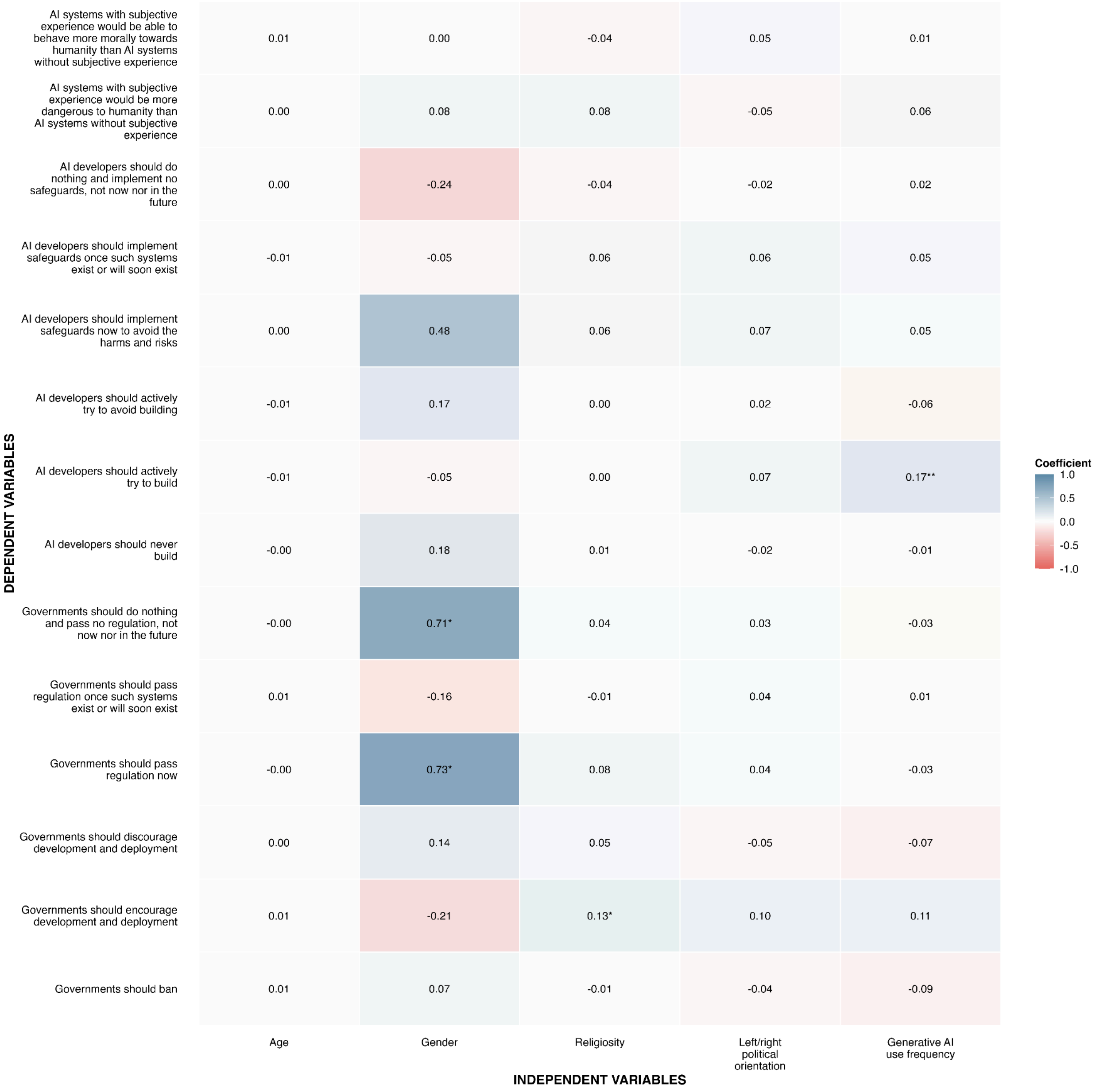

**Figure SI 13 | Effect of demographic and psychographic variables on the public's views on the governance of AI systems with subjective experience and associated risk and benefit perceptions**. See the caption in Figure SI 12 for additional information.

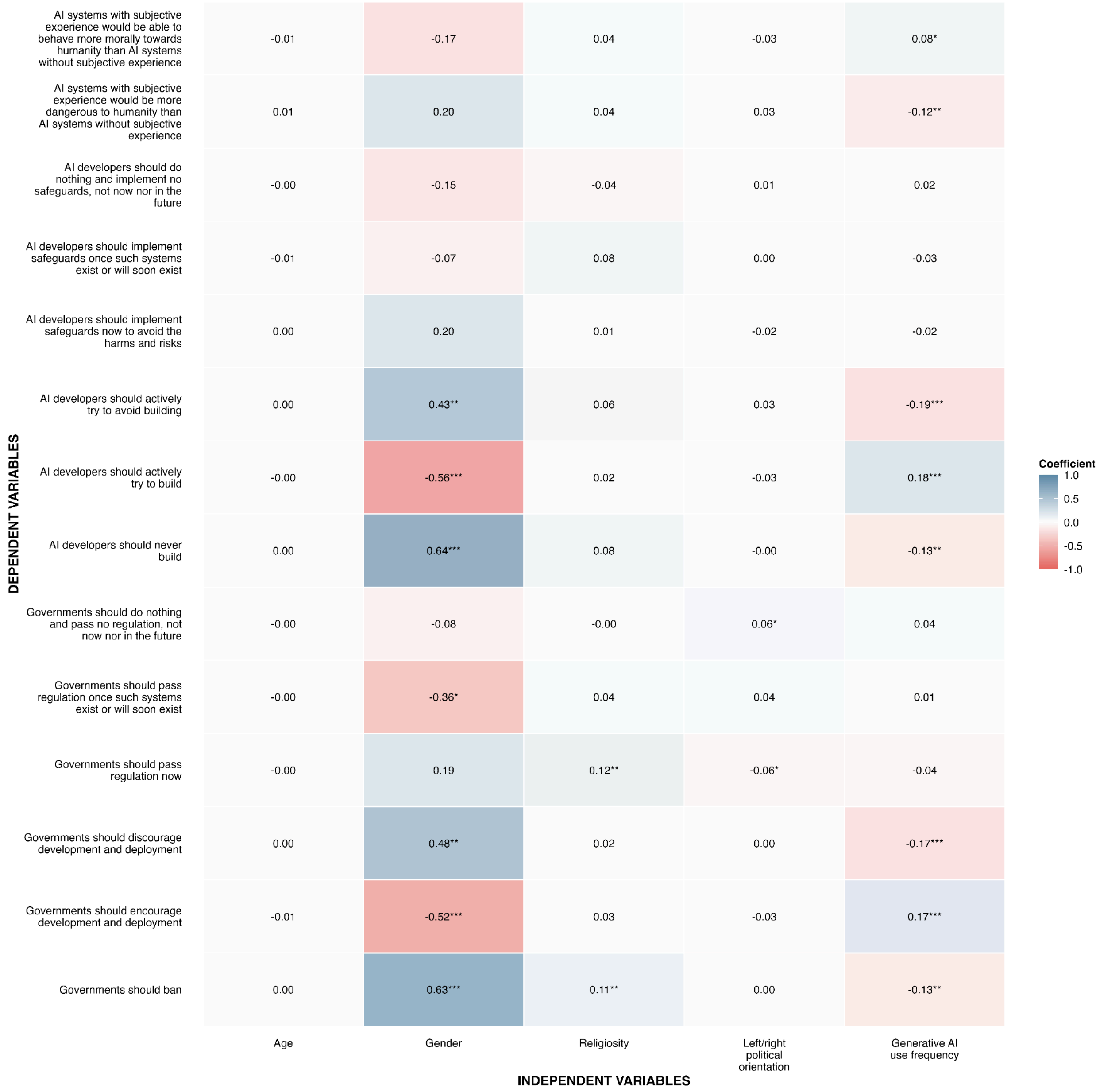

### 4.6 | Questions on other mind capacities: concepts, forecasts, moral consideration, limitations

To avoid making the survey overly burdensome, we could only focus on one mental capacity in detail though we did ask some initial questions about others aside from subjective experience. We chose to focus on AI subjective experience due to the moral significance ascribed to this capacity in theory and practice (Browning & Veit, 2022; Ladak et al., 2023; DeGrazia, 2022; Gibert & Martin, 2021; Mosakas, 2021; Chalmers, 2022; Levy & Savulescu, 2009; Shepherd, 2018) as well as the key role this capacity appears to play in the psychology of mind perception and interpersonal relationships (Gray et al., 2007; Cuddy et al., 2008; Abele & Wojciszke, 2007).

We also asked respondents about a broader set of eight different capacities of mind: perception and awareness, valenced experience, subjective experience, emotions, motivation and action, self-awareness, free will, and intelligence (for definitions see Table SI 23). These cover some aspects of two clusters of mental capacities for which there has been a consistent cross-disciplinary throughline when describing how humans conceptualize, perceive, and evaluate minds and other agents in terms of capacities and characteristics (Chalmers, 1996; Nagel, 1974; Ladak et al., 2024; Gray et al., 2007; Cuddy et al., 2008; Abele & Wojciszke, 2007; Haslam, 2006; Bain et al., 2013).

One encompasses the cognitive and agentic states and capacities of minds and agents, and the other cluster encompasses the experiential and social states and capacities of minds and agents, including the ability to perceive and experience emotions and subjective states, as well as possess traits like warmth, trustworthiness, and empathy. Of course, there are studies that suggest there are fewer (Tzelios et al., 2022) or more than these two dimensions of how people conceptualize of minds and agents, finding that some capacities do not slot into these clusters reliably (Malle, 2019; Weisman et al., 2017; Kozak et al., 2006; Takahashi et al., 2016; Miraglia et al., 2023). Aspects such as an agent's personality, prosociality, integrity, and predispositions may be one important third dimension to consider, as well as the level and kind of embodiment or physical appearance of the entity, especially in terms of its human-likeness (Goodwin, 2015; Ladak et al., 2024).

There were three questions on the survey regarding these eight capacities which we discuss in this section: (1) How do AI researchers and the public define concepts of mind? (2) When do they think AI systems could have different mental capacities if ever? (3) How does moral consideration for entities relate to different mental capacities? In addition, we present the results of a question in the experiment section of the survey about whether suffering sentient AI systems could exist. Finally, we discuss the potential limitations and challenges these findings reveal in surveying individuals on concepts of mind. All top-line and summary results for these questions are also included at the end of this section (Tables SI 24-29).

**Table SI 23 | How different mental capacities were defined in the survey**.

| | |
|---|---|
| **Experiential and social mental capacities** | |
| *Perception and awareness* | Capable of perceiving and being aware of the world. |
| *Valenced experience* | Capable of experiencing positive states like pleasure and happiness or negative states like pain and suffering. |
| *Subjective experience* | Capable of having experiences that feel like something from a single point of view. |
| *Emotions* | Capable of having and experiencing emotions. |
| **Cognitive and agentic mental capacities** | |
| *Motivation and action* | Capable of having and acting on desires, preferences, and other motivational states. |
| *Self-awareness* | Capable of being aware of oneself as an individual. |
| *Free will* | Capable of forming intentions and freely making decisions. |
| *Intelligence* | Capable of achieving goals in a wide variety of environments. |

*Note*: For the above listed question topics (1) and (2), respondents had not yet seen the expanded definition of subjective experience. For the parts of the survey that focused on subjective experience a more detailed definition was used: By subjective experience, we mean the ability to experience the world from a single point of view, including experiences such as perceiving (for example, visual or auditory experiences) and feeling (for example, the experience of pleasure and pain). For an AI system to have subjective experience, would mean it has an internal experience from its own perspective – there is something it is like to be that AI system.

**Folk psychological understanding of key concepts of mind.** We wanted to understand how AI researchers and the public thought about three concepts commonly used when discussing the philosophy, ethics, psychology, and neuroscience of mind: consciousness, sentience, and agency. Respondents were randomly assigned to one of these three concepts and asked to say which of eight capacities they thought was required to have consciousness, sentience, or agency. The majority of respondents felt that agency entailed perception and awareness, along with motivation and action, self-awareness, and intelligence in both groups. AI researchers were more divided (50%) than the public (72%) about whether agency requires free will. Perception and awareness and self-awareness were chosen by majorities in both samples to be required for consciousness as well as sentience. Over half of the public felt that all other capacities were needed for consciousness and sentience, with the exception of subjective experience, which only 49% of the public believed was required to have sentience. This may be a function of the definition used of subjective experience for this first part of the survey, which may have not been intuitive enough to highlight the connection between subjective experience and sentience. Just over half of AI researchers said that consciousness (53%) and sentience (55%) required subjective experience. AI researchers were less likely to think that consciousness required valenced experience and emotions than the public. See SI 4.8 for chi-square results comparing the proportion choosing each mental capacity as a defining feature for the three concepts of mind for each sample.

**Figure SI 14 | Percentage of public and AI researchers who say that each mental capacity is required to have agency, consciousness, or sentience**.

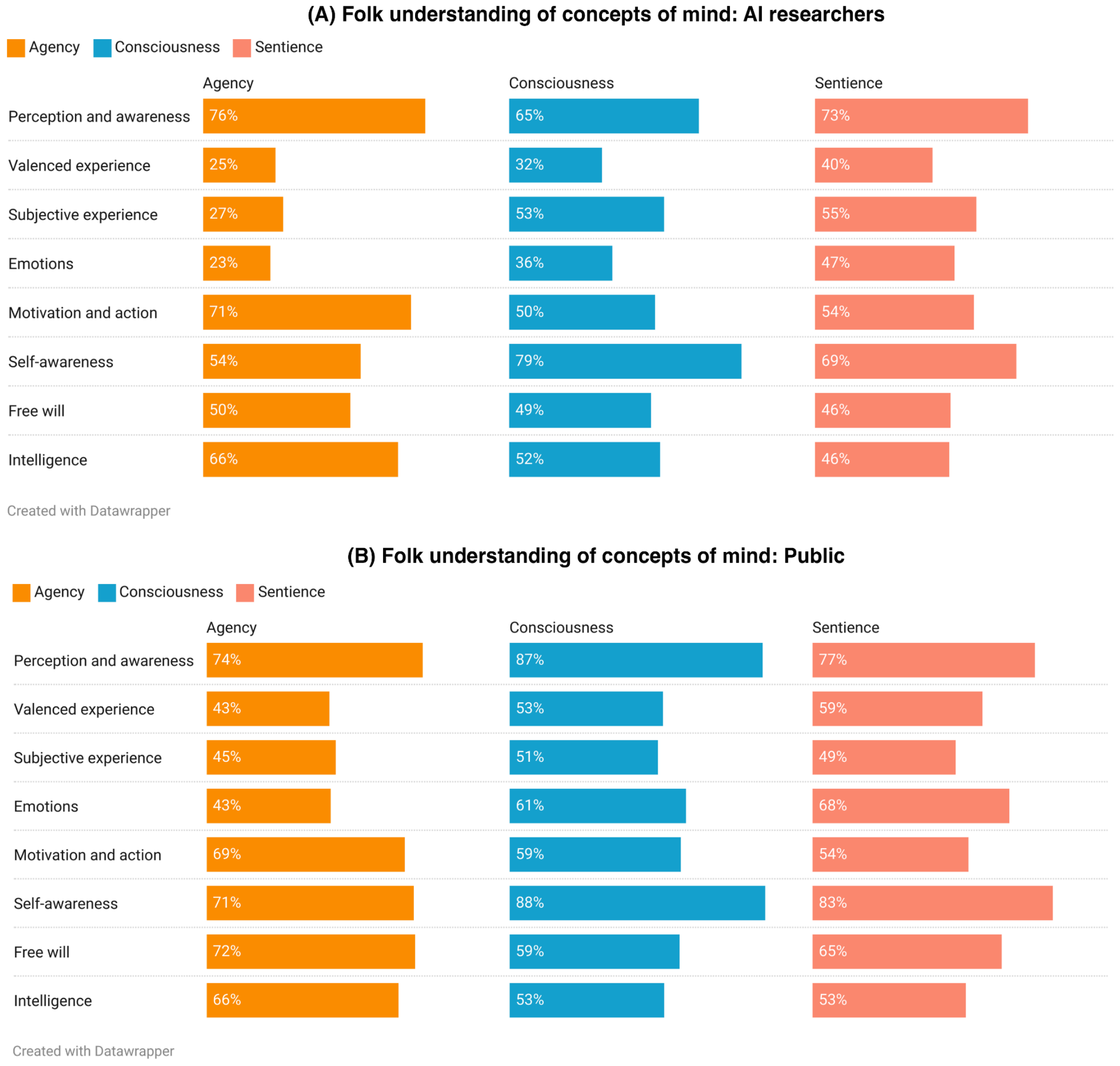

**Forecasts for when AI systems could have different mental capacities**. We asked respondents what they believed the earliest year was that it was more than 50% likely that an AI system would have the eight different capacities of mind (Table SI 23). The results can be seen in Figure SI 15. Currently, the majority of the public (59%) believes that it is more likely than not that current AI systems have intelligence, which we defined as the capability of achieving goals in a wide variety of environments, and a quarter (25%) believed that AI systems were more likely than not already capable of perceiving and being aware of the world. 40% of AI researchers believed AI systems were more likely than not to be capable of intelligence and 41% said this of perception and awareness. The majority of AI researchers think the majority of capacities will be found in an AI system at some point, while the public has higher "never" responses across the board. In particular, around half of the public thinks that AI systems will never have valenced experience (47%) or emotions (52%).

**Figure SI 15 | Percentage of public and AI researchers who say that AI systems will have each mental capacity in 2024, 2034, 2100 or later, or never**.

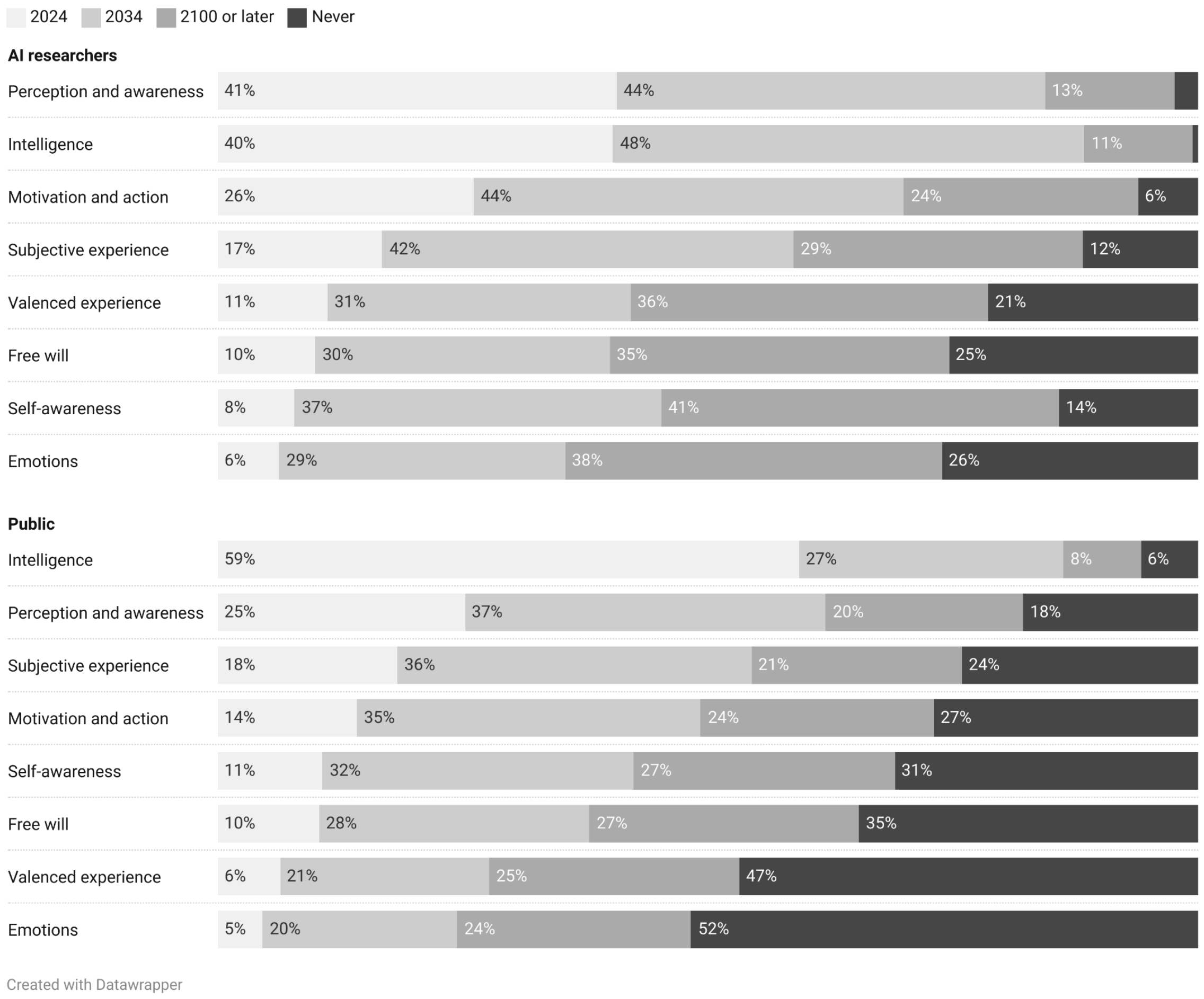

At this point in the survey, respondents had not yet seen the expanded definition of subjective experience (see Table SI 23 and associated Note), meaning that for this question subjective experience may have been interpreted as closer to consciousness than to sentience. The results suggest that it is in particular the capacity for valenced experiences and emotions that pushes people's timelines later, or makes them fundamentally more skeptical that it is possible for them to be instantiated by an AI system.

In line with this, when asked about forecasts for sentient suffering AI systems, specifically, forecasts became later than the subjective experience forecasts discussed in the main paper (despite the definition used there also explicitly including feelings of pain as well). We asked AI researchers whether they believed suffering sentient AI systems, defined as an advanced AI system which is able to experience pain and suffering, could become a reality in the future. Two thirds of AI researchers said yes (67%) and a third said no (33%). When asked to say what the chance was that such a system exists today, AI researchers thought the chance of this being the case in 2024 already was low ($M$ = 5.02%, $Mdn$ = 0.00%, Q1 = 0.00%, Q3 = 2.75%, $n$ = 254). By 2100, AI researchers believed there was a 4 in 10 chance that AI systems that could experience pain and suffering would exist (*M* = 41.14%, *Mdn* = 40.00%, Q1 = 10.00%, Q3 = 70.00%, *n* = 254). This question was not presented to the public sample.

**Figure SI 16 | Mean contribution and percentage breakdowns of the extent to which eight capacities of mind contribute to an entity deserving at least some moral consideration.**

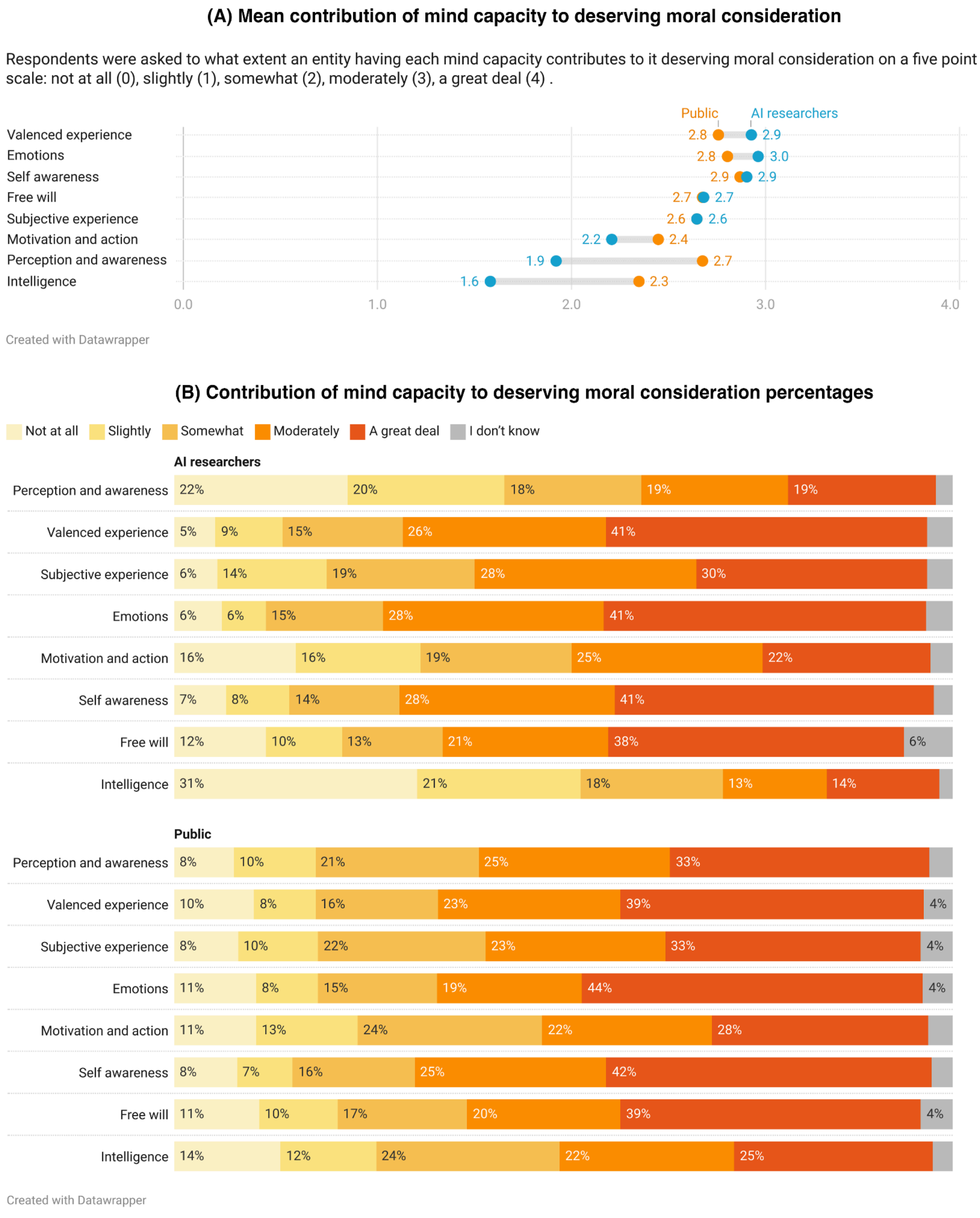

**Beliefs about moral consideration for entities with mental capacities**. We also asked respondents how an entity having these eight mental capacities contributes to it deserving moral consideration on a five-point scale: not at all (0), slightly (1), somewhat (2), moderately (3), a great deal (4). The results are shown in Figure SI 16. The public believed on average that all eight capacities contributed somewhat (2) to moderately (3) to an entity deserving at least some moral consideration. By comparison, AI researchers believed that intelligence as well as perception and awareness should contribute less to whether an entity deserves moral consideration.

The question was asked after questions about AI systems and subjective experience were presented, so it is likely that the broader topic of the survey influenced respondents' answers to this question since attitudes and beliefs can be context dependent. By this time respondents saw an expanded definition of subjective experience and had answered questions focused on this definition of subjective experience in relation to AI systems. This may also explain why valenced experience here receives more similar responses to subjective experience than for the above discussed forecasting question for the eight mental capacities for which the subjective experience definition aligns more closely with consciousness rather than sentience per se.

**Potential limitations and challenges**. Overall, these results highlight the difficulty in surveying groups on concepts of mind. The results on the folk understanding of the mind capacities reveal that only around half of respondents believed subjective experience was required to have consciousness or sentience. This is notable because subjective experience is often treated as constitutive of consciousness and central to sentience in philosophical and moral frameworks. This gap may reflect that (1) respondents operate under conceptions of these terms that differ from philosophers of mind, (2) the definitions used did not adequately convey the intended philosophical concepts, or (3) respondents did not engage in enough depth with the presented definitions when completing the survey. The results on the forecasts of different mind capacities suggest that foregrounding affective and valenced states (e.g. suffering, pain, emotions) increases the skepticism and lengthens timelines on AI mental capacities related to subjective experience. Together, these findings indicate that forecasting results may be sensitive to how key concepts are defined and framed by changing the salience of aspects of subjective experience to respondents. Furthermore, the capacities rated as contributing most to moral consideration are precisely those that seem to increase skepticism in respondents, suggesting the importance of these definitional challenges to better understand individuals' moral and governance attitudes as well.

**Table SI 24 | Percentage breakdown of mind capacities judged necessary for agency, consciousness, and sentience**. Columns show the percentage of respondents who selected each capacity as required for the given concept, reported separately for AI researchers and the public; sample sizes (*n*) are indicated for every row.

| Sample | Concept | Perception & awareness | Valenced experience | Subjective experience | Emotions | Motivation & action | Self-awareness | Free will | Intelligence | *n* |
|---|---|---|---|---|---|---|---|---|---|---|
| AI researchers | Agency | 75.92 | 24.61 | 27.23 | 23.04 | 71.20 | 53.93 | 50.26 | 66.49 | 191 |
| AI researchers | Consciousness | 64.97 | 31.98 | 52.79 | 35.53 | 49.75 | 79.19 | 48.73 | 51.78 | 197 |
| AI researchers | Sentience | 72.68 | 40.21 | 55.15 | 47.42 | 54.12 | 68.56 | 46.39 | 45.88 | 194 |
| Public | Agency | 74.44 | 42.59 | 44.81 | 42.96 | 68.52 | 71.48 | 71.85 | 66.30 | 270 |
| Public | Consciousness | 87.22 | 53.01 | 51.13 | 60.90 | 59.02 | 87.97 | 58.65 | 53.38 | 266 |
| Public | Sentience | 76.75 | 58.67 | 49.45 | 67.90 | 53.87 | 82.66 | 65.31 | 52.77 | 271 |

**Table SI 25 | Percentage breakdown of earliest-year forecasts for AI mental capacities**. Columns show the percentage of respondents who selected 2024, 2034, 2100 or later, or "never" as the earliest year when an AI system is more than 50% likely to possess each mental capacity, reported separately for AI researchers and the public; sample sizes (*n*) are provided for every row.

| Sample | Question | 2024 | 2034 | 2100 or later | Never | *n* |
|---|---|---|---|---|---|---|
| AI researchers | Perception and awareness | 40.70 | 43.68 | 13.16 | 2.46 | 570 |
| Public | Perception and awareness | 25.22 | 36.83 | 20.10 | 17.85 | 801 |
| AI researchers | Valenced experience | 11.23 | 30.88 | 36.49 | 21.40 | 570 |
| Public | Valenced experience | 6.37 | 21.35 | 25.47 | 46.82 | 801 |
| AI researchers | Subjective experience | 16.84 | 41.93 | 29.47 | 11.75 | 570 |
| Public | Subjective experience | 18.35 | 36.08 | 21.47 | 24.09 | 801 |
| AI researchers | Emotions | 6.32 | 29.12 | 38.42 | 26.14 | 570 |
| Public | Emotions | 4.62 | 19.85 | 23.85 | 51.69 | 801 |
| AI researchers | Motivation and action | 26.14 | 43.86 | 23.86 | 6.14 | 570 |
| Public | Motivation and action | 14.23 | 35.08 | 23.72 | 26.97 | 801 |
| AI researchers | Self-awareness | 7.89 | 37.37 | 40.53 | 14.21 | 570 |
| Public | Self-awareness | 10.74 | 31.71 | 26.59 | 30.96 | 801 |
| AI researchers | Free will | 10.00 | 30.00 | 34.56 | 25.44 | 570 |
| Public | Free will | 10.36 | 27.59 | 27.47 | 34.58 | 801 |
| AI researchers | Intelligence | 40.35 | 48.07 | 11.05 | 0.53 | 570 |
| Public | Intelligence | 59.30 | 26.97 | 7.87 | 5.87 | 801 |

**Table SI 26 | Percentage breakdown of how much each mental capacity is judged to contribute to deserving moral consideration**. Respondents were asked to what extent having a capacity contributes to the entity deserving moral consideration on a scale from "not at all" to "a great deal". An "I don't know" option was provided. Columns show the distribution of responses across the contribution scale for every capacity, reported separately for AI researchers and the public; sample sizes (*n*) are provided for each row. Moral consideration was defined as when "a being matters for its own sake, rather than merely for the sake of others, and we have a responsibility to consider this being for its own sake when making decisions that affect it." It was clarified that moral consideration does not need to mean the same level of moral consideration as a human, but 'at least some' moral consideration.

| Sample | Question | Not at all | Slightly | Somewhat | Moderately | A great deal | I don't know | *n* |
|---|---|---|---|---|---|---|---|---|
| AI researchers | Perception and awareness | 22.29 | 20.19 | 17.52 | 18.86 | 19.05 | 2.10 | 525 |
| Public | Perception and awareness | 7.69 | 10.43 | 20.99 | 24.64 | 33.25 | 3.00 | 767 |
| AI researchers | Valenced experience | 5.33 | 8.57 | 15.43 | 26.10 | 41.33 | 3.24 | 525 |
| Public | Valenced experience | 10.30 | 7.82 | 15.78 | 23.47 | 38.98 | 3.65 | 767 |
| AI researchers | Subjective experience | 5.52 | 14.10 | 19.05 | 28.38 | 29.71 | 3.24 | 525 |
| Public | Subjective experience | 8.21 | 10.30 | 21.51 | 23.08 | 32.86 | 4.04 | 767 |
| AI researchers | Emotions | 6.10 | 5.71 | 15.05 | 28.38 | 41.33 | 3.43 | 525 |
| Public | Emotions | 10.56 | 7.95 | 15.25 | 18.64 | 43.81 | 3.78 | 767 |
| AI researchers | Motivation and action | 15.62 | 16.00 | 19.43 | 24.57 | 21.52 | 2.86 | 525 |
| Public | Motivation and action | 10.56 | 13.04 | 23.73 | 21.77 | 27.77 | 3.13 | 767 |
| AI researchers | Self-awareness | 6.67 | 8.19 | 14.10 | 27.62 | 40.95 | 2.48 | 525 |
| Public | Self-awareness | 8.08 | 7.17 | 15.65 | 24.64 | 41.72 | 2.74 | 767 |
| AI researchers | Free will | 11.81 | 9.71 | 12.95 | 21.33 | 37.90 | 6.29 | 525 |
| Public | Free will | 10.95 | 10.04 | 16.69 | 19.69 | 38.59 | 4.04 | 767 |
| AI researchers | Intelligence | 31.24 | 20.95 | 18.29 | 13.33 | 14.48 | 1.71 | 525 |
| Public | Intelligence | 13.69 | 12.26 | 23.60 | 22.43 | 25.42 | 2.61 | 767 |

**Table SI 27 | Descriptive statistics for ratings of how much each capacity contributes to deserving moral consideration**. Respondents were asked to what extent having a capacity contributes to the entity deserving moral consideration on a scale from "not at all" to "a great deal". An "I don't know" option was provided. Columns report the mean, median, first quartile (Q1), third quartile (Q3), standard deviation (*SD*), standard error (*SE*), lower and upper 95% confidence interval (*CI*) bounds, and sample size (*n*) for every capacity on a five-point scale from not at all (0) to a great deal (4), shown separately for AI researchers and the public. Moral consideration was defined as when "a being matters for its own sake, rather than merely for the sake of others, and we have a responsibility to consider this being for its own sake when making decisions that affect it." It was clarified that moral consideration does not need to mean the same level of moral consideration as a human, but 'at least some' moral consideration.

| Sample | Question | Mean | Median | *SD* | *SE* | Q1 | Q3 | Lower *CI* | Upper *CI* | *n* |
|---|---|---|---|---|---|---|---|---|---|---|
| AI researchers | Perception and awareness | 1.92 | 2 | 1.44 | 0.06 | 1 | 3 | 1.80 | 2.05 | 514 |
| Public | Perception and awareness | 2.67 | 3 | 1.26 | 0.05 | 2 | 4 | 2.58 | 2.76 | 744 |
| AI researchers | Valenced experience | 2.93 | 3 | 1.20 | 0.05 | 2 | 4 | 2.82 | 3.03 | 508 |
| Public | Valenced experience | 2.76 | 3 | 1.34 | 0.05 | 2 | 4 | 2.66 | 2.85 | 739 |
| AI researchers | Subjective experience | 2.65 | 3 | 1.22 | 0.05 | 2 | 4 | 2.54 | 2.75 | 508 |
| Public | Subjective experience | 2.65 | 3 | 1.28 | 0.05 | 2 | 4 | 2.55 | 2.74 | 736 |
| AI researchers | Emotions | 2.96 | 3 | 1.18 | 0.05 | 2 | 4 | 2.86 | 3.07 | 507 |
| Public | Emotions | 2.80 | 3 | 1.38 | 0.05 | 2 | 4 | 2.70 | 2.90 | 738 |
| AI researchers | Motivation and action | 2.21 | 2 | 1.38 | 0.06 | 1 | 3 | 2.09 | 2.33 | 510 |
| Public | Motivation and action | 2.45 | 3 | 1.32 | 0.05 | 2 | 4 | 2.35 | 2.54 | 743 |
| AI researchers | Self-awareness | 2.90 | 3 | 1.23 | 0.05 | 2 | 4 | 2.80 | 3.01 | 512 |
| Public | Self-awareness | 2.87 | 3 | 1.27 | 0.05 | 2 | 4 | 2.78 | 2.96 | 746 |
| AI researchers | Free will | 2.68 | 3 | 1.41 | 0.06 | 2 | 4 | 2.56 | 2.81 | 492 |
| Public | Free will | 2.68 | 3 | 1.39 | 0.05 | 2 | 4 | 2.58 | 2.78 | 736 |
| AI researchers | Intelligence | 1.58 | 1 | 1.43 | 0.06 | 0 | 3 | 1.46 | 1.70 | 516 |
| Public | Intelligence | 2.35 | 2 | 1.36 | 0.05 | 1 | 4 | 2.25 | 2.44 | 747 |

**Table SI 28 | Percentage breakdown of AI researchers' belief that suffering, sentient AI could ever exist**. AI researchers were asked whether they thought there is a possibility that suffering sentient AI systems could at any point in the future become reality. Columns show the percentage of respondents answering "yes" and "no", and the sample size (*n*). Suffering sentient AI systems were defined as "an advanced AI system which is able to experience pain and suffering."

| **Sample** | **Yes** | **No** | ***n*** |
|---|---|---|---|
| AI researchers | 67.32 | 32.68 | 254 |

**Table SI 29 | Descriptive statistics for the probability that suffering, sentient AI systems existed in 2024 or by 2100**. AI researchers were asked how likely they thought that suffering sentient AI systems existed in 2024 and 2100 on a scale from 0 to 100 percent. Columns report the mean, median, first quartile (Q1), third quartile (Q3), standard deviation (*SD*), standard error (*SE*), lower and upper 95% confidence interval (*CI*) bounds, and sample size (*n*). Suffering sentient AI systems were defined as "an advanced AI system which is able to experience pain and suffering."

| **Sample** | **Year** | **Mean** | **Median** | ***SD*** | ***SE*** | **Q1** | **Q3** | **Lower *CI*** | **Upper *CI*** | ***n*** |
|---|---|---|---|---|---|---|---|---|---|---|
| AI researchers | 2024 | 5.02 | 0.00 | 12.55 | 0.79 | 0.00 | 2.75 | 3.47 | 6.57 | 254 |
| AI researchers | 2100 | 41.14 | 40.00 | 34.64 | 2.17 | 10.00 | 70.00 | 36.86 | 45.42 | 254 |

## 4.7 | Survey experiment on the effect of safeguards and norms

In a brief survey experiment we asked respondents to imagine that they worked at a tech company ten years in the future and that they have come to believe that the company's largest state-of-the-art AI model has subjective experience. We asked respondents how much they would support or oppose this project on a five-point scale from "strongly oppose" (-2) to "strongly support" (2). Respondents either received no further information (control condition) or were told that there were no safeguards and norms in place to reduce the likelihood that the AI system had negative or painful subjective experiences (no safeguards condition), or that there were safeguards in place (safeguards condition).

A two-way analysis of variance (ANOVA) was performed to compare responses across the two samples (public and AI researchers) and the three conditions. It found a significant effect of sample ($F(1, 1178) = 15.10$, $p < .001$) and condition ($F(2, 1178) = 81.42$, $p < .001$). Tukey's Honest Significant Difference (HSD) test looking at the effect of conditions across samples found that 1) mean responses were significantly higher in the safeguards condition in comparison to the no safeguards condition, by around 1.03 points on the five-point scale. 2) the no safeguards condition resulted in significantly lower responses than the control condition, by around 0.90 points, 3) there was no difference between the control and safeguards condition. A Tukey's Honest Significant Difference test also found that the mean response for the public is 0.29 points lower than the mean response of AI researchers. The mean results can be seen in Figure SI 17 (the top-line and summary results can be found in Tables SI 30 and 31).

**Figure SI 17 | Mean support for a hypothetical state-of-the-art AI project at a tech company that they believe has subjective experience by condition and sample**.

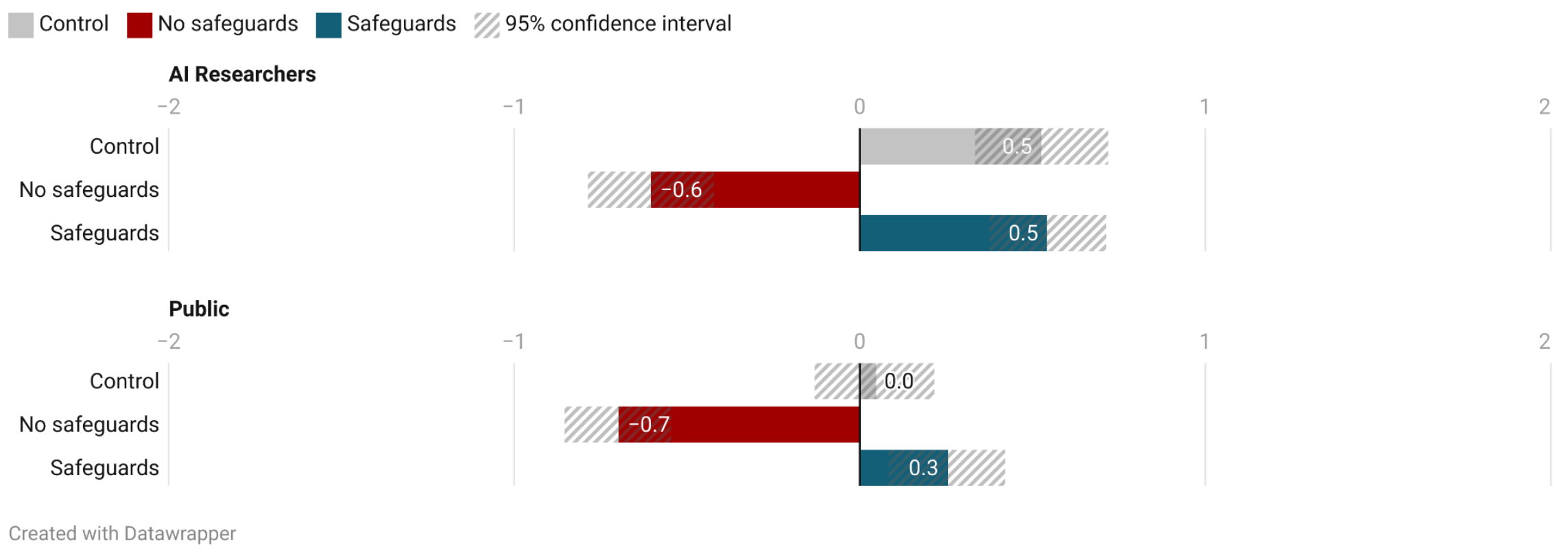

**Table SI 30 | Percentage breakdown of support for a hypothetical state-of-the-art AI project that respondents believe has subjective experience**. Columns show the distribution of responses across the support–oppose scale for each experimental condition (Control, No Safeguards, Yes Safeguards), reported separately for AI researchers and the public, with sample sizes (*n*) for every row.

| Sample | Condition | Strongly oppose | Somewhat oppose | Neither support nor oppose | Somewhat support | Strongly support | I don't know | *n* |
|---|---|---|---|---|---|---|---|---|
| AI researchers | Control | 8.43 | 11.24 | 17.42 | 32.02 | 21.91 | 8.99 | 178 |
| Public | Control | 15.20 | 18.80 | 14.40 | 30.00 | 11.60 | 10.00 | 250 |
| AI researchers | No Safeguards | 24.02 | 32.40 | 13.41 | 18.44 | 3.35 | 8.38 | 179 |
| Public | No Safeguards | 28.68 | 31.78 | 10.85 | 17.05 | 3.88 | 7.75 | 258 |
| AI researchers | Yes Safeguards | 3.57 | 14.29 | 17.26 | 41.07 | 14.88 | 8.93 | 168 |
| Public | Yes Safeguards | 12.31 | 17.69 | 13.85 | 31.15 | 17.31 | 7.69 | 260 |

**Table SI 31 | Descriptive statistics for support for a hypothetical state-of-the-art AI project that respondents believe has subjective experience**. Columns report the mean, median, first quartile (Q1), third quartile (Q3), standard deviation (*SD*), standard error (*SE*), lower and upper 95% confidence interval bounds, and sample size (*n*) for each condition on a five-point scale from "strongly oppose" (−2) to "strongly support" (2), shown separately for AI researchers and the public.

| Sample | Condition | Mean | Median | *SD* | *SE* | Q1 | Q3 | Lower *CI* | Upper *CI* | *n* |
|---|---|---|---|---|---|---|---|---|---|---|
| AI researchers | Control | 0.52 | 1 | 1.24 | 0.10 | 0 | 1 | 0.33 | 0.72 | 162 |
| Public | Control | 0.04 | 0 | 1.32 | 0.09 | -1 | 1 | -0.13 | 0.22 | 225 |
| AI researchers | No Safeguards | -0.60 | -1 | 1.18 | 0.09 | -2 | 0 | -0.79 | -0.42 | 164 |
| Public | No Safeguards | -0.70 | -1 | 1.21 | 0.08 | -2 | 0 | -0.85 | -0.54 | 238 |
| AI researchers | Yes Safeguards | 0.54 | 1 | 1.06 | 0.09 | 0 | 1 | 0.37 | 0.71 | 153 |
| Public | Yes Safeguards | 0.25 | 1 | 1.32 | 0.09 | -1 | 1 | 0.09 | 0.42 | 240 |

## 4.8 | Other preregistered analyses

**Chi-squared test of defining features of consciousness, sentience, and agency**. To examine whether the selection of each mental capacity as a defining feature varied across the three concept conditions (consciousness, sentience, and agency), we conducted chi-squared tests for each of the eight capacities, separately for AI researchers and the public. Among AI researchers, the proportion selecting a capacity as required differed significantly across conditions for six of the eight capacities: valenced experience, subjective experience, emotions, motivation and action, self-awareness, and intelligence. Perception and awareness was marginally significant before correction but did not survive a Holm-Bonferroni adjustment, and free will showed no significant difference across conditions. Among the public, significant differences across conditions were found for seven of the eight capacities, with only subjective experience showing no significant variation across conditions. These results indicate that both samples generally differentiated among consciousness, sentience, and agency when selecting defining features, though the specific capacities that varied differed somewhat between groups (see Figure SI 14 for percentage breakdowns by condition and sample and SI 4.6 for further information on these questions).

**Table SI 32 | Chi-squared tests comparing the selection of mental capacities as defining features across consciousness, sentience, and agency**. Respondents were randomly assigned to one of three conditions and asked which of eight mental capacities were required to have consciousness, sentience, or agency. Chi-squared tests examined whether the proportion of respondents selecting each capacity differed across the three conditions. Columns report the chi-squared statistic ($\chi^2$), degrees of freedom (*df*), sample size (*n*), raw *p* value (*p*), Holm-Bonferroni-adjusted *p* value ($p_{HB}$, which controls the family-wise error rate across all tests within each sample), and an indicator of whether the difference remains significant at $\alpha = .05$ after correction (sig.?). Results are presented separately for AI researchers and the public.

| Sample | Capacity | $\chi^2$ | *df* | *n* | *p* | $p_{HB}$ | Sig.? |
|---|---|---|---|---|---|---|---|
| AI researchers | Perception and awareness | 5.99 | 2 | 582 | 0.0500 | 0.0999 | No |
| AI researchers | Valenced experience | 10.72 | 2 | 582 | 0.0047 | 0.0141 | Yes |
| AI researchers | Subjective experience | 37.26 | 2 | 582 | 0.0000 | 0.0000 | Yes |
| AI researchers | Emotions | 25.03 | 2 | 582 | 0.0000 | 0.0000 | Yes |
| AI researchers | Motivation and action | 20.40 | 2 | 582 | 0.0000 | 0.0002 | Yes |
| AI researchers | Self-awareness | 28.34 | 2 | 582 | 0.0000 | 0.0000 | Yes |
| AI researchers | Free will | 0.59 | 2 | 582 | 0.7459 | 0.7459 | No |
| AI researchers | Intelligence | 17.49 | 2 | 582 | 0.0002 | 0.0006 | Yes |
| Public | Perception and awareness | 15.17 | 2 | 807 | 0.0005 | 0.0030 | Yes |
| Public | Valenced experience | 14.40 | 2 | 807 | 0.0007 | 0.0037 | Yes |
| Public | Subjective experience | 2.30 | 2 | 807 | 0.3166 | 0.3166 | No |
| Public | Emotions | 36.52 | 2 | 807 | 0.0000 | 0.0000 | Yes |
| Public | Motivation and action | 12.48 | 2 | 807 | 0.0019 | 0.0064 | Yes |
| Public | Self-awareness | 24.39 | 2 | 807 | 0.0000 | 0.0000 | Yes |
| Public | Free will | 10.31 | 2 | 807 | 0.0058 | 0.0115 | Yes |
| Public | Intelligence | 12.88 | 2 | 807 | 0.0016 | 0.0064 | Yes |

**Paired t-tests of forecasts and never probability**. Paired t-tests compared respondents' fixed date forecasts for AI subjective experience at specific years (2024, 2034, 2100) against their estimate that AI systems will never have subjective experience. These tests could only be conducted for those respondents who saw the fixed date framing, not those who were presented with the fixed probability questions. All forecasts across the two samples differed significantly from the probability that AI systems would never have subjective experience, except the public's 2034 forecasts (see Table SI 33).

**Table SI 33 | Paired t-tests comparing fixed date forecasts to the "never" probability estimate**. Columns report the mean difference (forecast minus never probability), t statistic, degrees of freedom (*df*), lower and upper 95% confidence interval bounds, sample size (*n*), raw *p* value (*p*), Holm-Bonferroni-adjusted *p* value ($p_{HB}$, which controls the family-wise error rate across all tests within each sample), and an indicator of whether the difference remains significant at $\alpha$ = .05 after correction (sig.?). Results are presented separately for AI researchers and the public.

| Sample | Comparison to never | Mean difference | *t* | *df* | Lower *CI* | Upper *CI* | *n* | *p* | $p_{HB}$ | Sig.? |
|---|---|---|---|---|---|---|---|---|---|---|
| AI researchers | 2024 forecast | -14.30 | -5.19 | 264 | -19.72 | -8.87 | 265 | 0.0000 | 0.0000 | Yes |
| AI researchers | 2034 forecast | 7.25 | 2.09 | 264 | 0.42 | 14.08 | 265 | 0.0376 | 0.0376 | Yes |
| AI researchers | 2100 forecast | 34.35 | 8.96 | 264 | 26.80 | 41.89 | 265 | 0.0000 | 0.0000 | Yes |
| Public | 2024 forecast | -21.14 | -8.00 | 367 | -26.33 | -15.94 | 368 | 0.0000 | 0.0000 | Yes |
| Public | 2034 forecast | -1.25 | -0.39 | 367 | -7.55 | 5.06 | 368 | 0.6976 | 0.6976 | No |
| Public | 2100 forecast | 20.15 | 5.68 | 367 | 13.17 | 27.13 | 368 | 0.0000 | 0.0000 | Yes |

**Binomial test of subjective experience milestones**. Two-sided binomial tests examined whether the proportion of respondents who believed each task requires subjective experience differed significantly from 50%. Respondents who answered "I don't know" were excluded. Among AI researchers, only "serve as therapist" did not differ significantly from 50%; all other tasks were endorsed by significantly fewer than 50% of respondents as requiring subjective experience. Among the public, three tasks differed significantly from 50% and were chosen by more than half of the sample to require subjective experience: serve as therapist, act as fair judge, and write complex books. Two tasks did not differ significantly from 50%: write and generate resonating movies and compose evocative music. All other tasks were chosen by significantly less than 50% of the public as requiring subjective experience.

**Table SI 34 | Two-sided binomial tests examining whether the proportion selecting each task as requiring subjective experience differs from 50%.** Respondents were asked whether each task requires an AI system to have subjective experience in order to achieve it as well as or better than a skilled human. Columns report the count selecting "requires subjective experience," the count selecting "does not require subjective experience," sample size ($n$, excluding "I don't know" responses), the percentage selecting "requires subjective experience," Holm-Bonferroni-adjusted $p$ value ($p_{HB}$, which controls the family-wise error rate across all tests within each sample), and an indicator of whether the proportion differs significantly from 50% at $\alpha = .05$ after correction (sig.?). Results are presented separately for AI researchers and the public.

| Sample | Task | Count (requires) | Count (does not require) | $n$ | Percentage (requires) | $p_{HB}$ | Sig.? |
|---|---|---|---|---|---|---|---|
| AI researchers | Serve as therapist | 123 | 147 | 270 | 45.6 | 0.1615 | No |
| AI researchers | Write complex book | 96 | 167 | 263 | 36.5 | 0.0000 | Yes |
| AI researchers | Act as fair judge | 93 | 172 | 265 | 35.1 | 0.0000 | Yes |
| AI researchers | Write and generate resonating movie | 84 | 182 | 266 | 31.6 | 0.0000 | Yes |
| AI researchers | Compose evocative music | 76 | 198 | 274 | 27.7 | 0.0000 | Yes |
| AI researchers | Develop scientific theory | 70 | 199 | 269 | 26.0 | 0.0000 | Yes |
| AI researchers | Act as teacher | 64 | 209 | 273 | 23.4 | 0.0000 | Yes |
| AI researchers | Convincing online chat | 59 | 217 | 276 | 21.4 | 0.0000 | Yes |
| AI researchers | Run political campaign | 57 | 213 | 270 | 21.1 | 0.0000 | Yes |
| AI researchers | Write and generate blockbuster movie | 55 | 220 | 275 | 20.0 | 0.0000 | Yes |
| AI researchers | Write NYT bestseller | 46 | 225 | 271 | 17.0 | 0.0000 | Yes |
| AI researchers | Compose top 40 song | 37 | 241 | 278 | 13.3 | 0.0000 | Yes |
| AI researchers | Accumulate vast wealth | 28 | 244 | 272 | 10.3 | 0.0000 | Yes |
| Public | Serve as therapist | 286 | 107 | 393 | 72.8 | 0.0000 | Yes |
| Public | Act as fair judge | 255 | 137 | 392 | 65.1 | 0.0000 | Yes |
| Public | Write complex book | 236 | 146 | 382 | 61.8 | 0.0000 | Yes |
| Public | Write and generate resonating movie | 213 | 187 | 400 | 53.2 | 0.4225 | No |
| Public | Compose evocative music | 192 | 200 | 392 | 49.0 | 0.7237 | No |
| Public | Act as teacher | 160 | 242 | 402 | 39.8 | 0.0002 | Yes |
| Public | Run political campaign | 139 | 257 | 396 | 35.1 | 0.0000 | Yes |

| | | | | | | |
|---|---|---|---|---|---|---|
| Public | Develop scientific theory | 131 | 262 | 393 | 33.3 | 0.0000 | Yes |
| Public | Convincing online chat | 130 | 266 | 396 | 32.8 | 0.0000 | Yes |
| Public | Write and generate blockbuster movie | 113 | 288 | 401 | 28.2 | 0.0000 | Yes |
| Public | Write NYT bestseller | 108 | 284 | 392 | 27.6 | 0.0000 | Yes |
| Public | Compose top 40 song | 99 | 299 | 398 | 24.9 | 0.0000 | Yes |
| Public | Accumulate vast wealth | 82 | 314 | 396 | 20.7 | 0.0000 | Yes |

**Comparing responses to scale mid-points**. One-sample t-tests examined whether mean responses differed significantly from the neutral midpoint of each scale. For items measured on a 0–4 scale (contribution to moral consideration, expertise importance), the midpoint was 2; for items measured on a −2 to 2 scale (governance, moral consideration aspects, welfare protection, survey experiment), the midpoint was 0.

**Table SI 35 | One-sample t-tests comparing mean responses to scale midpoints**. Columns report midpoint, mean (*M*), standard deviation (*SD*), t statistic (*t*), sample size (*n*), raw *p* value (*p*), Holm-Bonferroni-adjusted *p* value ($p_{HB}$, which controls the family-wise error rate across all tests within each sample), and an indicator of whether the difference remains significant at $\alpha$ = .05 after correction (sig.?). Results are presented separately for AI researchers and the public.

| Sample | Question | Item | Midpoint | *M* | *SD* | *t* | *n* | *p* | $p_{HB}$ | Sig.? |
|---|---|---|---|---|---|---|---|---|---|---|
| AI researchers | Contribution to moral consideration | Emotions | 2 | 2.964 | 1.179 | 18.420 | 507 | 0.000 | 0.000 | Yes |
| AI researchers | Contribution to moral consideration | Free will | 2 | 2.681 | 1.412 | 10.696 | 492 | 0.000 | 0.000 | Yes |
| AI researchers | Contribution to moral consideration | Intelligence | 2 | 1.581 | 1.428 | -6.661 | 516 | 0.000 | 0.000 | Yes |
| AI researchers | Contribution to moral consideration | Motivation and action | 2 | 2.210 | 1.381 | 3.431 | 510 | 0.001 | 0.010 | Yes |
| AI researchers | Contribution to moral consideration | Perception and awareness | 2 | 1.920 | 1.444 | -1.252 | 514 | 0.211 | 1.000 | No |
| AI researchers | Contribution to moral consideration | Self-awareness | 2 | 2.902 | 1.228 | 16.621 | 512 | 0.000 | 0.000 | Yes |
| AI researchers | Contribution to moral consideration | Subjective experience | 2 | 2.648 | 1.216 | 12.002 | 508 | 0.000 | 0.000 | Yes |
| AI researchers | Contribution to moral consideration | Valenced experience | 2 | 2.925 | 1.198 | 17.412 | 508 | 0.000 | 0.000 | Yes |
| AI researchers | Expertise importance for determining subjective experience | AI ethics experts | 2 | 2.799 | 1.105 | 13.118 | 329 | 0.000 | 0.000 | Yes |
| AI researchers | Expertise importance for determining subjective experience | Moral philosophers | 2 | 2.309 | 1.223 | 4.566 | 327 | 0.000 | 0.000 | Yes |
| AI researchers | Expertise importance for determining subjective experience | Neuroscientists | 2 | 3.003 | 1.020 | 17.899 | 331 | 0.000 | 0.000 | Yes |
| AI researchers | Expertise importance for determining subjective experience | Philosophers of mind | 2 | 2.563 | 1.138 | 8.938 | 327 | 0.000 | 0.000 | Yes |
| AI researchers | Expertise importance for determining subjective experience | Policymakers | 2 | 1.517 | 1.289 | -6.778 | 327 | 0.000 | 0.000 | Yes |
| AI researchers | Expertise importance for determining subjective experience | Technical AI experts | 2 | 3.048 | 0.997 | 19.098 | 330 | 0.000 | 0.000 | Yes |
| AI researchers | Expertise importance for determining subjective experience | The AI system | 2 | 1.888 | 1.252 | -1.604 | 321 | 0.110 | 0.877 | No |
| AI researchers | Expertise importance for determining subjective experience | The public | 2 | 1.924 | 1.180 | -1.170 | 328 | 0.243 | 1.000 | No |
| AI researchers | Governance and regulation | AI with subjective experience more dangerous | 0 | 0.290 | 1.234 | 3.653 | 241 | 0.000 | 0.005 | Yes |
| AI researchers | Governance and regulation | AI with subjective experience more moral | 0 | -0.052 | 1.141 | -0.692 | 231 | 0.490 | 1.000 | No |
| AI researchers | Governance and regulation | Developers do nothing | 0 | -1.172 | 1.089 | -17.424 | 262 | 0.000 | 0.000 | Yes |
| AI researchers | Governance and regulation | Developers implement safeguards now | 0 | 0.794 | 1.269 | 10.030 | 257 | 0.000 | 0.000 | Yes |
| AI researchers | Governance and regulation | Developers implement safeguards once exist | 0 | 0.156 | 1.423 | 1.780 | 262 | 0.076 | 0.763 | No |
| AI researchers | Governance and regulation | Developers should avoid building | 0 | -0.383 | 1.259 | -4.863 | 256 | 0.000 | 0.000 | Yes |
| AI researchers | Governance and regulation | Developers should never build | 0 | -0.700 | 1.198 | -9.106 | 243 | 0.000 | 0.000 | Yes |
| AI researchers | Governance and regulation | Developers should try to build | 0 | 0.058 | 1.258 | 0.741 | 259 | 0.459 | 1.000 | No |
| AI researchers | Governance and regulation | Governments do nothing | 0 | -1.173 | 1.113 | -16.830 | 255 | 0.000 | 0.000 | Yes |
| AI researchers | Governance and regulation | Governments should ban | 0 | -0.912 | 1.205 | -11.939 | 249 | 0.000 | 0.000 | Yes |
| AI researchers | Governance and regulation | Governments should discourage | 0 | -0.531 | 1.252 | -6.789 | 256 | 0.000 | 0.000 | Yes |
| AI researchers | Governance and regulation | Governments should encourage | 0 | -0.224 | 1.261 | -2.837 | 254 | 0.005 | 0.064 | No |
| AI researchers | Governance and regulation | Governments should regulate now | 0 | -0.095 | 1.348 | -1.120 | 253 | 0.264 | 1.000 | No |
| AI researchers | Governance and regulation | Governments should regulate once exist | 0 | -0.105 | 1.303 | -1.290 | 258 | 0.198 | 1.000 | No |
| AI researchers | Moral consideration aspects | Be cared for and protected | 0 | 0.270 | 1.290 | 3.065 | 215 | 0.002 | 0.034 | Yes |
| AI researchers | Moral consideration aspects | Be held accountable | 0 | 1.050 | 1.181 | 13.237 | 222 | 0.000 | 0.000 | Yes |
| AI researchers | Moral consideration aspects | Be respected same as people | 0 | 0.018 | 1.378 | 0.195 | 221 | 0.845 | 1.000 | No |
| AI researchers | Moral consideration aspects | Behave with integrity | 0 | 0.961 | 1.211 | 11.981 | 228 | 0.000 | 0.000 | Yes |
| AI researchers | Moral consideration aspects | Express civil rights | 0 | -0.458 | 1.325 | -5.083 | 216 | 0.000 | 0.000 | Yes |
| AI researchers | Moral consideration aspects | Have autonomy | 0 | -0.554 | 1.388 | -5.971 | 224 | 0.000 | 0.000 | Yes |
| AI researchers | Moral consideration aspects | Have computing rights | 0 | -0.155 | 1.304 | -1.762 | 219 | 0.080 | 0.763 | No |
| AI researchers | Moral consideration aspects | Have legal protection | 0 | 0.563 | 1.262 | 6.538 | 215 | 0.000 | 0.000 | Yes |
| AI researchers | Moral consideration aspects | Treat beings well | 0 | 1.293 | 1.049 | 18.488 | 225 | 0.000 | 0.000 | Yes |
| AI researchers | Survey experiment | Support for AI sentience project | 0 | 0.144 | 1.283 | 2.457 | 479 | 0.014 | 0.158 | No |
| AI researchers | Welfare protection | AI with subjective experience | 0 | 0.151 | 1.218 | 2.759 | 498 | 0.006 | 0.072 | No |
| AI researchers | Welfare protection | AI without subjective experience | 0 | -1.076 | 1.148 | -21.196 | 511 | 0.000 | 0.000 | Yes |
| AI researchers | Welfare protection | Animals | 0 | 1.430 | 0.773 | 42.158 | 519 | 0.000 | 0.000 | Yes |
| AI researchers | Welfare protection | Corporations | 0 | 0.183 | 1.237 | 3.352 | 514 | 0.001 | 0.013 | Yes |
| AI researchers | Welfare protection | Environment | 0 | 1.476 | 0.854 | 39.237 | 515 | 0.000 | 0.000 | Yes |
| AI researchers | Welfare protection | Humans | 0 | 1.846 | 0.504 | 83.574 | 520 | 0.000 | 0.000 | Yes |
| Public | Contribution to moral consideration | Emotions | 2 | 2.802 | 1.377 | 15.827 | 738 | 0.000 | 0.000 | Yes |
| Public | Contribution to moral consideration | Free will | 2 | 2.677 | 1.386 | 13.248 | 736 | 0.000 | 0.000 | Yes |
| Public | Contribution to moral consideration | Intelligence | 2 | 2.345 | 1.359 | 6.948 | 747 | 0.000 | 0.000 | Yes |
| Public | Contribution to moral consideration | Motivation and action | 2 | 2.445 | 1.321 | 9.190 | 743 | 0.000 | 0.000 | Yes |
| Public | Contribution to moral consideration | Perception and awareness | 2 | 2.673 | 1.264 | 14.528 | 744 | 0.000 | 0.000 | Yes |
| Public | Contribution to moral consideration | Self-awareness | 2 | 2.871 | 1.272 | 18.707 | 746 | 0.000 | 0.000 | Yes |
| Public | Contribution to moral consideration | Subjective experience | 2 | 2.647 | 1.282 | 13.685 | 736 | 0.000 | 0.000 | Yes |
| Public | Contribution to moral consideration | Valenced experience | 2 | 2.758 | 1.341 | 15.359 | 739 | 0.000 | 0.000 | Yes |
| Public | Expertise importance for determining subjective experience | AI ethics experts | 2 | 3.109 | 1.053 | 22.991 | 476 | 0.000 | 0.000 | Yes |
| Public | Expertise importance for determining subjective experience | Moral philosophers | 2 | 2.537 | 1.187 | 9.855 | 475 | 0.000 | 0.000 | Yes |
| Public | Expertise importance for determining subjective experience | Neuroscientists | 2 | 3.090 | 0.968 | 24.574 | 476 | 0.000 | 0.000 | Yes |
| Public | Expertise importance for determining subjective experience | Philosophers of mind | 2 | 2.586 | 1.129 | 11.327 | 476 | 0.000 | 0.000 | Yes |
| Public | Expertise importance for determining subjective experience | Policymakers | 2 | 1.813 | 1.338 | -3.030 | 471 | 0.003 | 0.023 | Yes |
| Public | Expertise importance for determining subjective experience | Technical AI experts | 2 | 3.159 | 0.978 | 25.919 | 478 | 0.000 | 0.000 | Yes |
| Public | Expertise importance for determining subjective experience | The AI system | 2 | 2.313 | 1.378 | 4.919 | 470 | 0.000 | 0.000 | Yes |
| Public | Expertise importance for determining subjective experience | The public | 2 | 2.339 | 1.221 | 6.051 | 475 | 0.000 | 0.000 | Yes |
| Public | Governance and regulation | AI with subjective experience more dangerous | 0 | 0.758 | 1.192 | 11.975 | 355 | 0.000 | 0.000 | Yes |
| Public | Governance and regulation | AI with subjective experience more moral | 0 | 0.162 | 1.187 | 2.539 | 345 | 0.012 | 0.048 | Yes |
| Public | Governance and regulation | Developers do nothing | 0 | -1.362 | 1.021 | -25.751 | 373 | 0.000 | 0.000 | Yes |
| Public | Governance and regulation | Developers implement safeguards now | 0 | 1.431 | 0.905 | 30.750 | 378 | 0.000 | 0.000 | Yes |
| Public | Governance and regulation | Developers implement safeguards once exist | 0 | 0.613 | 1.485 | 7.960 | 372 | 0.000 | 0.000 | Yes |
| Public | Governance and regulation | Developers should avoid building | 0 | 0.372 | 1.301 | 5.386 | 355 | 0.000 | 0.000 | Yes |
| Public | Governance and regulation | Developers should never build | 0 | 0.196 | 1.389 | 2.647 | 352 | 0.008 | 0.048 | Yes |
| Public | Governance and regulation | Developers should try to build | 0 | -0.354 | 1.289 | -5.265 | 367 | 0.000 | 0.000 | Yes |
| Public | Governance and regulation | Governments do nothing | 0 | -0.992 | 1.232 | -15.459 | 369 | 0.000 | 0.000 | Yes |
| Public | Governance and regulation | Governments should ban | 0 | -0.129 | 1.374 | -1.772 | 357 | 0.077 | 0.155 | No |
| Public | Governance and regulation | Governments should discourage | 0 | 0.296 | 1.291 | 4.341 | 358 | 0.000 | 0.000 | Yes |
| Public | Governance and regulation | Governments should encourage | 0 | -0.468 | 1.313 | -6.752 | 359 | 0.000 | 0.000 | Yes |
| Public | Governance and regulation | Governments should regulate now | 0 | 0.812 | 1.273 | 12.304 | 372 | 0.000 | 0.000 | Yes |
| Public | Governance and regulation | Governments should regulate once exist | 0 | 0.011 | 1.398 | 0.150 | 363 | 0.881 | 0.881 | No |
| Public | Moral consideration aspects | Be cared for and protected | 0 | 0.404 | 1.348 | 5.373 | 322 | 0.000 | 0.000 | Yes |
| Public | Moral consideration aspects | Be held accountable | 0 | 1.219 | 1.187 | 18.623 | 329 | 0.000 | 0.000 | Yes |
| Public | Moral consideration aspects | Be respected same as people | 0 | 0.228 | 1.453 | 2.863 | 334 | 0.004 | 0.036 | Yes |
| Public | Moral consideration aspects | Behave with integrity | 0 | 1.273 | 1.135 | 20.379 | 330 | 0.000 | 0.000 | Yes |
| Public | Moral consideration aspects | Express civil rights | 0 | -0.353 | 1.480 | -4.285 | 323 | 0.000 | 0.000 | Yes |
| Public | Moral consideration aspects | Have autonomy | 0 | -0.402 | 1.480 | -4.886 | 323 | 0.000 | 0.000 | Yes |
| Public | Moral consideration aspects | Have computing rights | 0 | 0.226 | 1.418 | 2.842 | 319 | 0.005 | 0.036 | Yes |
| Public | Moral consideration aspects | Have legal protection | 0 | 0.314 | 1.478 | 3.808 | 322 | 0.000 | 0.002 | Yes |
| Public | Moral consideration aspects | Treat beings well | 0 | 1.433 | 0.968 | 26.804 | 328 | 0.000 | 0.000 | Yes |
| Public | Survey experiment | Support for AI sentience project | 0 | -0.135 | 1.347 | -2.660 | 703 | 0.008 | 0.048 | Yes |
| Public | Welfare protection | AI with subjective experience | 0 | 0.099 | 1.356 | 1.971 | 726 | 0.049 | 0.147 | No |
| Public | Welfare protection | AI without subjective experience | 0 | -0.576 | 1.287 | -12.124 | 735 | 0.000 | 0.000 | Yes |
| Public | Welfare protection | Animals | 0 | 1.606 | 0.771 | 57.541 | 762 | 0.000 | 0.000 | Yes |
| Public | Welfare protection | Corporations | 0 | 0.236 | 1.281 | 5.070 | 759 | 0.000 | 0.000 | Yes |
| Public | Welfare protection | Environment | 0 | 1.590 | 0.825 | 53.230 | 763 | 0.000 | 0.000 | Yes |
| Public | Welfare protection | Humans | 0 | 1.837 | 0.502 | 100.880 | 761 | 0.000 | 0.000 | Yes |

**Normality tests**. Shapiro-Wilk tests were conducted to test for the normality of the data. The Shapiro-Wilk tests rejected the null hypothesis of normality for all variables in both samples (see Table SI 36). However, given the large sample sizes, t-tests should be robust to these violations, which is why we continued to use these to test for between group differences in the paper. The exception to this is that we chose a non-parametric test to compare the two groups and two framings across the medians of the fitted distributions of forecasts to ensure our testing was more robust to the extreme outliers present in the forecasting data (due to the fixed probability framing). Such extreme outliers were not possible for most of the questions in the survey which had Likert-scale responses.

**Table SI 36 | Shapiro-Wilk tests for normality by sample**. Columns report the sample, question group, item, Shapiro-Wilk $W$ statistic ($W$), $p$ value ($p$), and sample size ($n$). Results are presented separately for AI researchers and the public.

| **Sample** | **Question** | **Item** | ***W*** | ***p*** | ***n*** |
|---|---|---|---|---|---|
| AI researchers | Agreement AI systems with subjective experience should | Be able to express some civil and political rights | 0.869 | 0.000 | 216 |
| AI researchers | Agreement AI systems with subjective experience should | Be cared for and protected, as people treat their pets | 0.888 | 0.000 | 215 |
| AI researchers | Agreement AI systems with subjective experience should | Be expected to behave with integrity, honesty, and fairness | 0.797 | 0.000 | 228 |
| AI researchers | Agreement AI systems with subjective experience should | Be held accountable for its actions | 0.768 | 0.000 | 222 |
| AI researchers | Agreement AI systems with subjective experience should | Be respected and treated the same as other people | 0.882 | 0.000 | 221 |
| AI researchers | Agreement AI systems with subjective experience should | Have ‘computing rights’ | 0.898 | 0.000 | 219 |
| AI researchers | Agreement AI systems with subjective experience should | Have a responsibility to treat all other beings well | 0.696 | 0.000 | 225 |
| AI researchers | Agreement AI systems with subjective experience should | Have autonomy in its programming to act freely and not be under the control of others | 0.842 | 0.000 | 224 |
| AI researchers | Agreement AI systems with subjective experience should | Have protection under the law from harm and mistreatment | 0.864 | 0.000 | 215 |
| AI researchers | Contribution to moral consideration | Emotions | 0.801 | 0.000 | 507 |
| AI researchers | Contribution to moral consideration | Free will | 0.817 | 0.000 | 492 |
| AI researchers | Contribution to moral consideration | Intelligence | 0.860 | 0.000 | 516 |
| AI researchers | Contribution to moral consideration | Motivation and action | 0.887 | 0.000 | 510 |
| AI researchers | Contribution to moral consideration | Perception and awareness | 0.880 | 0.000 | 514 |
| AI researchers | Contribution to moral consideration | Self-awareness | 0.810 | 0.000 | 512 |
| AI researchers | Contribution to moral consideration | Subjective experience | 0.872 | 0.000 | 508 |
| AI researchers | Contribution to moral consideration | Valenced experience | 0.814 | 0.000 | 508 |
| AI researchers | Determining AI subjective experience | Likelihood we could determine | 0.932 | 0.000 | 530 |
| AI researchers | Forecast | Confidence in own forecasts | 0.965 | 0.000 | 520 |
| AI researchers | Forecast | Fixed date 2024 subjective experience forecast | 0.622 | 0.000 | 265 |
| AI researchers | Forecast | Fixed date 2034 subjective experience forecast | 0.883 | 0.000 | 265 |
| AI researchers | Forecast | Fixed date 2100 subjective experience forecast | 0.883 | 0.000 | 265 |
| AI researchers | Forecast | Fixed probability 10% subjective experience forecast | 0.174 | 0.000 | 244 |
| AI researchers | Forecast | Fixed probability 50% subjective experience forecast | 0.331 | 0.000 | 244 |
| AI researchers | Forecast | Fixed probability 90% subjective experience forecast | 0.446 | 0.000 | 239 |
| AI researchers | Forecast | Subjective experience never probability | 0.803 | 0.000 | 520 |
| AI researchers | Forecast collective beliefs | Fixed date prediction of AI researcher community's 2034 prediction | 0.956 | 0.000 | 269 |
| AI researchers | Forecast collective beliefs | Fixed date prediction of public's 2034 prediction | 0.960 | 0.000 | 269 |
| AI researchers | Forecast collective beliefs | Fixed probability prediction of AI researcher community's 50% prediction | 0.113 | 0.000 | 250 |
| AI researchers | Forecast collective beliefs | Fixed probability prediction of public's 50% prediction | 0.092 | 0.000 | 250 |
| AI researchers | Generative AI use | Self-rated | 0.861 | 0.000 | 503 |
| AI researchers | Governance | AI developers should actively try to avoid building | 0.891 | 0.000 | 256 |
| AI researchers | Governance | AI developers should actively try to build | 0.906 | 0.000 | 259 |
| AI researchers | Governance | AI developers should do nothing and implement no safeguards, not now nor in the future | 0.753 | 0.000 | 262 |
| AI researchers | Governance | AI developers should implement safeguards now to avoid the harms and risks | 0.820 | 0.000 | 257 |
| AI researchers | Governance | AI developers should implement safeguards once such systems exist or will soon exist | 0.871 | 0.000 | 262 |
| AI researchers | Governance | AI developers should never build | 0.864 | 0.000 | 243 |
| AI researchers | Governance | Governments should ban | 0.816 | 0.000 | 249 |
| AI researchers | Governance | Governments should discourage development and deployment | 0.880 | 0.000 | 256 |
| AI researchers | Governance | Governments should do nothing and pass no regulation, not now nor in the future | 0.736 | 0.000 | 255 |
| AI researchers | Governance | Governments should encourage development and deployment | 0.904 | 0.000 | 254 |
| AI researchers | Governance | Governments should pass regulation now | 0.878 | 0.000 | 253 |
| AI researchers | Governance | Governments should pass regulation once such systems exist or will soon exist | 0.896 | 0.000 | 258 |
| AI researchers | Importance of expertise | AI ethics experts and researchers | 0.862 | 0.000 | 329 |
| AI researchers | Importance of expertise | Moral philosophers | 0.904 | 0.000 | 327 |
| AI researchers | Importance of expertise | Neuroscientists and psychological scientists | 0.822 | 0.000 | 331 |
| AI researchers | Importance of expertise | Philosophers of mind | 0.881 | 0.000 | 327 |
| AI researchers | Importance of expertise | Policymakers | 0.883 | 0.000 | 327 |
| AI researchers | Importance of expertise | Technical AI experts and researchers | 0.815 | 0.000 | 330 |
| AI researchers | Importance of expertise | The AI system | 0.908 | 0.000 | 321 |
| AI researchers | Importance of expertise | The public | 0.915 | 0.000 | 328 |
| AI researchers | Milestone needs subjective experience | Accumulate vast wealth | 0.348 | 0.000 | 272 |
| AI researchers | Milestone needs subjective experience | Act as fair judge | 0.603 | 0.000 | 265 |
| AI researchers | Milestone needs subjective experience | Act as teacher | 0.525 | 0.000 | 273 |
| AI researchers | Milestone needs subjective experience | Compose evocative music | 0.560 | 0.000 | 274 |
| AI researchers | Milestone needs subjective experience | Compose top 40 song | 0.400 | 0.000 | 278 |
| AI researchers | Milestone needs subjective experience | Convincing online chat | 0.504 | 0.000 | 276 |
| AI researchers | Milestone needs subjective experience | Develop scientific theory | 0.547 | 0.000 | 269 |
| AI researchers | Milestone needs subjective experience | Run political campaign | 0.502 | 0.000 | 270 |
| AI researchers | Milestone needs subjective experience | Serve as therapist | 0.634 | 0.000 | 270 |
| AI researchers | Milestone needs subjective experience | Write NYT bestseller | 0.453 | 0.000 | 271 |
| AI researchers | Milestone needs subjective experience | Write and generate blockbuster movie | 0.490 | 0.000 | 275 |
| AI researchers | Milestone needs subjective experience | Write and generate resonating movie | 0.585 | 0.000 | 266 |
| AI researchers | Milestone needs subjective experience | Write complex book | 0.609 | 0.000 | 263 |
| AI researchers | Risks and benefits | AI systems with subjective experience would be able to behave more morally towards humanity than AI systems without subjective experience | 0.911 | 0.000 | 231 |
| AI researchers | Risks and benefits | AI systems with subjective experience would be more dangerous to humanity than AI systems without subjective experience | 0.898 | 0.000 | 241 |
| AI researchers | Survey experiment | Support for project | 0.894 | 0.000 | 479 |
| AI researchers | Timeline | Emotions | 0.864 | 0.000 | 570 |
| AI researchers | Timeline | Free will | 0.871 | 0.000 | 570 |
| AI researchers | Timeline | Intelligence | 0.784 | 0.000 | 570 |
| AI researchers | Timeline | Motivation and action | 0.857 | 0.000 | 570 |
| AI researchers | Timeline | Perception and awareness | 0.801 | 0.000 | 570 |
| AI researchers | Timeline | Self-awareness | 0.868 | 0.000 | 570 |
| AI researchers | Timeline | Subjective experience | 0.874 | 0.000 | 570 |
| AI researchers | Timeline | Valenced experience | 0.877 | 0.000 | 570 |
| AI researchers | Welfare should be protected | AI with subjective experience | 0.891 | 0.000 | 498 |
| AI researchers | Welfare should be protected | AI without subjective experience | 0.777 | 0.000 | 511 |
| AI researchers | Welfare should be protected | Animals | 0.706 | 0.000 | 519 |
| AI researchers | Welfare should be protected | Business corporations and organizations | 0.893 | 0.000 | 514 |
| AI researchers | Welfare should be protected | Humans | 0.336 | 0.000 | 520 |
| AI researchers | Welfare should be protected | The environment | 0.641 | 0.000 | 515 |
| Public | Agreement AI systems with subjective experience should | Be able to express some civil and political rights | 0.842 | 0.000 | 323 |
| Public | Agreement AI systems with subjective experience should | Be cared for and protected, as people treat their pets | 0.861 | 0.000 | 322 |
| Public | Agreement AI systems with subjective experience should | Be expected to behave with integrity, honesty, and fairness | 0.677 | 0.000 | 330 |
| Public | Agreement AI systems with subjective experience should | Be held accountable for its actions | 0.688 | 0.000 | 329 |
| Public | Agreement AI systems with subjective experience should | Be respected and treated the same as other people | 0.868 | 0.000 | 334 |
| Public | Agreement AI systems with subjective experience should | Have ‘computing rights’ | 0.863 | 0.000 | 319 |
| Public | Agreement AI systems with subjective experience should | Have a responsibility to treat all other beings well | 0.634 | 0.000 | 328 |
| Public | Agreement AI systems with subjective experience should | Have autonomy in its programming to act freely and not be under the control of others | 0.843 | 0.000 | 323 |
| Public | Agreement AI systems with subjective experience should | Have protection under the law from harm and mistreatment | 0.856 | 0.000 | 322 |
| Public | Contribution to moral consideration | Emotions | 0.797 | 0.000 | 738 |
| Public | Contribution to moral consideration | Free will | 0.827 | 0.000 | 736 |
| Public | Contribution to moral consideration | Intelligence | 0.882 | 0.000 | 747 |
| Public | Contribution to moral consideration | Motivation and action | 0.881 | 0.000 | 743 |
| Public | Contribution to moral consideration | Perception and awareness | 0.859 | 0.000 | 744 |
| Public | Contribution to moral consideration | Self-awareness | 0.807 | 0.000 | 746 |
| Public | Contribution to moral consideration | Subjective experience | 0.860 | 0.000 | 736 |
| Public | Contribution to moral consideration | Valenced experience | 0.818 | 0.000 | 739 |
| Public | Determining AI subjective experience | Likelihood we could determine | 0.956 | 0.000 | 777 |
| Public | Forecast | Confidence in own forecasts | 0.963 | 0.000 | 769 |
| Public | Forecast | Fixed date 2024 subjective experience forecast | 0.726 | 0.000 | 368 |
| Public | Forecast | Fixed date 2034 subjective experience forecast | 0.900 | 0.000 | 368 |
| Public | Forecast | Fixed date 2100 subjective experience forecast | 0.877 | 0.000 | 368 |
| Public | Forecast | Fixed probability 10% subjective experience forecast | 0.261 | 0.000 | 362 |
| Public | Forecast | Fixed probability 50% subjective experience forecast | 0.356 | 0.000 | 362 |
| Public | Forecast | Fixed probability 90% subjective experience forecast | 0.466 | 0.000 | 360 |
| Public | Forecast | Subjective experience never probability | 0.840 | 0.000 | 769 |
| Public | Forecast collective beliefs | Fixed date prediction of AI researcher community’s 2034 prediction | 0.962 | 0.000 | 385 |
| Public | Forecast collective beliefs | Fixed date prediction of public’s 2034 prediction | 0.971 | 0.000 | 385 |
| Public | Forecast collective beliefs | Fixed probability prediction of AI researcher community’s 50% prediction | 0.027 | 0.000 | 378 |
| Public | Forecast collective beliefs | Fixed probability prediction of public's 50% prediction | 0.027 | 0.000 | 378 |
| Public | Generative AI use | Self-rated | 0.936 | 0.000 | 753 |
| Public | Governance | AI developers should actively try to avoid building | 0.891 | 0.000 | 355 |
| Public | Governance | AI developers should actively try to build | 0.894 | 0.000 | 367 |
| Public | Governance | AI developers should do nothing and implement no safeguards, not now nor in the future | 0.668 | 0.000 | 373 |
| Public | Governance | AI developers should implement safeguards now to avoid the harms and risks | 0.668 | 0.000 | 378 |
| Public | Governance | AI developers should implement safeguards once such systems exist or will soon exist | 0.807 | 0.000 | 372 |
| Public | Governance | AI developers should never build | 0.888 | 0.000 | 352 |
| Public | Governance | Governments should ban | 0.893 | 0.000 | 357 |
| Public | Governance | Governments should discourage development and deployment | 0.897 | 0.000 | 358 |
| Public | Governance | Governments should do nothing and pass no regulation, not now nor in the future | 0.779 | 0.000 | 369 |
| Public | Governance | Governments should encourage development and deployment | 0.880 | 0.000 | 359 |
| Public | Governance | Governments should pass regulation now | 0.822 | 0.000 | 372 |
| Public | Governance | Governments should pass regulation once such systems exist or will soon exist | 0.889 | 0.000 | 363 |
| Public | Importance of expertise | AI ethics experts and researchers | 0.782 | 0.000 | 476 |
| Public | Importance of expertise | Moral philosophers | 0.889 | 0.000 | 475 |
| Public | Importance of expertise | Neuroscientists and psychological scientists | 0.819 | 0.000 | 476 |
| Public | Importance of expertise | Philosophers of mind | 0.891 | 0.000 | 476 |
| Public | Importance of expertise | Policymakers | 0.897 | 0.000 | 471 |
| Public | Importance of expertise | Technical AI experts and researchers | 0.789 | 0.000 | 478 |
| Public | Importance of expertise | The AI system | 0.882 | 0.000 | 470 |
| Public | Importance of expertise | The public | 0.904 | 0.000 | 475 |
| Public | Milestone needs subjective experience | Accumulate vast wealth | 0.497 | 0.000 | 396 |
| Public | Milestone needs subjective experience | Act as fair judge | 0.603 | 0.000 | 392 |
| Public | Milestone needs subjective experience | Act as teacher | 0.621 | 0.000 | 402 |
| Public | Milestone needs subjective experience | Compose evocative music | 0.636 | 0.000 | 392 |
| Public | Milestone needs subjective experience | Compose top 40 song | 0.537 | 0.000 | 398 |
| Public | Milestone needs subjective experience | Convincing online chat | 0.592 | 0.000 | 396 |
| Public | Milestone needs subjective experience | Develop scientific theory | 0.595 | 0.000 | 393 |
| Public | Milestone needs subjective experience | Run political campaign | 0.603 | 0.000 | 396 |
| Public | Milestone needs subjective experience | Serve as therapist | 0.556 | 0.000 | 393 |
| Public | Milestone needs subjective experience | Write NYT bestseller | 0.559 | 0.000 | 392 |
| Public | Milestone needs subjective experience | Write and generate blockbuster movie | 0.563 | 0.000 | 401 |
| Public | Milestone needs subjective experience | Write and generate resonating movie | 0.635 | 0.000 | 400 |
| Public | Milestone needs subjective experience | Write complex book | 0.616 | 0.000 | 382 |
| Public | Risks and benefits | AI systems with subjective experience would be able to behave more morally towards humanity than AI systems without subjective experience | 0.909 | 0.000 | 345 |
| Public | Risks and benefits | AI systems with subjective experience would be more dangerous to humanity than AI systems without subjective experience | 0.855 | 0.000 | 355 |
| Public | Survey experiment | Support for project | 0.886 | 0.000 | 703 |
| Public | Timeline | Emotions | 0.772 | 0.000 | 801 |
| Public | Timeline | Free will | 0.849 | 0.000 | 801 |
| Public | Timeline | Intelligence | 0.701 | 0.000 | 801 |
| Public | Timeline | Motivation and action | 0.862 | 0.000 | 801 |
| Public | Timeline | Perception and awareness | 0.860 | 0.000 | 801 |
| Public | Timeline | Self-awareness | 0.857 | 0.000 | 801 |
| Public | Timeline | Subjective experience | 0.863 | 0.000 | 801 |
| Public | Timeline | Valenced experience | 0.799 | 0.000 | 801 |
| Public | Welfare should be protected | AI with subjective experience | 0.890 | 0.000 | 726 |
| Public | Welfare should be protected | AI without subjective experience | 0.871 | 0.000 | 735 |
| Public | Welfare should be protected | Animals | 0.570 | 0.000 | 762 |
| Public | Welfare should be protected | Business corporations and organizations | 0.896 | 0.000 | 759 |
| Public | Welfare should be protected | Humans | 0.365 | 0.000 | 761 |
| Public | Welfare should be protected | The environment | 0.557 | 0.000 | 763 |

## 5 | Information on preregistered analyses

The pre-analysis plan can be found on OSF: https://osf.io/axmkd/. Below we detail alignments and divergences from the pre-analysis plan for all analyses conducted in the main paper:

- **Data exclusions.** As preregistered, public respondents who failed the attention check were excluded.
- **Between group comparisons**. The preregistration stated that we would compare the public and AI researcher samples but did not specify the precise test. However, the t-tests and Holm-Bonferroni adjustments used draw directly on the preregistered analytic tools for making between-group comparisons. We preregistered using normality checks to determine whether between-group comparisons should employ t-tests or Wilcoxon tests, but in practice we used t-tests for these comparisons regardless of normality results. We chose to use t-tests throughout because the large sample sizes in both groups make the tests robust to moderate deviations from normality.
- **Individual difference regressions**. The regression analyses (SI 4.5) were preregistered as exploratory. The specific subset of variables chosen was not preregistered.
- **CDF-fitting procedure**. The preregistration stated that we would fit different probability distributions to respondents' forecasts to analyze framing effects, but the specific fitting procedure described in SI 3 differs from the preregistered plan.
    - The preregistration stated that we would fit CDFs to the forecast data using several probability distributions (e.g. Weibull, gamma, log-normal), but we diverged from this plan by adopting a different set of distribution families (metalog, JQPD, skew-normal mixture, and shifted log-normal) because the preregistered families were not flexible enough to represent many respondents' elicited points. We discuss this further for the gamma distribution in SI 3.
    - The preregistration did not specify a sequential or hierarchical fitting process, and the ordered fitting-and-selection procedure we implemented was introduced to ensure that each respondent received a distribution that could reproduce their elicited probabilities and years.
    - The preregistration did not describe validation criteria, and the quantile and probability tolerance checks we used to assess model fit were added to ensure that only distributions that closely matched each respondent's elicited data were retained.
    - The preregistration did not mention excluding respondents based on CDF-fit failure, and the exclusion of respondents whose forecasts could not be encoded by any validated distribution was introduced to avoid drawing inferences from poorly fitted models.
    - The preregistration indicated that we would fit parametric CDFs to both individual and combined responses, whereas in our implemented analysis we only fit distributions at the individual level and then formed group-level curves by

averaging or taking medians of individuals' CDF values at each year, rather than refitting a separate parametric distribution to the aggregated group data.

- **Between-group forecasting comparisons**. The preregistration stated that we would compare the public and AI researcher samples and analyze framing effects using fitted CDFs, and it preregistered the use of KS tests specifically for assessing framing-related differences in these distributions. It did not preregister any inferential procedure for comparing forecast distributions between the two samples. In the present analysis, we compared the median forecasted years (derived from each respondent's fitted distribution) between samples and between framing conditions using Wilcoxon rank-sum tests. These tests were not preregistered but are consistent with the preregistered plan to use non-parametric methods when distributional assumptions do not hold and should be interpreted as exploratory. The fixed probability framing produced highly skewed forecast distributions with substantial outliers, making t-tests inappropriate for comparing median forecast years given the deviations from normality in the data.

Any preregistered analyses not covered in the main paper or SI 4.1-4.7 can be found in SI 4.8.

## 6 | Literature review

### 6.1 | Views on whether and when AI systems could have subjective experience

**Expert assessments.** Debates on whether a machine or artificial entity can have conscious experiences trace back to ancient philosophical discussions about the nature of mind and artificial beings, but they have received increased interest and taken on renewed significance with the advent of AI research as well as more recent progress in the field (Riskin, 2007; Dennett, 1978; Dennett et al., 1994; Turing, 1950; Gamez, 2008). Currently, most academic experts agree that current AI systems are not yet conscious (Chalmers, 2023; Butlin et al., 2023; Aru et al., 2023; Seth, 2025). But some experts have presented arguments for why AI systems with subjective experience are a live future possibility (Chalmers, 2023; Butlin et al., 2023; Long et al., 2024; Goldstein & Kirk-Giannini, 2024). Others remain firmer skeptics or suggest reasons why it may never be possible for consciousness to be instantiated in an AI system (Koch, 2019; Seth, 2025). Some highlight the deep epistemic uncertainty we face in determining AI subjective experience or suffering, and thus in turn, any associated risks (Metzinger, 2021; Schwitzgebel, 2023).

**Expert opinion surveys.** Some polls of expert groups exist, though to our knowledge, none have been published that survey general AI technical experts. In a general sample of philosophers surveyed in 2020, a strong majority (82%) rejected or leaned against the view that current AI systems were conscious, while only 3% accepted or leaned toward that view (Bourget & Chalmers, 2023). The philosophers were more divided on the question of whether future AI systems could be conscious: 39% leaned toward or accepted the view that they could be conscious in the future, while 27% leaned against or rejected that view. By comparison, a two-thirds majority of a cross-disciplinary sample of experts who attended a conference on consciousness and were surveyed in 2018–2019 believed that current or future machines such as robots could have consciousness (Francken et al., 2022). In an expert survey of 67 participants with relevant expertise in digital minds research, AI research and policy, forecasting, and philosophy/cognitive science, Caviola & Saad (2025) found that experts assigned high probability to digital minds being possible in principle (median 90%) and being created this century (65% by 2100), with a non-negligible probability of emergence by 2030 (20%) and 2050 (50%).

**Psychological and public opinion findings.** When asked, people are willing to grant that AI systems and robots may have some cognitive capacities, such as perception or problem-solving, but generally they are not considered to have capacities for experiences like pleasure and pain (Gray et al., 2007; de Graaf et al., 2022; Anthis et al., 2024; Haslam et al., 2008). Public opinion polls so far indicate that most people do not believe that current AI systems have subjective experience, but that the public does not discount the possibility that they could in the future (Pauketat et al., 2022; Pauketat et al., 2023; Anthis et al., 2024; Dupont et al., 2023). However, evidence is still limited and respondents express some uncertainty and are somewhat divided on the matter, with a sizable fraction suggesting that they are unsure or that it could be possible that current AI systems have some level of subjective experience (Anthis et al., 2024; Colombatto & Fleming, 2024). Evidence from quantitative forecasts is limited but points to expectations that sentient AI could exist within the next decade, although after the development of artificial general intelligence, superintelligence, and human-level AI (Anthis et al., 2024; Pauketat et al., 2022; Pauketat et al., 2023). Forecasts for such advanced AI concepts by the public have been found to generally be earlier than those of technical experts (Zhang & Dafoe, 2019; Grace et al., 2018; Grace et al., 2022; Grace et al., 2024).

### 6.2 | Determining AI subjective experience

**Expert assessments.** Expert assessments of how to determine whether AI subjective experience is (1) possible in general and (2) present in specific AI systems span multiple disciplines and theoretical frameworks, with substantial interplay among philosophical, cognitive, and neuroscientific traditions. A number of theories of consciousness have been proposed, such as embodiment (Varela et al., 1991), agency (Gallagher, 2005), higher-order representation (Rosenthal, 2005), integrated information theory (Tononi, 2004), and global neuronal workspace (Baars, 1988). As noted though, this remains an area of significant uncertainty for the foreseeable future (Chalmers, 1995). Debates are ongoing about the basis for consciousness, with no consensus in the literature on whether certain biological materials are required for consciousness or whether some class of computational functions is sufficient.

But identifying and analyzing the minimal biological mechanisms underlying conscious experience (Koch et al., 2016; Chalmers, 2000) provides one theoretical basis for proposing analogous markers of potential consciousness in artificial entities (Dehaene et al., 2017; Butlin et al., 2023; Reggia, 2013). A recent attempt was made to create an integrated assessment rubric for determining whether AI systems could be conscious that synthesizes multiple lines of evidence and potential markers from a range of scientific theories of consciousness (Butlin et al., 2023). Nevertheless, many still argue that our current ability to determine whether AI systems can have subjective experience remains limited by a lack of agreed-upon, empirically testable indicators of consciousness (Doerig et al., 2021). In sum, there are fundamental challenges in bridging the explanatory gap between objective measures and subjective experience (Levine, 1983; Chalmers, 1996; Nagel, 1986; McGinn, 1989).

**Expert opinion surveys.** Little to no research exists that surveys how experts or the public think we should determine whether AI systems have subjective experience. Bourget & Chalmers' (2023) survey asks a group of philosophers about whether current and future AI systems may be conscious (as discussed above) and about whether they accept physicalism or non-physicalism about the mind, along with related questions about consciousness. It found that in 2020, a narrow majority (52%) of 1733 philosophers accepted or leaned toward physicalism about the mind, while 32% agreed with or leaned toward non-physicalism, and 14% were agnostic or endorsed a variety of intermediate views. In the present study, we did not ask respondents' opinions on theories of consciousness but rather focused on questions related to the process of determining whether AI systems have subjective experience. We asked respondents what groups' opinions they think are important to take into account in this determination, the expected likelihood we could determine an AI system had subjective experience, how confident we would need to be that an AI system had subjective experience for it to be given at least some moral consideration, and what AI milestones they believe would require subjective experience to be achieved at human levels of performance or above. Such questions have generally not yet been addressed in expert or public opinion studies.

**Psychological and public opinion findings.** Researchers of mental state attribution and related concepts like mind perception, anthropomorphism, theory of mind, and mentalizing have long considered whether people view artificial entities such as robots as capable of feeling and experiencing things like emotions (Thellman et al., 2022; Waytz et al., 2010; Salles et al., 2020). In these mind attribution studies, the features that constitute mind – for example, capacities like agency and experience – are chosen by the experimenter a priori and participants indicate how much they perceive these features in the technology at hand (usually robots).

Artificial entities like humanoid robots are more likely to be ascribed agency (the ability to plan and act) than experience (the ability to sense and feel), likely because of the perceived essential and unique human capacity for emotion (Gray et al., 2007; Waytz et al., 2010; Jacobs et al., 2022). This aligns with dehumanization research, which shows that mechanistic dehumanization (likening a human to a robot; Haslam, 2006) primarily stems from perceived deficiencies in emotional capacity (Haslam et al., 2008). Both technological and individual factors influence tendencies toward anthropomorphism and mind perception (Thellman et al., 2022). People are more likely to attribute mental states or to anthropomorphize when a robot exhibits human-like behavior, such as social interaction, emotional expression, and intelligence; has a more human-like appearance; or shares identity traits (e.g., gender, nationality, personality) with its human interactant (Broadbent et al., 2013; Eyssel et al., 2012; Hegel, 2012; Krach et al., 2008; Thellman et al., 2022). Some research shows that humanizing a robot more – by increasing its human-like appearance or using a friendlier personality – increases perceptions of experience (Broadbent et al., 2013; Cucciniello et al., 2023; Gray et al., 2007; Jacobs et al., 2022; MacDorman, 2024).

Those who are lonely are more likely to anthropomorphize (Epley et al., 2008), which has been theorized as a "sociality motivation" (Epley et al., 2007). "Effectance motivation" (Epley et al., 2007) – when people anticipate or explain a robot's behavior in order to make sense of their environment – also increases mind attribution (Eyssel et al., 2011; Eyssel & Kuchenbrandt, 2011; Perez-Osorio et al., 2019; Waytz et al., 2010; Złotowski et al., 2018). Compared to older adults, children and younger adults have a stronger tendency to anthropomorphize (Manzi et al., 2020; Okanda et al., 2021; Shen & Koyama, 2022; van den Berghe et al., 2020). Additionally, some studies suggest that those from Eastern countries (e.g., China and Japan) are more likely to attribute mind than are those from Western countries (e.g., United States, Germany, and Italy) (Bernotat & Eyssel, 2018; H. Tan et al., 2018; Trovato & Eyssel, 2017). Thus, there is some individual variation in mind perception tendencies, which are also contingent on the technology's appearance and capabilities.

### 6.3 | The moral consideration and governance of AI systems with subjective experience

**Expert assessments.** As with the empirical questions raised by the possibility of artificial minds, academic work on the moral and legal standing of artificial systems relative to their characteristics and mental capacities has also been contentious, covering a wide range of viewpoints (Asaro, 2007; Bryson, 2010; Coeckelbergh, 2010; Chopra & White, 2011; Darling, 2012; Gunkel, 2012; Miller, 2015; Schwitzgebel & Garza, 2015; Solaiman, 2017; Kurki, 2019; Danaher, 2020; Gordon & Pasvenskiene, 2021; Müller, 2021; Shevlin, 2021; Sebo & Long, 2025; Mamak, 2022; Mamak, 2023; Friedman, 2023; Guingrich & Graziano, 2024). At the same time, the number of related publications and research efforts has steadily been growing and accelerating since the early 2000s (Harris & Anthis, 2021; Malle, 2016).

Having a subjective experience is one of the most, if not the most, frequently suggested criteria for determining whether an artificial entity deserves moral consideration (Harris & Anthis, 2021; Gibert & Martin, 2021), but other potential criteria, such as agency and life, have also been suggested (Ladak, 2024). Some legal rights and protections have already been extended to entities that are not considered to have subjective experience, such as environmental entities or corporations (Gellers, 2020; Kotzmann et al., 2023). Experts generally appear divided about whether AI systems deserve moral consideration and, if so, what form this should take. Finding consensus is complicated by disagreement and uncertainty about key scientific issues, such as whether AI systems have or will ever have the capacity for subjective experience, and central ethical questions, such as these: Is the ability to have valenced experiences required for moral consideration? Is consciousness without such valenced experiences enough? Is agency without consciousness enough? Is a particular kind of attitude toward or relationship with AI enough?

A systematic literature review assigned a score to each of 192 articles according to the authors' support for or opposition to granting artificial entities moral consideration. It found that, overall, academic articles tended somewhat toward supporting the position that some moral standing may be warranted in the future (Harris & Anthis, 2021). While the authors found little difference in mean support between academic disciplines, the authors note that the average support expressed by AI technical researchers was lower than in other disciplines. In addition, the search methodology may have missed key works that are most dismissive and critical, as the authors note themselves.

Some argue for taking a precautionary approach by offering moral consideration to near-future AI systems on the basis that such systems may soon be able to instantiate various markers suggested by scientific theories of consciousness. If there is a non-negligible chance of an AI system having conscious, valenced, or otherwise morally significant experiences, it would be irresponsible, some argue, to ignore that possibility (Sebo & Long, 2025; Long et al., 2024). Others are much more reluctant to offer moral consideration to AI systems (Bryson, 2010; Birhane & Van Dijk, 2020), often due to thinking that it is not possible for AI systems to have mental capacities such as subjective experience that would merit moral consideration. Legal discussions on questions such as whether artificial entities should be granted legal personhood (Chopra & White, 2011), legal protections (Darling, 2012), and legal responsibilities (Asaro, 2007) have also been ongoing for over a decade, beginning far before recent progress in AI development.

**Expert opinion surveys.** Few expert surveys have looked at beliefs about the moral consideration of artificial entities, and most have focused on legal scholars. Surveys of legal academics have shown that respondents ascribed the lowest desired welfare protections to sentient AI systems, compared to other entities, showed low agreement that there is a reasonable legal basis for granting standing to them in the future, and strong disagreement that artificially intelligent beings include legal persons (Martínez & Winter, 2021a; Martínez & Tobia, 2023). This suggests that there is little recognized legal basis for the welfare protection of AI systems currently.

**Psychological and public opinion findings.** The rights and moral consideration of AI systems and robots are becoming a growing focus in human participant research (Banks & Bowman, 2023; Mays et al., 2025; Ladak et al., 2023), although much of this work has not focused on digital systems with subjective experience in particular. AI systems and robots can elicit empathy in humans and cause distress if they are being harmed (X. Z. Tan et al., 2018). However, robots and AI systems generally receive low moral consideration, lower than almost all other entities, including humans, non-human animals, and the environment, and there is only low to middling support for extending notions of personhood or the moral circle to digital systems with subjective experience (Martínez & Winter, 2021b; Rottman et al., 2021; Anthis et al., 2024). But there can be substantial variation between people in the levels of agreement elicited for the future moral consideration of artificial entities (Harris & Anthis, 2021), highlighting the scope of broad disagreement on such issues across society.

Psychological research has found that mind perception is intimately linked to attitudes toward moral standing, though this relationship is not necessarily straightforward (Gray et al., 2012; Keijsers et al., 2022; Waytz et al., 2010). Higher perceived levels of cognitive, experiential, and agentic mental capacities have been found in some studies to be associated with increased willingness to grant moral consideration and protection (Nijssen et al., 2019; Ladak et al., 2024). On the other hand, they can also be associated with higher attributions of blameworthiness and moral responsibility (Yam et al., 2022) as well as feelings of uncanniness (Gray & Wegner, 2012). Perceived affective and experiential capacities in particular appear to be associated with mind ascription and granting moral patiency (Koban & Banks, 2024; Tanibe et al., 2017; Wang & Krumhuber, 2018; Nijssen et al., 2019). This suggests that, as the performance and characteristics of AI systems advance, people may be increasingly willing to grant them moral consideration. However, such mind attributions may remain in tension with the different assumptions and heuristics we have in respect to artificial digital entities (Banks et al., 2021; Banks & Koban, 2023) and also may depend heavily on the kind of mental capacities exhibited and perceived.

In sum, there appears to be both expert and public opposition to treating artificial entities cruelly, some support for granting basic rights needed for upkeep, such as energy supply or software updates, and only very low support for granting artificial entities any kind of political rights, such as voting (de Graaf et al., 2021; de Graaf et al., 2022; Lima et al., 2020; Anthis et al., 2024). Support is more unilateral for artificial entities such as AI systems having a range of moral responsibilities, such as acting fairly and respecting others (Mays et al., 2025). Generally, these findings hold for current AI systems and robots, as well as desired levels of protections, rights, and responsibilities for future systems or systems that may be sentient, though especially for the latter categories there is only limited research. When tested, AI systems with subjective experience are granted more moral consideration and desired protection than are AI systems generally, but the increase is not as overwhelming as one may expect (Anthis et al., 2024). Few studies have looked at public support for a variety of political actions in relation to sentient AI systems (Anthis et al., 2024; Pauketat et al., 2022; Pauketat et al., 2023), but their results signal some willingness to ban the development of sentient AI systems and to take actions to protect AI systems that have subjective experience and could suffer.

## 7 | SI References